\documentclass[12pt]{article}  
\usepackage{epsfig}
\usepackage{wrapfig}
\usepackage{xspace}
\usepackage{cite}
\newcounter{enumct}
\newenvironment{Enumerate}{\begin{list}{\arabic{enumct}.}%
{\usecounter{enumct}\setlength{\topsep}{0.2mm}%
\setlength{\partopsep}{0.2mm}\setlength{\itemsep}{0.2mm}%
\setlength{\parsep}{0.2mm}}}{\end{list}}
\newcommand{\pz     }{\ensuremath{\phantom{-}}\xspace}
\newcommand{\qsq    }{\ensuremath{Q^{2}}\xspace}
\newcommand{\wsq    }{\ensuremath{W^{2}}\xspace}
\newcommand{\psq    }{\ensuremath{P^{2}}\xspace}
\newcommand{\ft     }{\ensuremath{F_{2}^{\gamma}}\xspace}
\newcommand{\ftxq   }{\ensuremath{\ft(x,\qsq)}\xspace}
\newcommand{\der    }{\ensuremath{{\mathrm d}}\xspace}
\newcommand{\kt     }{\ensuremath{k_{\mathrm{t}}}\xspace}
\newcommand{\ktsq   }{\ensuremath{k_{\mathrm{t}}^2}\xspace}
\newcommand{\ktsqm  }{\ensuremath{k_{\mathrm{t,max}}^2}\xspace}
\newcommand{\pt     }{\ensuremath{p_{\mathrm{t}}}\xspace}
\newcommand{\kn     }{\ensuremath{k_0}\xspace}
\newcommand{\nch    }{\ensuremath{N_{\mathrm{trk}}}\xspace}
\newcommand{\ptch   }{\ensuremath{p_{\mathrm{t,trk}}}\xspace}
\newcommand{\Wres   }{\ensuremath{W_{\mathrm{res}}}\xspace}
\newcommand{\etout  }{\ensuremath{E_{\mathrm{t,out}}}\xspace}
\newcommand{\flow   }{\ensuremath{1/\Nevt\cdot\der\,E/\der\,\eta}\xspace}
\newcommand{\Xdm    }{\ensuremath{\mathrm{X_{data}^{meas}}}\xspace}
\newcommand{\Xd     }{\ensuremath{\mathrm{X_{data}^{corr}}}\xspace}
\newcommand{\Xmm    }{\ensuremath{\mathrm{X_{MC}^{meas}}}\xspace}
\newcommand{\Xm     }{\ensuremath{\mathrm{X_{MC}}}\xspace}
\newcommand{\Nevt   }{\ensuremath{N}\xspace}
\newcommand{\Nex    }{\ensuremath{N_\mathrm{ex}}\xspace}
\newcommand{\dsdw   }{\ensuremath{\der\sigma/\der\Wres}\xspace}
\newcommand{\dsde   }{\ensuremath{\der\sigma/\der\etout}\xspace}
\newcommand{\dsdn   }{\ensuremath{\der\sigma/\der\nch}\xspace}
\newcommand{\dsdp   }{\ensuremath{\der\sigma/\der\ptch}\xspace}
\newcommand{\scale  }{\ensuremath{\sqrt{\chi^2/(\Nex - 1)}}\xspace}
\newcommand{\lumi   }{\ensuremath{{{\mathcal L}_{int}}}\xspace}
\newcommand{\ttag   }{\ensuremath{\theta_{\rm tag}}\xspace}
\begin{document}                           
\begin{titlepage}
\begin{center}{\large EUROPEAN ORGANISATION FOR NUCLEAR RESEARCH}
\end{center}\bigskip
\begin{flushright}
 CERN-EP-2000-109 \\ July 3, 2000
\end{flushright}\bigskip\bigskip\bigskip
\begin{center}
 {\huge\bf 
  Comparison of Deep Inelastic Electron-Photon Scattering Data
  with the Herwig and Phojet Monte Carlo Models
 }
\end{center}\bigskip\bigskip
\begin{center}
 {\LARGE ALEPH, L3 and OPAL Collaborations\footnote{For the members of 
  the Collaborations see for example~\cite{OPAL,L3C-9803L3C-9804,ALEPH}}.
 \\}
 {\LARGE The LEP Working Group for Two-Photon Physics\footnote{
  Contributions from 
  P.~Achard,
  V.~Andreev,
  S.~Braccini,
  M.~Chamizo,
  G.~Cowan,
  A. De Roeck,
  J.H.~Field,
  A.J.~Finch, 
  C.H.~Lin, 
  J.A.~Lauber,
  M.~Lehto, 
  M.N.~Kienzle-Focacci,
  D.J.~Miller,
  R.~Nisius,
  S.~Saremi,
  S.~S{\"o}ldner-Rembold,
  B.~Surrow,
  R.J.~Taylor,
  M.~Wadhwa
  and
  A.E.~Wright.  
 }}
\end{center}\bigskip\bigskip
\bigskip\begin{center}{\large  Abstract}\end{center}
 Deep inelastic electron-photon scattering is studied in the \qsq 
 range from 1.2 to 30~GeV$^2$ using the  
 LEP1 data taken with the ALEPH, L3 and OPAL detectors
 at centre-of-mass energies close to the mass of the Z boson.
 Distributions of the measured hadronic final state are corrected to the 
 hadron level and compared to the predictions of the HERWIG and PHOJET
 Monte Carlo models. 
 For large regions in most of the distributions studied
 the results of the different experiments agree with one another.
 However, significant differences are found between the data and the models.
 Therefore the combined LEP data serve as an 
 important input to improve on the Monte Carlo models.
 \bigskip\bigskip\bigskip\bigskip
 \begin{center}{\large Submitted to European Physical Journal C}\end{center}
\end{titlepage}
\section{Introduction}
 The measurement of the hadronic structure function \ft crucially depends on
 the accurate description of the hadronic final state by Monte Carlo models.
 The available models do not properly account for all features 
 observed in the data, and therefore, at present, the accuracy of the
 measurement of \ft is mainly limited by the imperfect description of the
 hadronic final state by the Monte Carlo models.
 In previous analyses of the individual LEP 
 experiments~\cite{OPAL,DELPHI,L3C-9803L3C-9804,ALEPH},
 it had been shown that there are discrepancies in several distributions 
 of the hadronic final state between the various QCD models and the data. 
 It has also be seen that the data are precise enough to further constrain
 the models.
 The purpose of this paper is to combine the ALEPH, L3 and OPAL data to
 establish a consistent and significant measurement, which can be used
 to optimise the models. 
 \par
 In this paper the reaction 
 ${\rm e}^+{\rm e}^-  \rightarrow {\rm e}^+{\rm e}^- \gamma^\star\gamma 
 \rightarrow {\rm e}^+{\rm e}^- $ hadrons, proceeding via the exchange of
 two photons, is studied in the single tag configuration, 
 where one scattered electron\footnote{Electrons and positrons are 
 referred to as electrons.} is detected. 
 The differential cross-section
 for the deep inelastic electron-photon scattering reaction, 
 $\rm e(k)\,\gamma(p)\rightarrow\rm e(k')\,\gamma(p)\,\gamma^*(q)
 \rightarrow\rm e\,X$, where the terms in brackets denote the four-momentum
 vectors of the particles, is proportional to 
 \ftxq~\cite{BER-8701}.
 Here $\qsq=-q^2=-(k-k')^2$ and $x=\qsq /2p\cdot q$.
 Experimentally, in the single tag configuration, the value of $x$ 
 is obtained using
%
 \begin{equation}
  x = \frac{\qsq}{\qsq+\wsq},
 \label{eqn:Xval}
 \end{equation}
%
 where \wsq is the hadronic invariant mass squared, and $\psq=-p^2$ is 
 neglected in calculating $x$.
 \par
 The hadronic structure function \ft receives contributions both from the
 point-like part and from the hadron-like part of the photon structure.
 The point-like part can be calculated in perturbative QCD.
 At low \qsq the hadron-like part is usually modelled
 based on the Vector Meson Dominance model.
 The combined contributions are evolved using the DGLAP evolution 
 equation.
 \par
 Combining the results of three of the LEP experiments 
 not only reduces the statistical errors compared to the individual results,
 but the difference between the results also gives
 a reliable estimate of the systematic, detector dependent, errors.
 The experimental data are fully corrected for trigger inefficiencies and 
 background has been subtracted.
 The experimental distributions presented are also corrected for detector
 effects  using different Monte Carlo models, and can directly be compared
 to the model predictions based on generated quantities only, i.e.,
 without the simulation of the detector response, provided that a well defined
 set of hadron level cuts defined below is applied.
 This experimental information serves as a basis for improvements 
 on the models.
 \par
 A set of variables is chosen to compare the corrected data to the hadron 
 level predictions of the Monte Carlo models.
 The variables used are:\\
 $\bullet$
 the reconstructed invariant hadronic mass, \Wres, 
 within a restricted range in polar angles with respect to the beam axis,\\
 $\bullet$
 the transverse energy out of the plane defined by the beam direction and the 
 direction of the tagged electron, \etout,\\
 $\bullet$
 the number of charged particles, \nch,\\
 $\bullet$
 the transverse momenta of charged particles with respect to the beam axis, 
 \ptch,\\
 $\bullet$
 and the hadronic energy flow, \flow, as a function of the pseudorapidity
 $\eta=-\ln(\tan(\theta/2))$ with respect to the beam axis, where \Nevt
 denotes the number of events.
 The complete definition of how these variables are calculated 
 is given in Section~\ref{sec:evsel}.
 \par
 For \Wres, \etout and \nch the result is a differential event 
 cross-section, which means, the distributions have one entry per event,
 whereas for \ptch the distribution is a one-particle inclusive
 cross-section.
 The hadronic energy flow is shown as an average energy flow per
 event, $\sum E/\Nevt$, where the sum runs over all objects and over all
 events in a given bin of pseudorapidity.
 \par
 The analysis presented here is based on data of the individual
 experiments taken at centre-of-mass energies close to the mass of 
 the $Z$ boson.
 The approximate ranges in $Q^2$ and $x$ of the different experiments used
 in this analysis, and the integrated luminosities are listed in 
 Table~\ref{tab:lumi}.
 The \qsq ranges are calculated from the requirements on the energy and angle
 of the  deeply inelastically scattered electron, and the $x$ ranges are
 derived from Eq.~\ref{eqn:Xval}, using the range in \qsq 
 and the approximate reach in hadronic invariant mass of $3<W<35$~GeV.
%
\begin{table}[h]
\begin{center}
\begin{tabular}{|c|c|c|c|} \hline
             & ALEPH           & L3               & OPAL            \\ 
\hline
 $Q^2$ range & 1.2--30 GeV$^2$ & 1.2--6.3 GeV$^2$ & 1.2--30 GeV$^2$ \\
 $x$ range   & 0.001--0.77     & 0.001--0.4       & 0.001--0.77     \\
  \lumi      & 144 pb$^{-1}$   & 140 pb$^{-1}$    & 87 pb$^{-1}$    \\
\hline
\end{tabular}
\caption{
         The approximate ranges in $Q^2$ and $x$ of the different 
         experiments and the integrated luminosities \lumi used in this
         analysis.
        }
\label{tab:lumi}
\end{center}
\end{table}
%
%
\section{Monte Carlo Models}
\label{sec:MC}
 The two Monte Carlo models studied are
 HERWIG5.9~\cite{Herwig} and PHOJET 1.05c~\cite{Phojet},
 using the leading order GRV~\cite{GRV} parton distribution functions.
 \par
 In the HERWIG model the hard interaction is simulated as
 $\rm{e\, q} \rightarrow \rm {e\, q}$ scattering, where the incoming quark
 is generated according to a set of parton distribution functions
 of the photon.
 The incoming quark is subject to an initial state parton shower
 which contains the $\gamma\rightarrow\rm{q\, q}$ vertex, and
 the outgoing partons undergo final state parton showers as in the
 case of ${\rm e}^+{\rm e}^-$ annihilations.
 The hadronisation is based on the cluster model.
 The initial state parton shower is designed in such a way that the hardest
 emission is matched to the sum of the matrix elements for the
 resolved processes, $\rm{g}\rightarrow\rm{q\,q}$, 
 $\rm{q}\rightarrow\rm{q\,g}$ and
 the point-like $\gamma\rightarrow\rm{q\,q}$ process.
 The parton shower uses an angular evolution parameter, and so it 
 obeys angular ordering.
 For point-like events the transverse momentum of the partons
 with respect to the direction of the incoming photon is given by 
 perturbation theory.
 In contrast, for hadron-like events, the photon remnant gets a transverse
 momentum \kt with respect to the direction of the incoming photon, where
 originally the transverse momentum was generated from a Gaussian
 distribution.
 \par
 The HERWIG version used for event simulation is HERWIG5.9+\kt, where the 
 label \kt denotes that the \kt distribution has been altered from the 
 program default.
 This change is made to improve the agreement for the high-\qsq region
 between the ALEPH and OPAL data and the original HERWIG prediction,
 where the original program is denoted as HERWIG default.
 The default Gaussian behaviour is replaced by a power-law function of 
 the form $\der\ktsq/(\ktsq+\kn^2)$~\cite{ktdyn} with 
 $\kn=0.66$~GeV.
 The change is motivated by the observation made in photoproduction studies
 at HERA~\cite{Zeus-B354}, where the power-law function gave a better 
 description of the data.
 This is a good example of how information from two different, but related
 reactions can be used to improve on a general purpose Monte Carlo program.
 It is interesting to note that the same value of \kn is chosen to describe
 both, photoproduction and deep inelastic electron-photon scattering events.
 The upper limit of \ktsq in HERWIG+\kt is fixed at $\ktsqm=25$~GeV$^2$,
 which is almost the upper limit of \qsq for the \qsq region studied, as 
 shown in Table~\ref{tab:lumi}.
 \par
 Based on the distributions presented in Section~\ref{sec:cor} the fixed
 cut-off has been replaced by dynamically adjusting the upper limit on \kt
 on an event by event basis~\cite{MikePrivatComm}.
 In this scheme the maximum transverse momentum \ktsqm is set to 
 the hardest virtuality scale in the event, which is of order \qsq.
 The distributions produced with this procedure are denoted as
 HERWIG+\kt(dyn).
 \par
 The PHOJET Monte Carlo is based on the Dual Parton Model~\cite{CAP-9401}.
 It is designed for hadron-hadron, photon-hadron and photon-photon
 collisions, where originally only real or quasi-real photons were considered.
 It has recently been extended to simulate the deep inelastic electron-photon
 scattering case, where one of the photons is highly virtual.
 For the case of deep inelastic scattering the program is not based on
 the DIS formula using \ft,
 but the $\gamma^\star\gamma$ cross-section is calculated from
 the $\gamma\gamma$ cross-section by extrapolating in \qsq on the basis
 of the Generalised Vector Dominance model. 
 The events are generated from soft and hard partonic  processes, where a
 cut-off of 2.5~GeV on the transverse momentum of the scattered partons in the
 photon-photon centre-of-mass system is used to separate the two classes
 of events.
 The hard processes are sub-divided into direct processes, where the photon as
 a whole takes part in the hard interactions, and resolved processes.
 In resolved processes either one or both photons fluctuate into a hadronic
 state, and a quark or gluon of one hadronic state interacts either with 
 the other photon, or with a quark or gluon of the second hadronic state.
 Also virtual photons can interact as resolved states, however, the parton
 distribution functions of the photons are suppressed as a function of
 the photon virtualities.
 The sum of the processes is matched to the deep inelastic scattering
 cross-section.
 Initial state parton showers are simulated with a backward evolution
 algorithm using the transverse momentum as evolution scale. 
 Final state parton showers are generated with the Lund code
 JETSET~\cite{SJO-9401}. 
 Both satisfy angular ordering implied by coherence effects.
 The hadronisation is based on the Lund string model as implemented
 in JETSET.
%
%
\section{Experimental Method}
\label{sec:evsel}
 Large data sets are generated with the PHOJET and HERWIG+\kt
 programs respectively, for $\sqrt{s}=M_{\rm Z}$. 
 All unstable particles with lifetimes of less than 1~ns are allowed
 to decay in the event generation.
 In this way the particles of the final state correspond
 approximately to those actually seen in the detectors. 
 The corresponding integrated luminosities for the PHOJET and HERWIG+\kt
 samples are 831 pb$^{-1}$ and 683 pb$^{-1}$ 
 respectively\footnote{The hadronisation parameters used for the
 PHOJET and HERWIG+\kt simulation have been determined from 
 hadronic decays of the $Z$ boson by the L3~\cite{L3had} 
 and OPAL~\cite{OPALhad} collaborations respectively.}.
 \par
\begin{figure}[htb]
\begin{center}
\epsfig{file=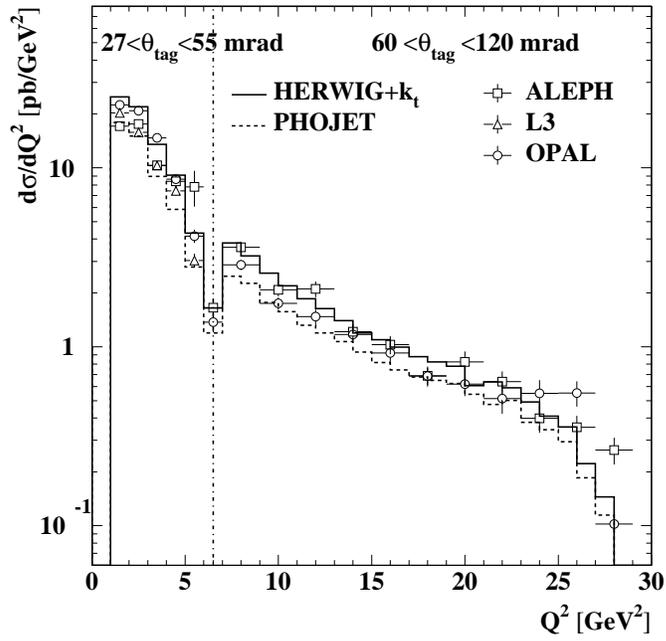,,width=0.6\linewidth}
\caption{\label{fig:q2} 
         The differential cross-section, $\der\sigma/\der\qsq$,
         observed in the two ranges 
         in scattering angles \ttag studied, compared to the predictions
         of the HERWIG+\kt and PHOJET models. The cross-sections are
         given in the kinematical ranges described in the text.
         The errors are statistical only.
        }
\end{center}
\end{figure}

 The definitions of the phase space and observables include cuts at
 generator level both on the events and on the particles within the
 events.
 The cuts are chosen such that the individual detectors have good 
 acceptance and therefore detector related uncertainties are expected
 to be small.
 To select the events at this stage, the following cuts are 
 applied to the generated hadron level quantities:
%
 \begin{Enumerate}
  \item The energy of the deeply inelastically scattered electron has
        to be larger than 35~GeV.
  \item The polar angle \ttag of the  deeply inelastically scattered electron
        with respect to either beam direction
        has to be in the ranges $27 - 55$~mrad (low-\qsq region) or
        $60 - 120$~mrad (high-\qsq region).
        The two ranges in scattering angles studied 
        correspond to \qsq ranges of about $1.2<\qsq<6.3$~GeV$^2$ and 
        $5.7<\qsq<30$~GeV$^2$.
  \item The events are required to contain no electron with 
        energy of more than 35~GeV and polar angle above 25~mrad 
        with respect to the beam direction in the hemisphere opposite 
        to the one containing the deeply inelastically scattered electron.
  \item The number of charged particles \nch, calculated by summing over all
        charged particles which have a transverse momentum
        \pt with respect to the beam axis of more than $200$~MeV and polar 
        angles, $\theta$, with respect to the beam axis, 
        in the range $20 < \theta < 160^\circ$,
        has to be greater than or equal to 3. 
        The values chosen closely resemble the acceptance of the tracking 
        detectors.
  \item The invariant mass \Wres, calculated by summing over all charged and 
        neutral particles
        fulfilling $\pt>200$~MeV and $20 < \theta < 160^\circ$,
        corresponding to $|\eta|<1.735$, has to be larger than 3~GeV.
 \end{Enumerate}
%
 This set of particles and cuts defines the {\it hadron level}
 and the data are corrected to this level.
 The hadron level predictions of \Wres and \etout are calculated from 
 the charged and neutral particles defined above, and the 
 distributions of \nch and \ptch from the charged particles alone.
 The only exception is the hadronic energy flow.
 For the hadronic energy flow, the same event selection has been applied,
 but all particles are included in the distribution without applying a
 cut on transverse momentum.
 Figure~\ref{fig:q2} shows the differential cross-section
 $\der\sigma/\der\qsq$ within the cuts listed above and corrected for 
 detector effects.
 The vertical line roughly separates the low-\qsq and high-\qsq regions.
 Since \qsq depends on the energy and angle of the deeply inelastically
 scattered electron there is a slight overlap in \qsq, but due
 to the cut on \ttag the two samples are statistically independent.
 In the kinematic region studied the cross-section prediction 
 of HERWIG+\kt is about 40$\%$ higher than the prediction
 based on PHOJET as shown in Figure~\ref{fig:q2}.
 \par
 To study the experimentally observable distributions at the detector level
 samples of 60k HERWIG+\kt and 120k PHOJET events, which are statistically 
 independent from the samples mentioned above, are passed through 
 the detector simulation programs of the ALEPH, L3 and OPAL 
 collaborations, ensuring that all experiments use identical events. 
 The objects reconstructed after the detector simulation are energy 
 clusters, measured in the electromagnetic and hadronic calorimeters, 
 and tracks, measured in the tracking devices.
 Identical event selection cuts at the detector level are applied by all
 experiments, closely reflecting the cuts applied at the hadron level
 as described above: 
%
 \begin{Enumerate}
 \item A cluster of at least 35~GeV is required in one of the 
       small angle electromagnetic luminosity monitors.
 \item The polar angle with respect to either beam direction
       of the cluster has to be in the range from 
       $27-55$~mrad or $60-120$~mrad.
 \item The most energetic cluster in the hemisphere opposite to the
       tagged electron has to have energy less than 35~GeV.
 \item At least 3 tracks, fulfilling a set of quality criteria,
       are required to be observed in the tracking devices
       with $20^\circ < \theta < 160^\circ$, and with \ptch of at
       least 200~MeV.
 \item The invariant mass, \Wres, calculated from
       all tracks and clusters with $\pt>200$~MeV and
       $20<\theta<160^\circ$ is required to be greater than 3~GeV.
 \end{Enumerate}
%
 These objects define the {\it detector level}. They are used 
 for the \Wres, \etout, \nch and \ptch distributions.
 The only exception is the energy flow \flow, where again no cut 
 on the transverse momentum has been applied.
 \par
 With this strategy it is ensured that the hadron level distributions,
 which are obtained from the large size samples without detector simulation,
 and the detector level distributions,
 which are obtained from the samples of smaller size with a detailed 
 detector simulation for each individual experiment,
 are statistically independent.
 Both samples are used in the correction procedure applied to the data
 described in the next section.
%
%
\section{Corrections for Detector Effects}
\begin{figure}[htb]
\begin{center}
\epsfig{file=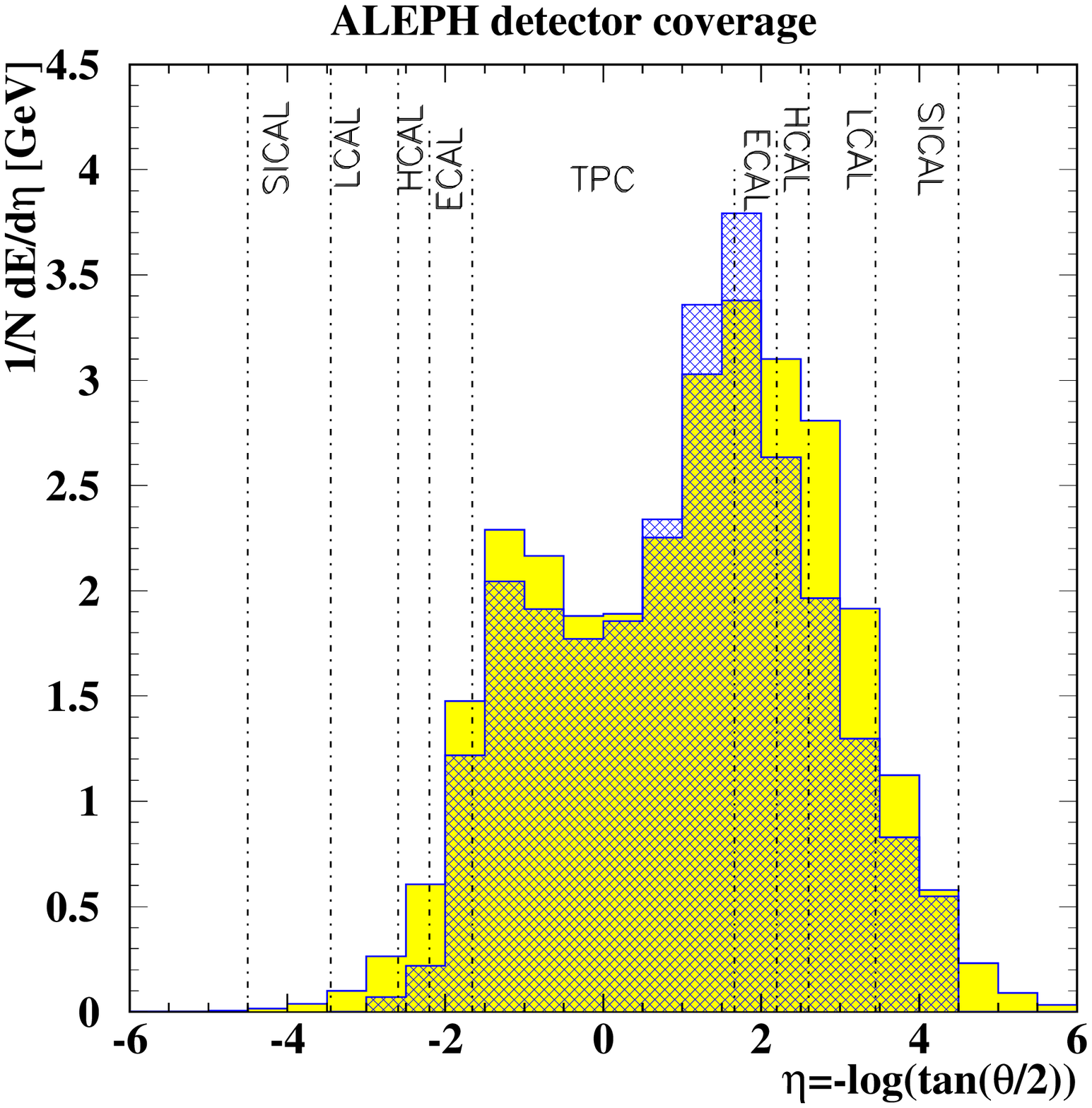,width=0.49\linewidth}
\epsfig{file=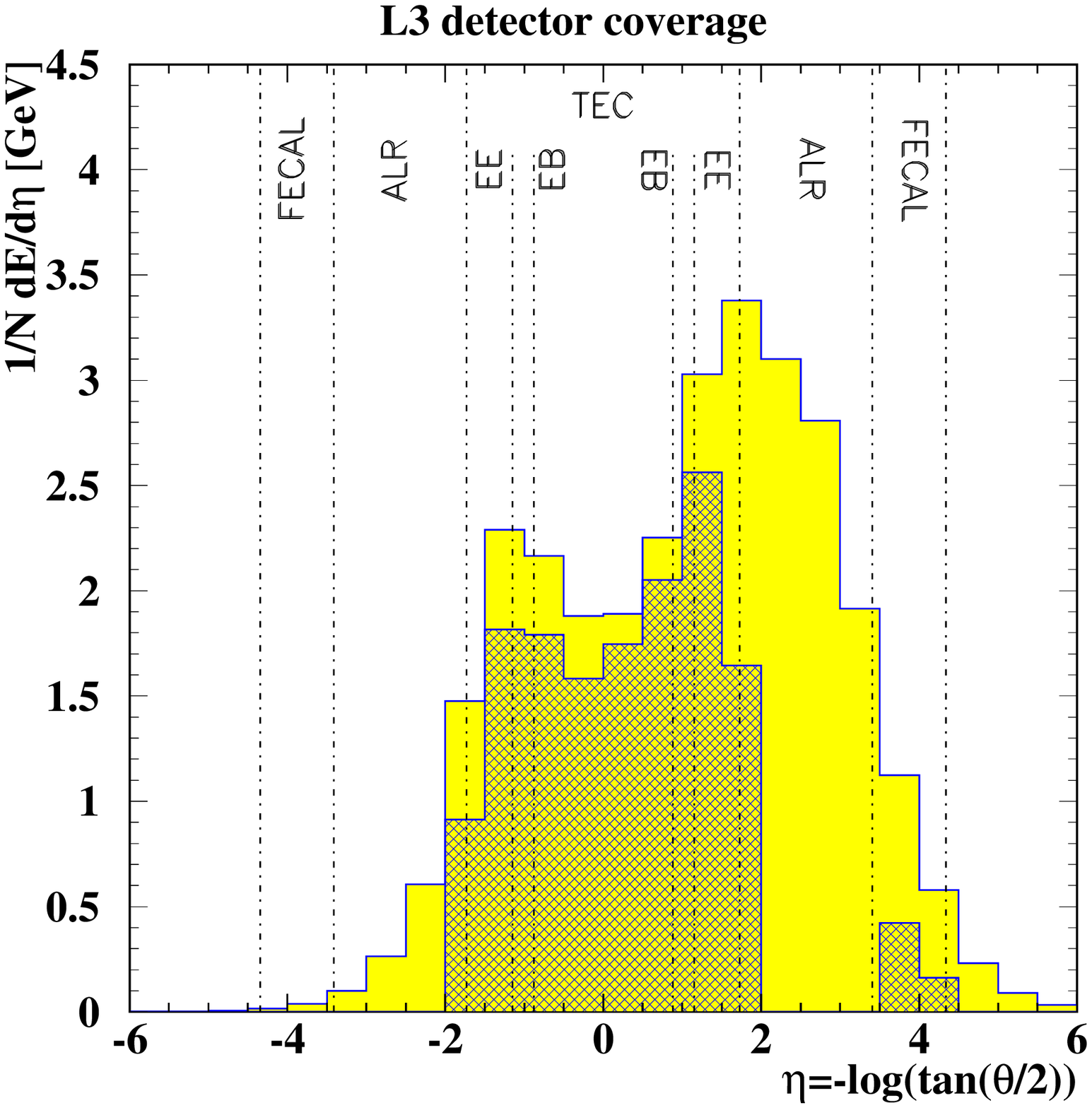,width=0.49\linewidth}
\epsfig{file=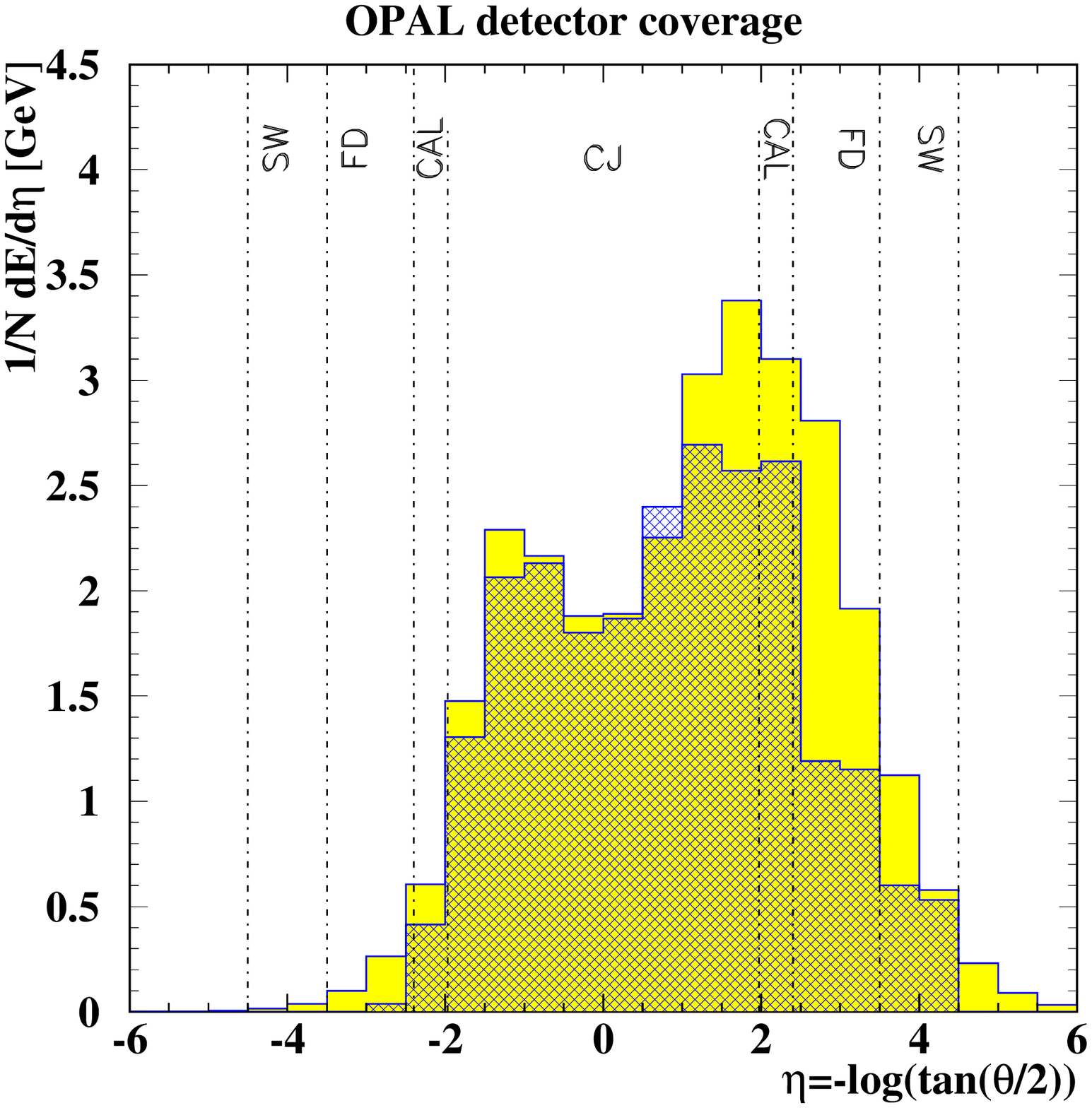,width=0.49\linewidth}
\epsfig{file=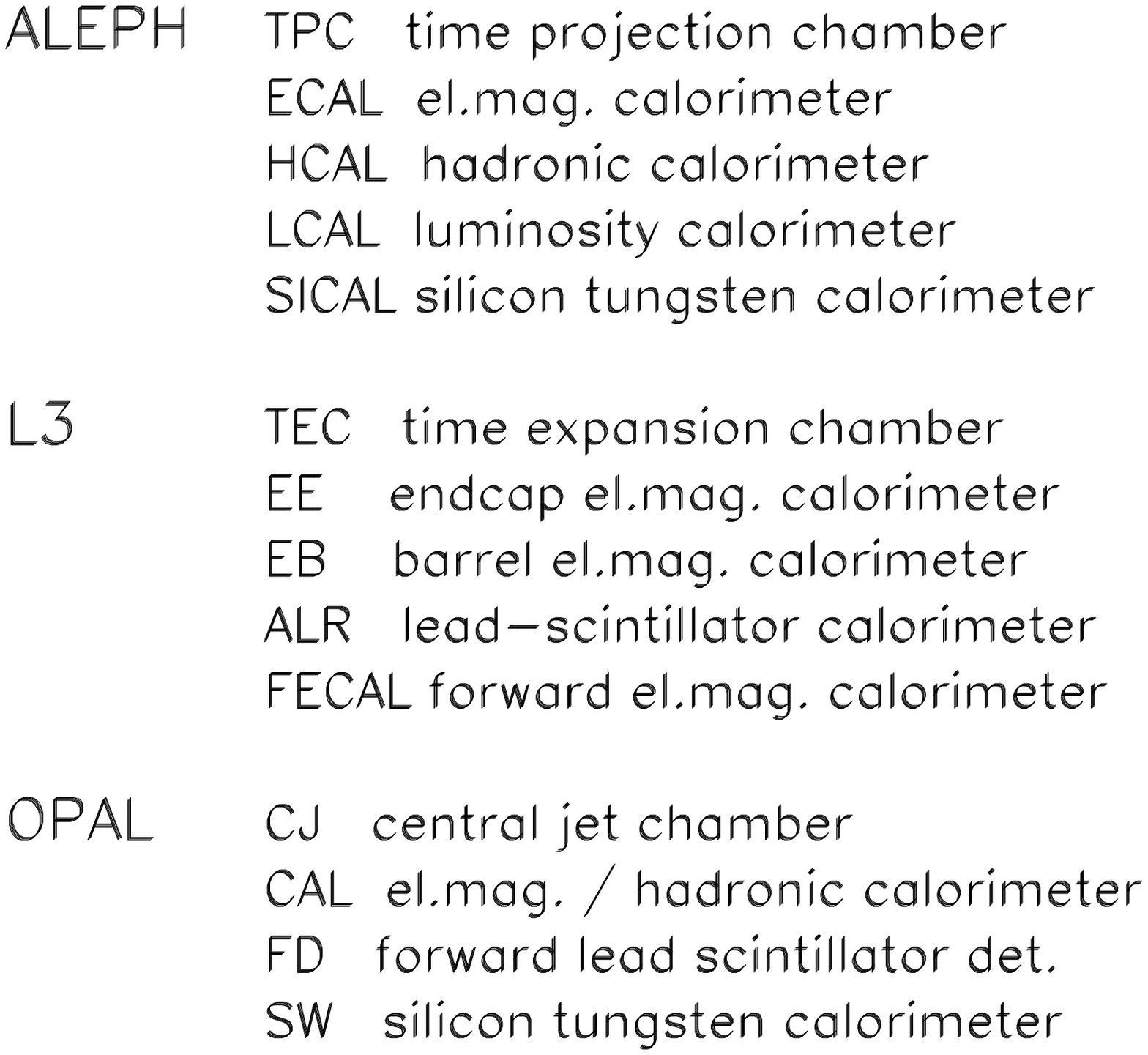,width=0.49\linewidth}
\caption{\label{fig:dets} 
         The HERWIG+\kt energy flow, \flow,  
         at the hadron level (lightly shaded) as well as on
         detector level (darkly shaded), as measured by the three detectors.
         The coverage of the different detector components are indicated
         by dot-dashed lines.
         In the case of L3 the ALR is not used in this analysis.
         }
\end{center}
\end{figure}

 Before a measured quantity can be compared to theoretical predictions or
 to the results of other measurements it must first be corrected for various
 detector related effects, such as geometrical acceptance, detector
 inefficiency and resolution, decays, particle interactions with the
 material of the detector and the effects of the event and track selections.
 Figure~\ref{fig:dets} shows the energy flow, \flow, predicted by the
 HERWIG+\kt model at the hadron level as well as at the detector level,
 for the three detectors.
 The events are entered in the figure such that the deeply inelastically 
 scattered electron is always at negative rapidities, but not shown.
 Also shown in the figure is the coverage in $\eta$ of the various 
 sub-detectors used in this analysis. 
 A detailed description of the ALEPH, L3 and OPAL detectors can be found
 in~\cite{ADET},~\cite{LDET}, and~\cite{ODET} respectively.
 The central region of all the detectors with $-1.735<\eta<1.735$
 is covered by tracking and calorimetry.
 The forward $\eta>1.735$, and backward $\eta<-1.735$ regions are 
 covered by electromagnetic calorimeters for luminosity measurements and 
 typically extend out to $\eta=\pm 4.3$.
 The electromagnetic calorimeter of L3 between $1.735<|\eta|<3.411$
 is not used in the present analysis.
 \par
 For the analysis presented here, the correction of the data to the hadron
 level is done with multiplicative factors, $f$, relating the measured
 value \Xdm of a  quantity X, such as a bin content, to the corrected value,
 \Xd using the relation: 
%
 \begin{eqnarray}
 \Xd & = & \Xdm \cdot f =  \Xdm \cdot \frac{\Xm}{\Xmm} \,\, .
 \label{eqn:cor}
 \end{eqnarray}
%
 For distributions, the correction factors are computed bin by bin,
 e.g. for the energy flow, $f$ is the ratio of the hadron
 level (lightly shaded) and the detector level (darkly shaded)
 distributions shown in Figure~\ref{fig:dets}.
 In this way of correcting the data the assumption is made that
 within the restricted angular range there is little smearing of the
 variables between bins, hence a simple correction factor is justified,
 and therefore no attempt to use an unfolding procedure has been made. 
 The Monte Carlo was used to verify the accuracy of this assumption.
 Application of this correction results in measurements corrected to a 
 well-defined kinematical region and particle composition,  
 as defined in Section~\ref{sec:evsel}. 
 \par
\begin{figure}[htb]
\begin{center}
\epsfig{file=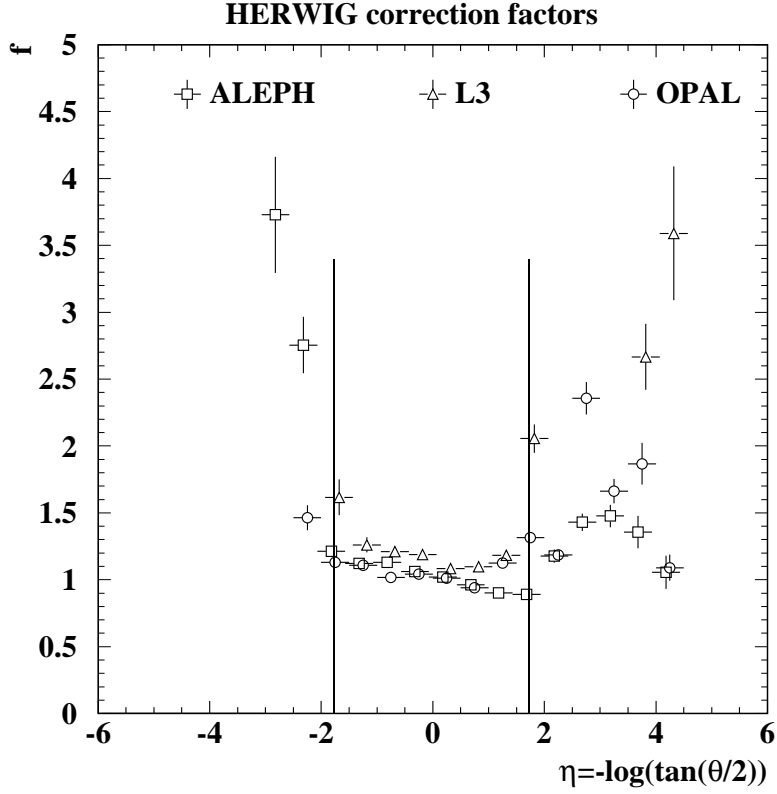,width=0.70\linewidth}
\epsfig{file=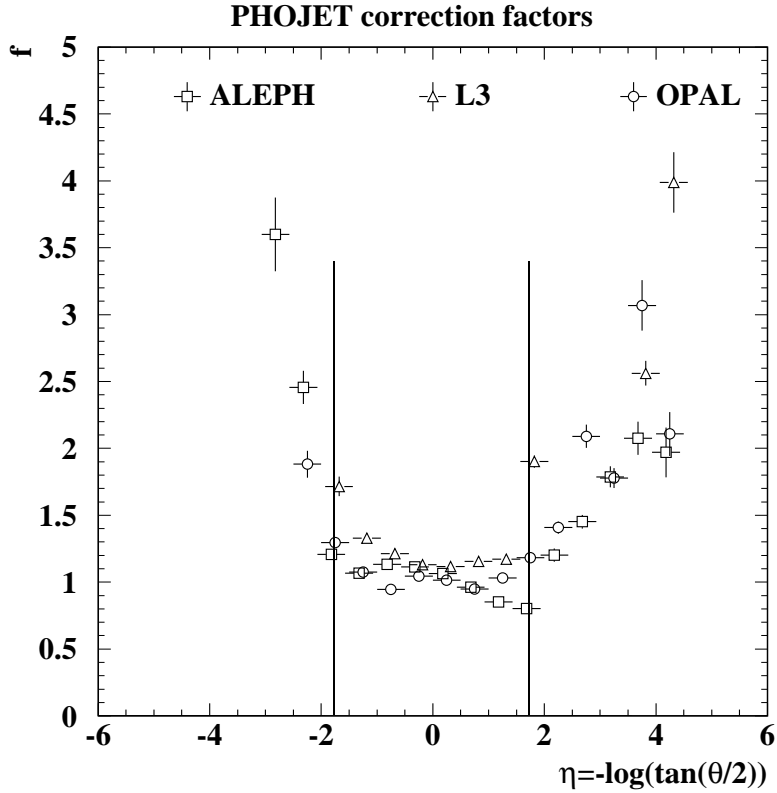,width=0.70\linewidth}
\caption{\label{fig:corflow} 
         The HERWIG+\kt and PHOJET correction factors, $f$, for the  
         ALEPH, L3 and OPAL energy flow for the low-\qsq region.        
         The symbols are slightly displaced for better visibility.
         The vertical lines indicate the central rapidity 
         region, $|\eta|<1.735$.
        }
\end{center}
\end{figure}

 The correction factors for the energy flow in the low-\qsq region
 are shown in Figure~\ref{fig:corflow}.  
 The correction factors are near one in the central region
 of pseudorapidity where identical cuts have  been applied
 and they are similar for the three experiments.
 However, there is a much larger spread in the region of larger 
 pseudorapidity $\eta>1.735$, where the experiments have different 
 sub-detectors, different angular coverage, and apply different cuts.
 For ALEPH and OPAL the clusters in the forward detectors are required to have
 an energy of at least 1~GeV, while for L3 this requirement is at least 4~GeV. 
 This leads to larger correction factors for L3 in that region.  
 In addition, for L3, the clusters measured in the forward detectors 
 on the side of the tagged electron, i.e. at $\eta<-2$, are not 
 considered in this study.  
 Another difference in the treatment of clusters measured in the forward
 detectors is that OPAL uses a correction function, obtained from Monte Carlo 
 studies of the detector response to hadrons, to correct for losses 
 in the measurement of the hadronic energy in the forward detector, thereby
 also reducing the correction factors.
 It should be stressed that for most of the distributions used for the 
 comparisons in the following section, only the central detector part 
 for polar angles with respect to the beam axis in the range 
 $20<\theta<160^\circ$, that is $|\eta|<1.735$, is exploited. 
 The only exception is the energy flow where no cuts on the angle are applied. 
 \par
\begin{figure}[htb]
\begin{center}
\epsfig{file=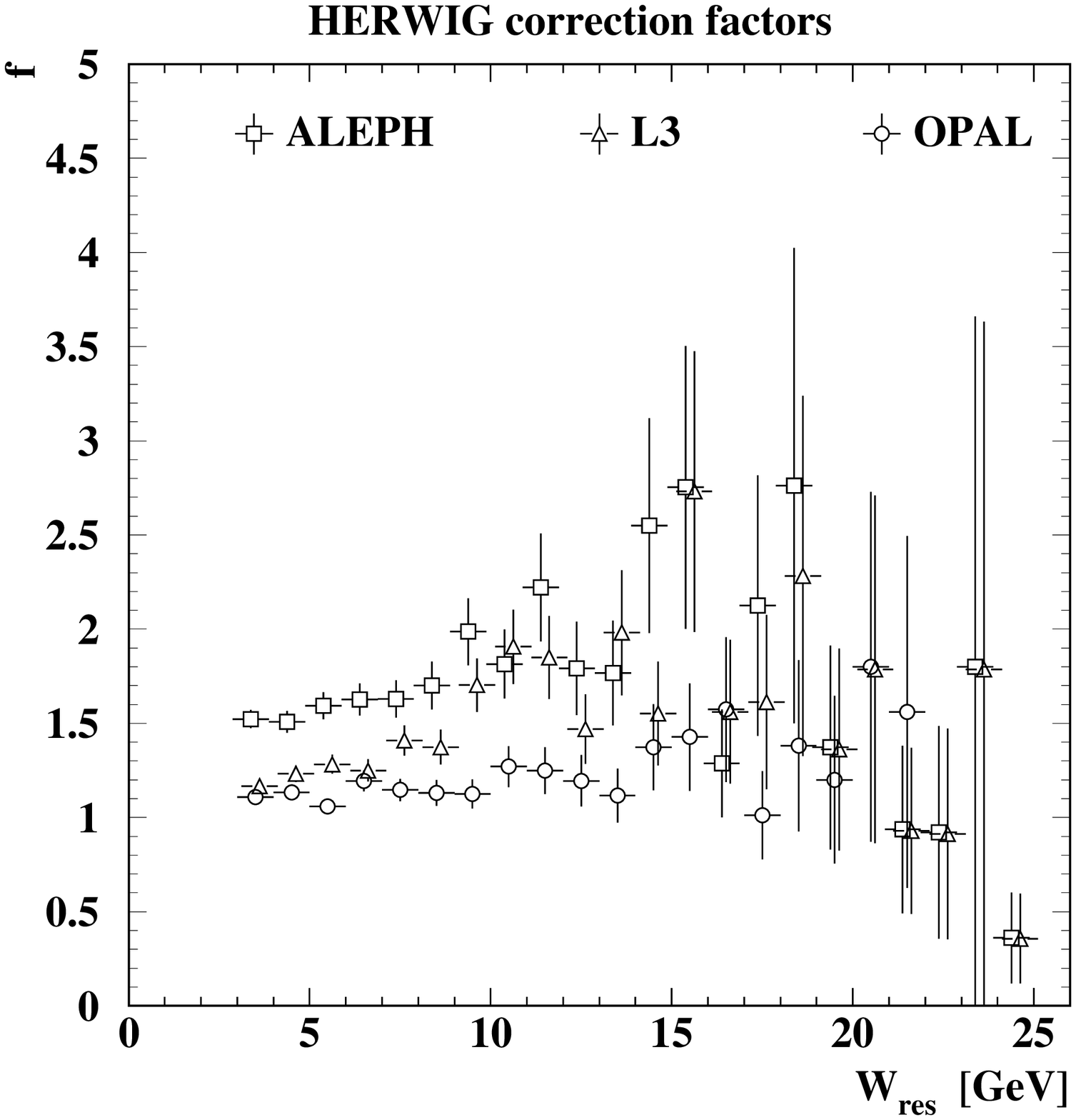,width=0.49\linewidth}
\epsfig{file=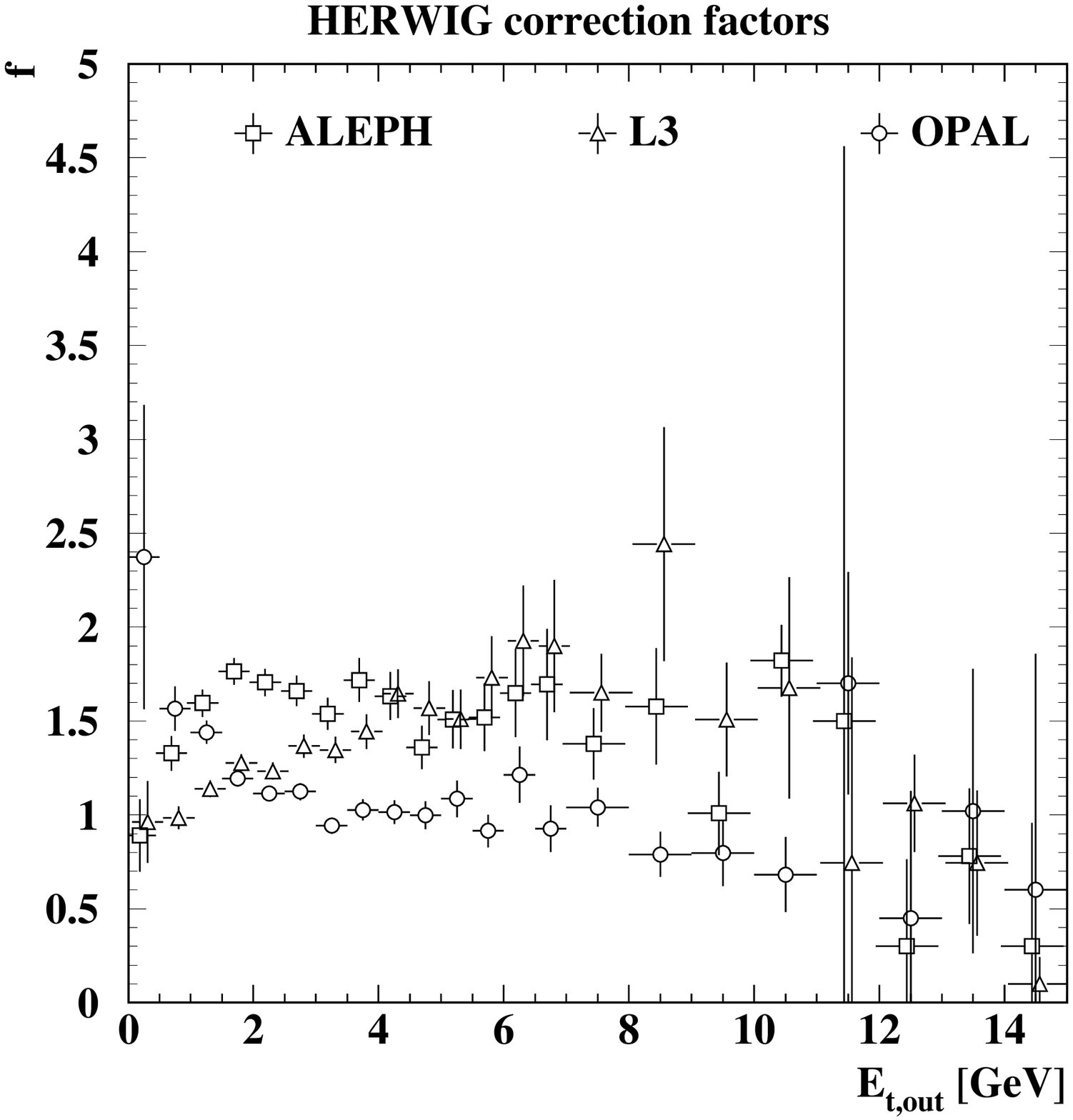,width=0.49\linewidth}
\epsfig{file=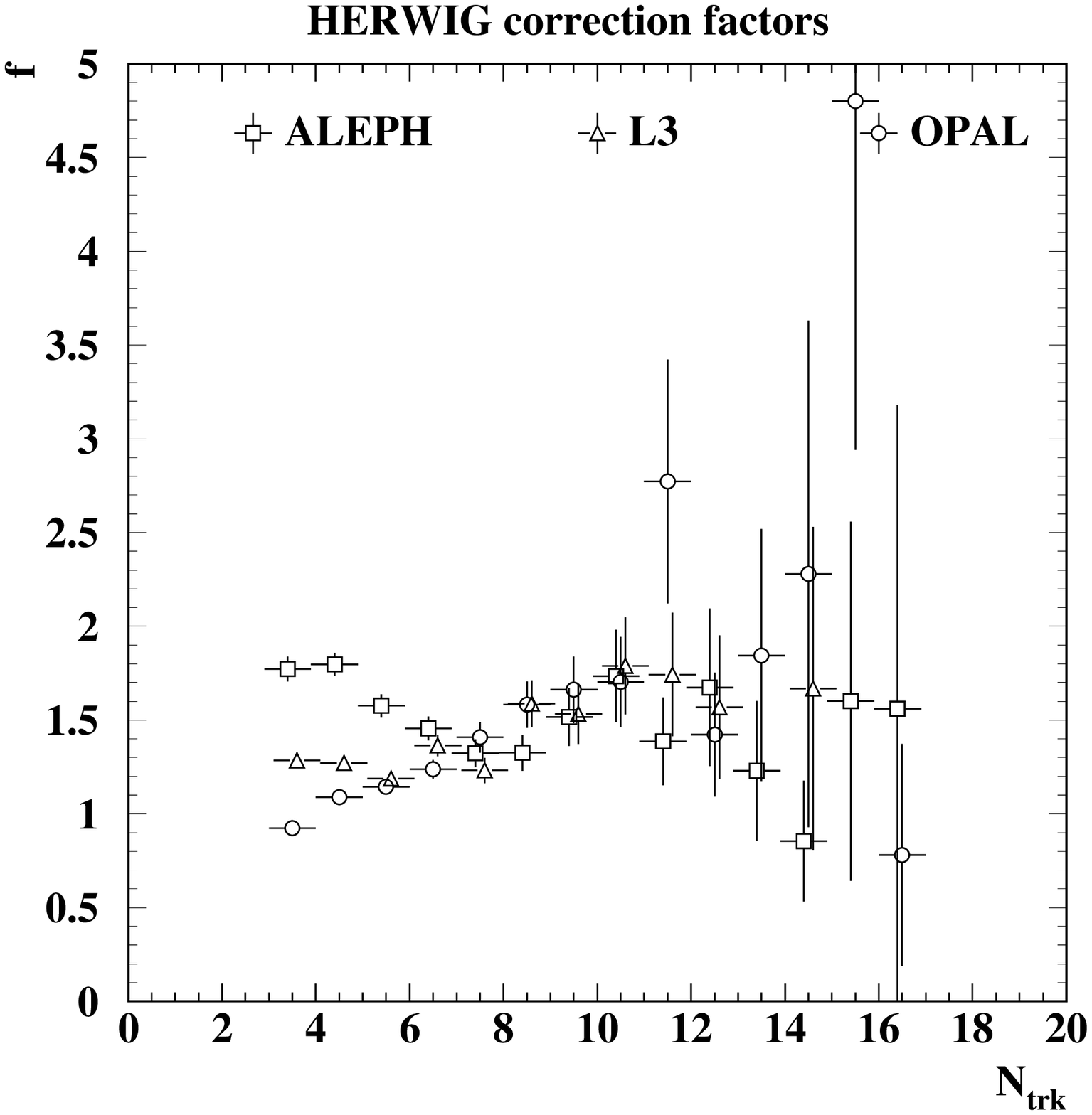,width=0.49\linewidth}
\epsfig{file=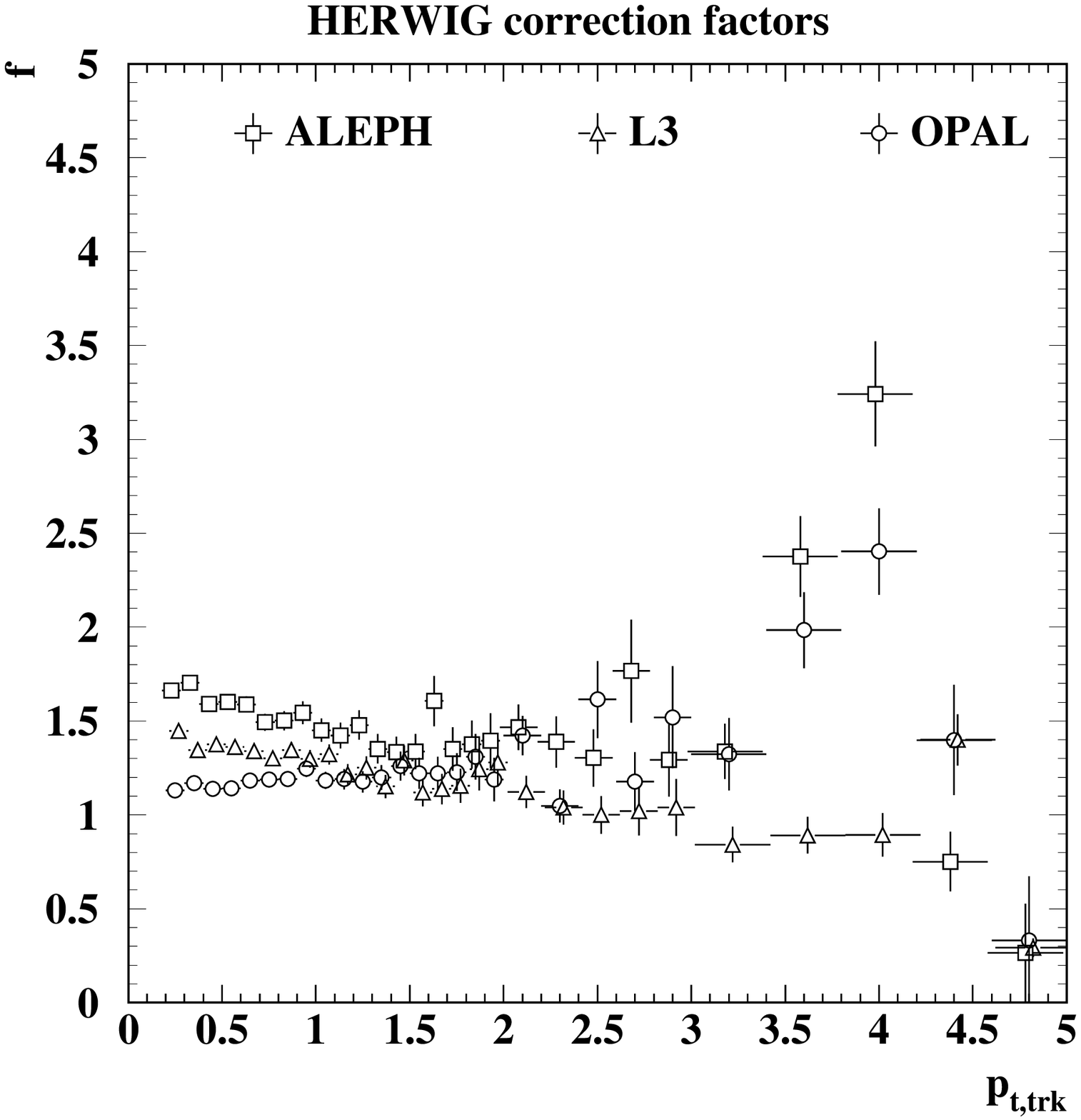,width=0.49\linewidth}
\caption{\label{fig:cor1} 
         The HERWIG+\kt correction factors, $f$, for the  ALEPH, L3 and 
         OPAL \Wres, \etout, \nch and \ptch distributions for the 
         low-\qsq region.
         The symbols are slightly displaced for better visibility.
        }
\end{center}
\end{figure}
\begin{figure}[htb]
\begin{center}
\epsfig{file=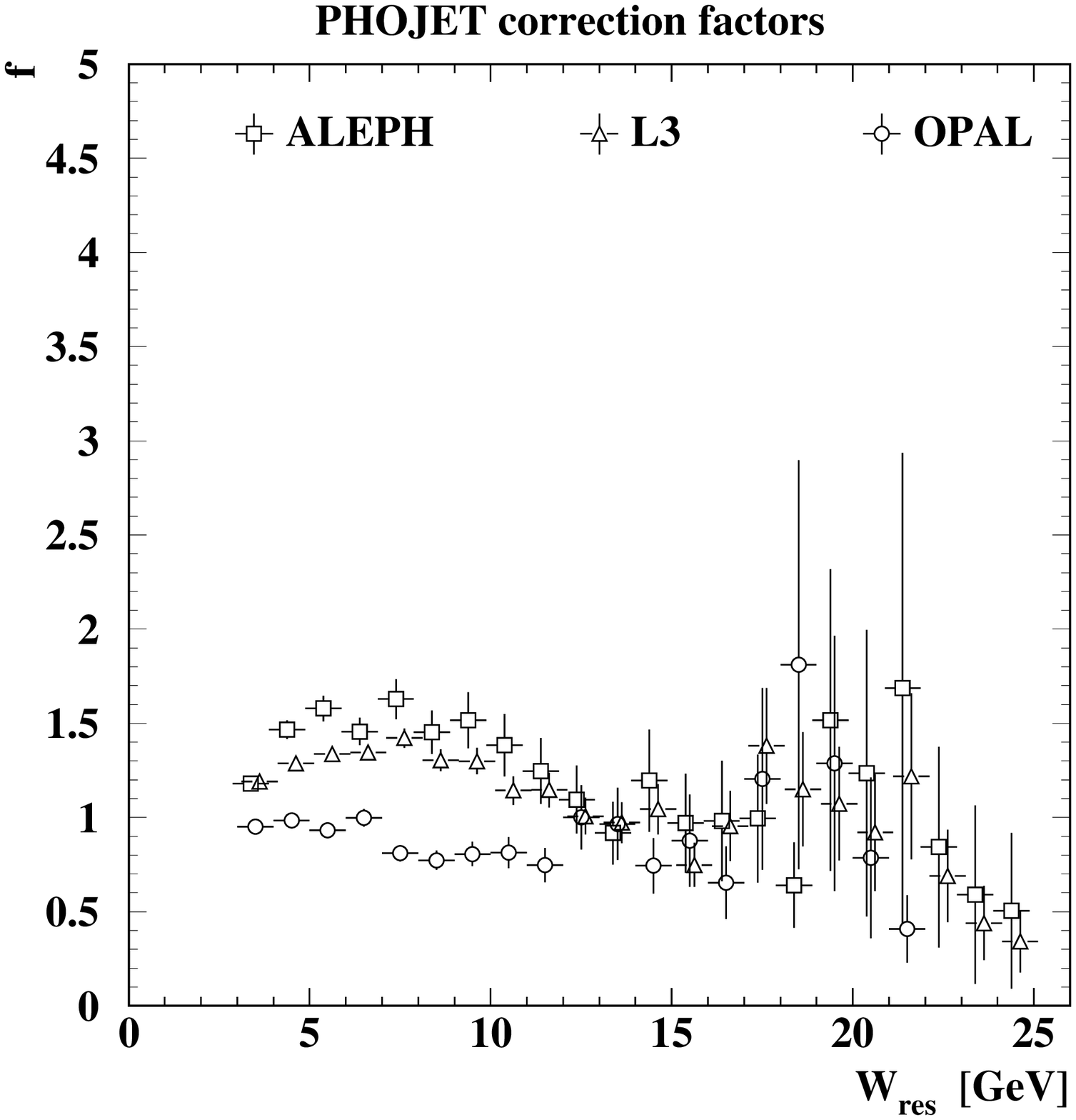,width=0.49\linewidth}
\epsfig{file=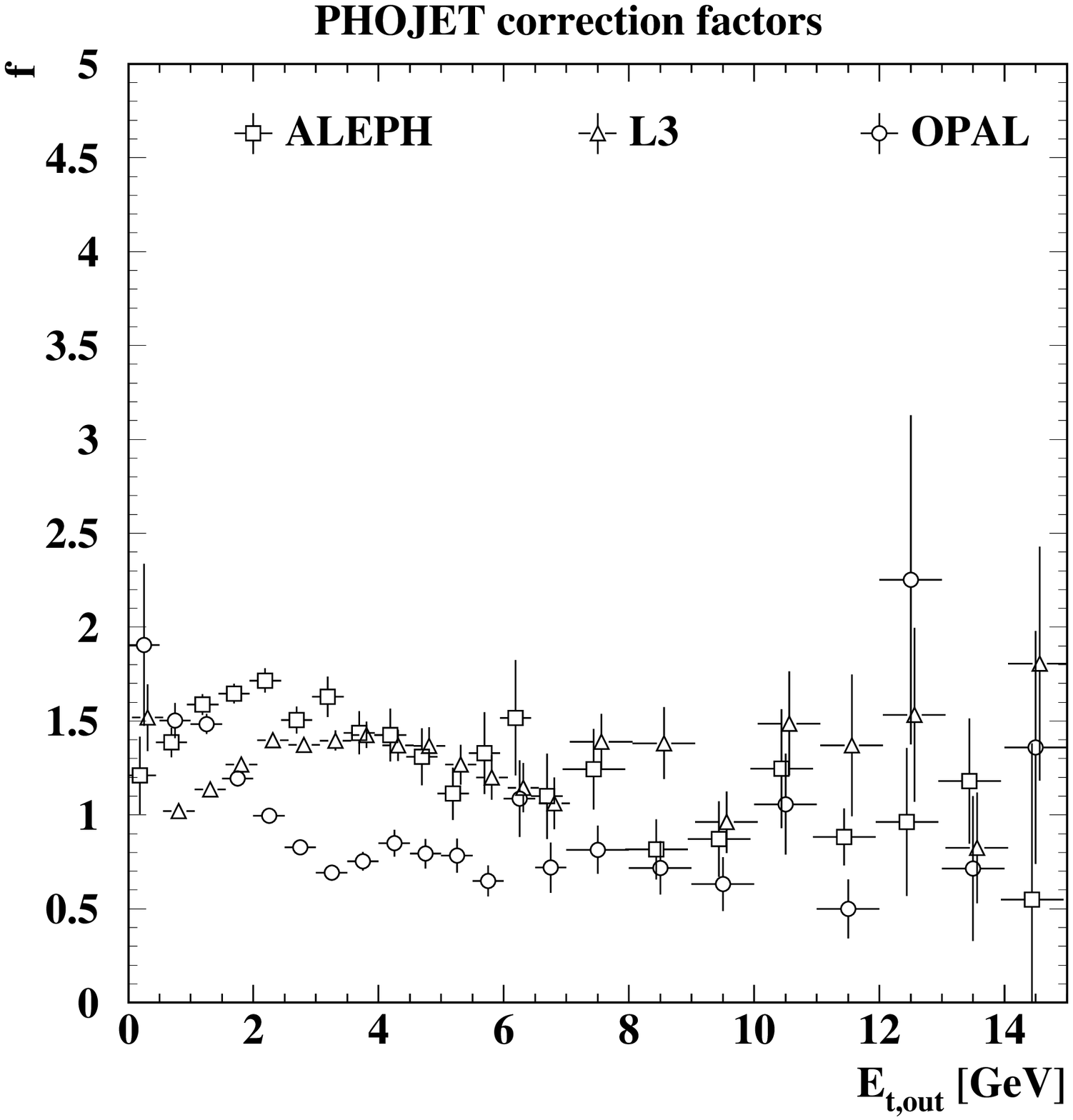,width=0.49\linewidth}
\epsfig{file=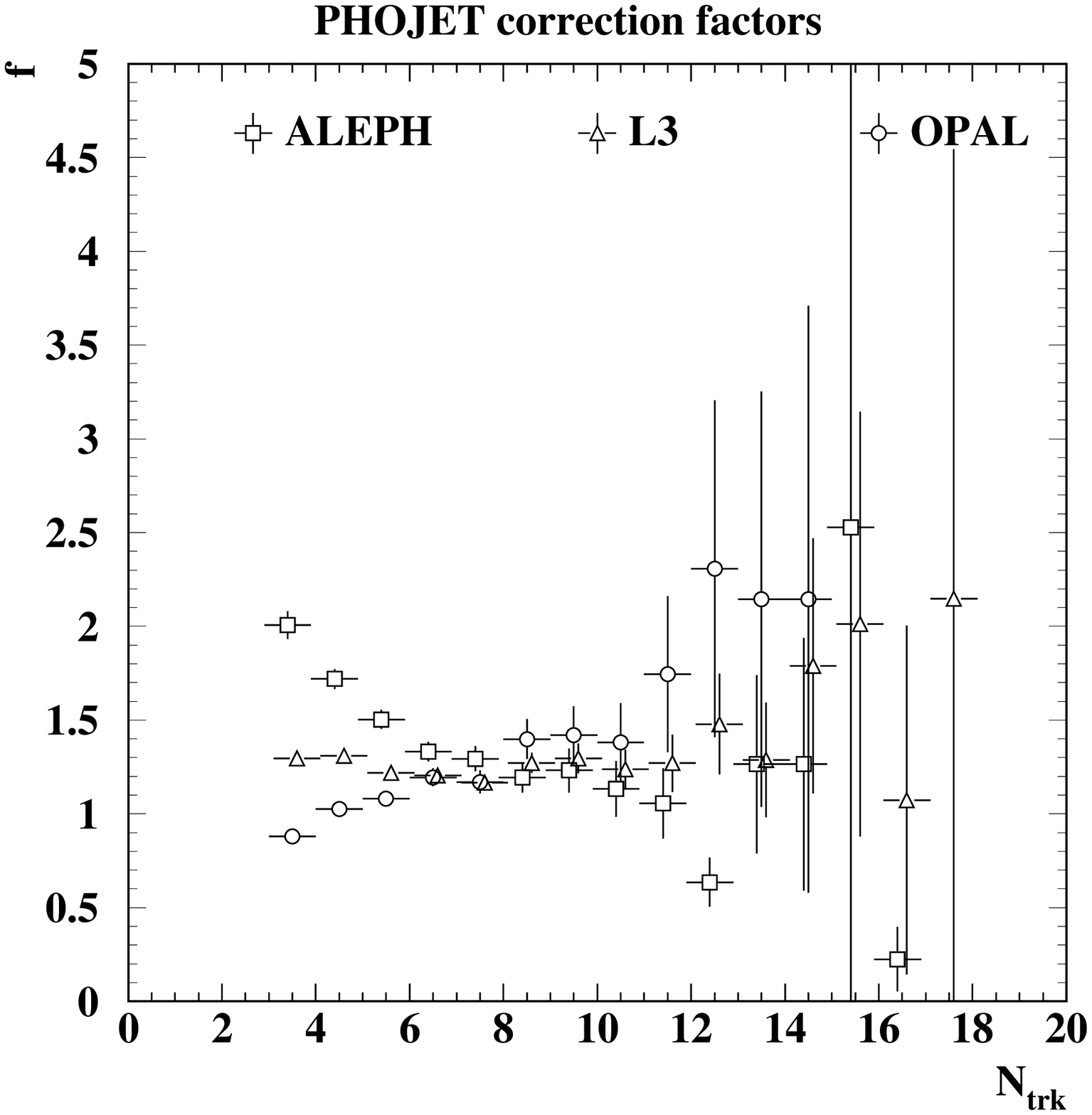,width=0.49\linewidth}
\epsfig{file=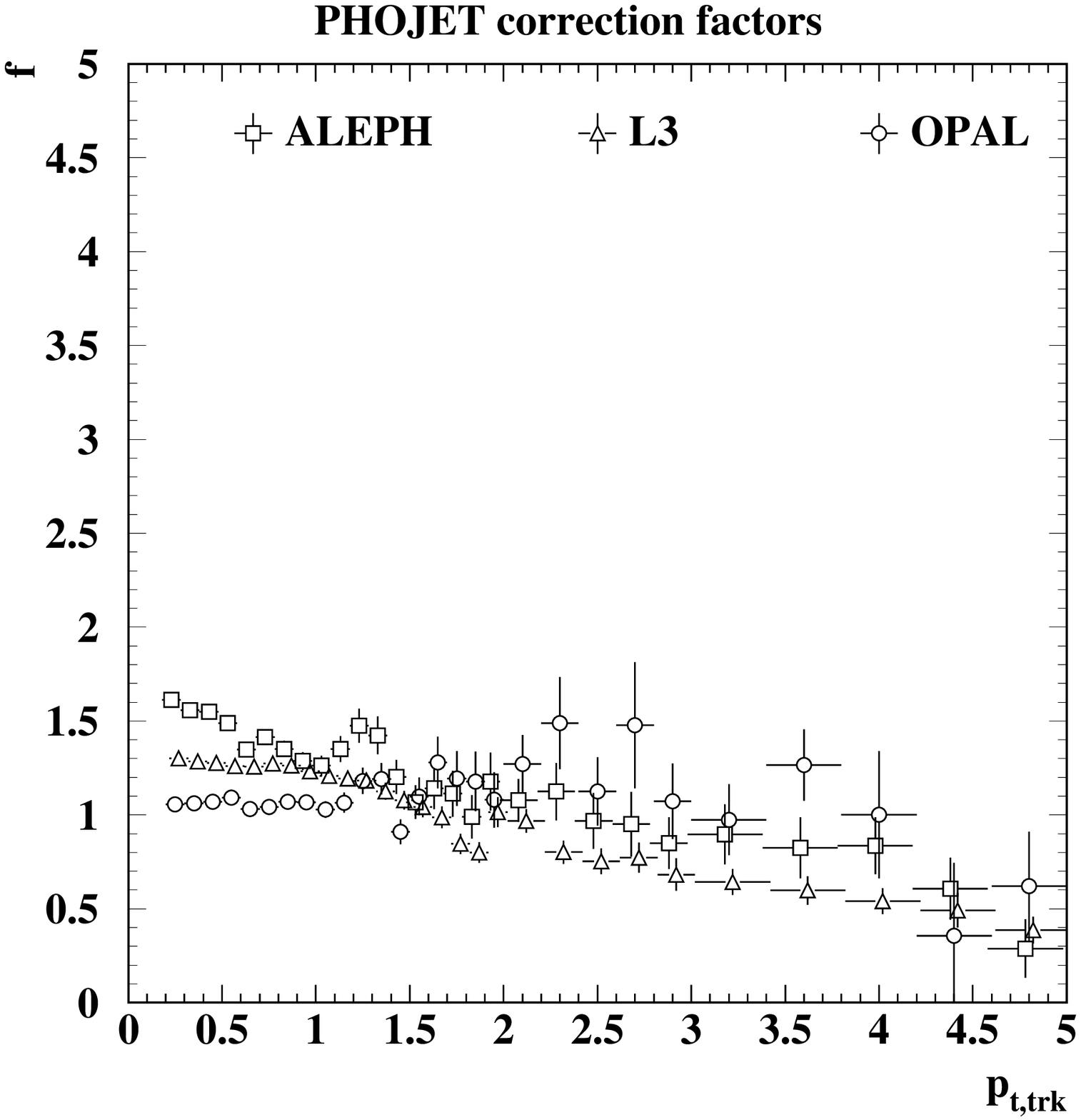,width=0.49\linewidth}
\caption{\label{fig:cor2} 
         The PHOJET correction factors, $f$, for the  
         ALEPH, L3 and OPAL \Wres, \etout, \nch and \ptch
         distributions for the low-\qsq region.
         The symbols are slightly displaced for better visibility.
        }
\end{center}
\end{figure}

 The correction factors in the low-\qsq region
 for the other chosen variables and for the ALEPH, 
 L3 and OPAL experiments using the HERWIG+\kt and PHOJET models,
 are shown in Figures~\ref{fig:cor1} and~\ref{fig:cor2}.
 The quoted errors of the correction factors are the combined statistical 
 uncertainties of the generated hadron level and the simulated 
 detector level quantities.
 While the differences between the correction factors obtained from
 HERWIG+\kt and PHOJET are very small for the energy flow,
 they can vary significantly for other variables. 
 For example, in the case of OPAL, for \Wres the HERWIG+\kt correction 
 factors are on average about 20$\%$ higher than the factors obtained 
 with PHOJET.
 The correction factors for the low-\qsq and high-\qsq regions 
 typically differ by less than 10$\%$. 
%
%
\section{Corrected Data Comparisons}
\label{sec:cor}
 The discussion of the comparison is subdivided into three parts. 
 First the corrected data from the individual experiments are compared
 to each other and to the Monte Carlo models.
 In this comparison only statistical errors are used and no attempt has
 been made to obtain estimates of systematic errors for the individual
 experiments.
 Based on the above comparison a modified version of the HERWIG+\kt model
 has been developed which is described next.
 Finally the data are combined and compared to the Monte Carlo models.
 In the combination of the data the spread of the experiments is used as 
 an estimate of the systematic uncertainty of the measured distributions.
 The numerical results are listed in Tables~\ref{tab:Wall}--\ref{tab:pefave} 
 and can be obtained in electronic form~\cite{Numbers}.
%
%
 \subsection{Data from individual experiments}
 \label{sec:indiv}
\begin{figure}
\begin{center}
\epsfig{file=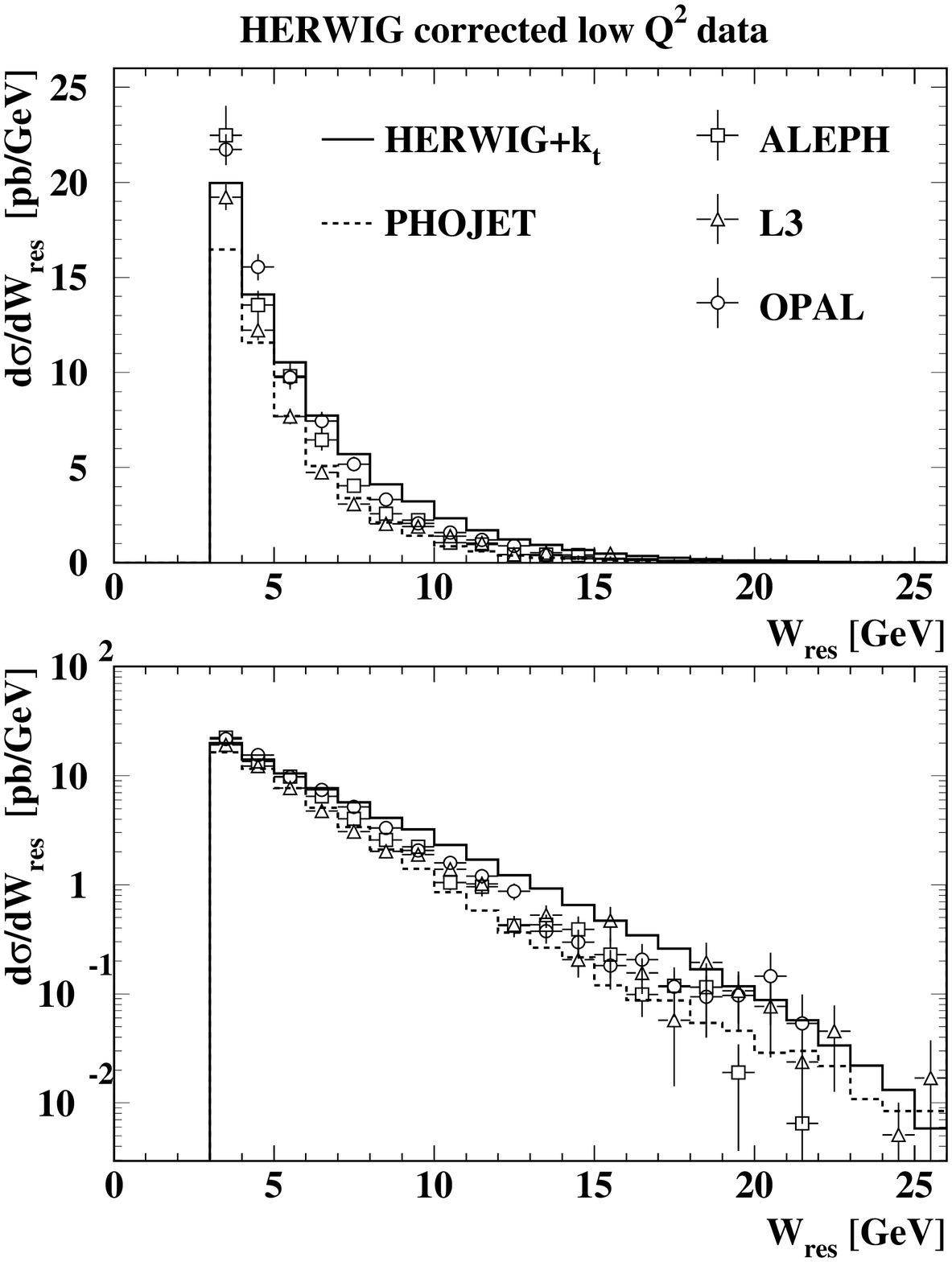,width=0.49\linewidth}
\epsfig{file=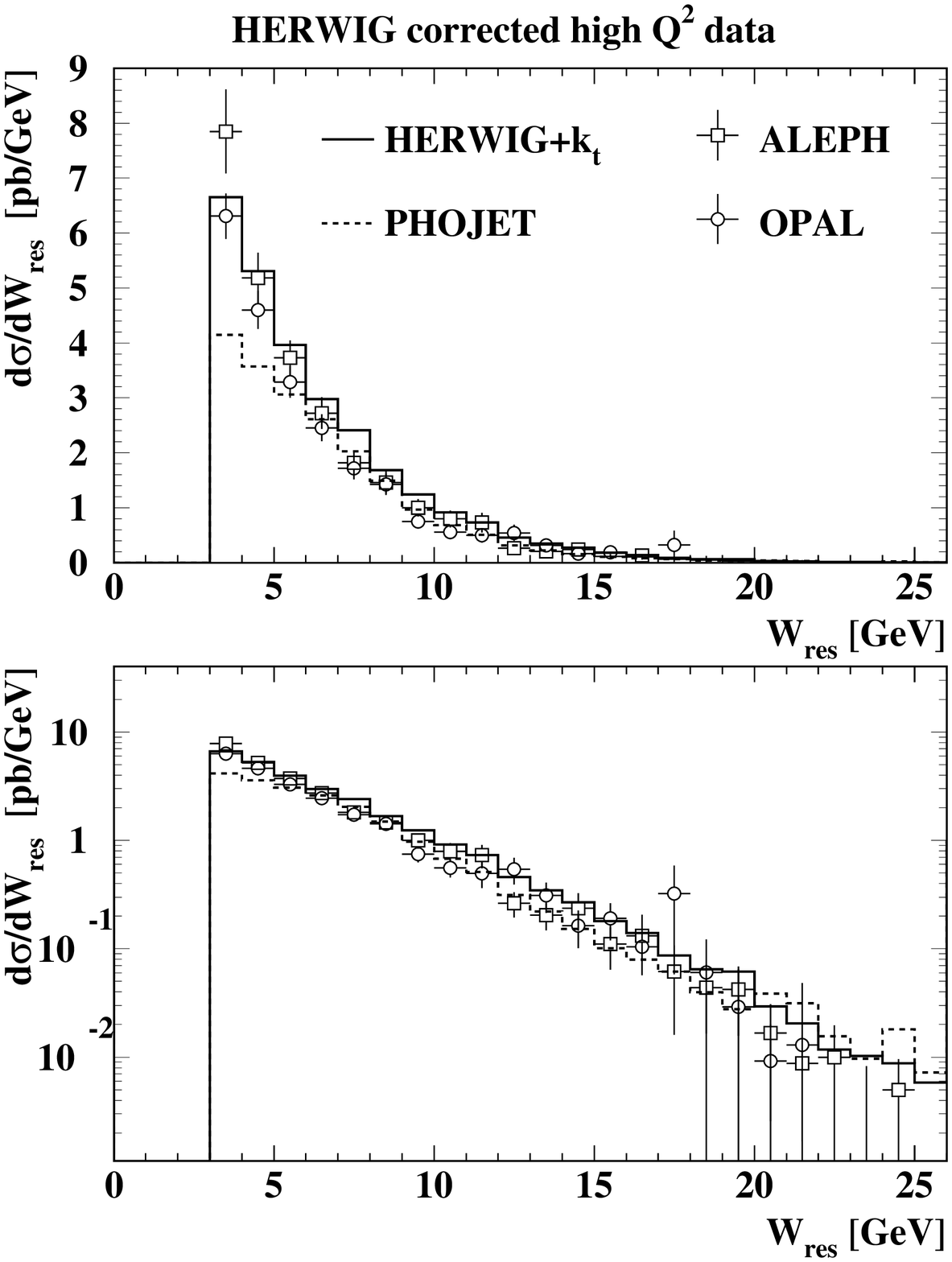,width=0.49\linewidth}
\caption{\label{fig:data1} 
      The \Wres distributions from ALEPH, L3 and OPAL
      for the low-\qsq (left) and high-\qsq region (right), corrected with 
      the HERWIG+\kt model on a linear scale (top) and on a 
      log scale (bottom).}
\end{center}
\end{figure}

\begin{figure}
\begin{center}
\vspace{-0.5cm}
\epsfig{file=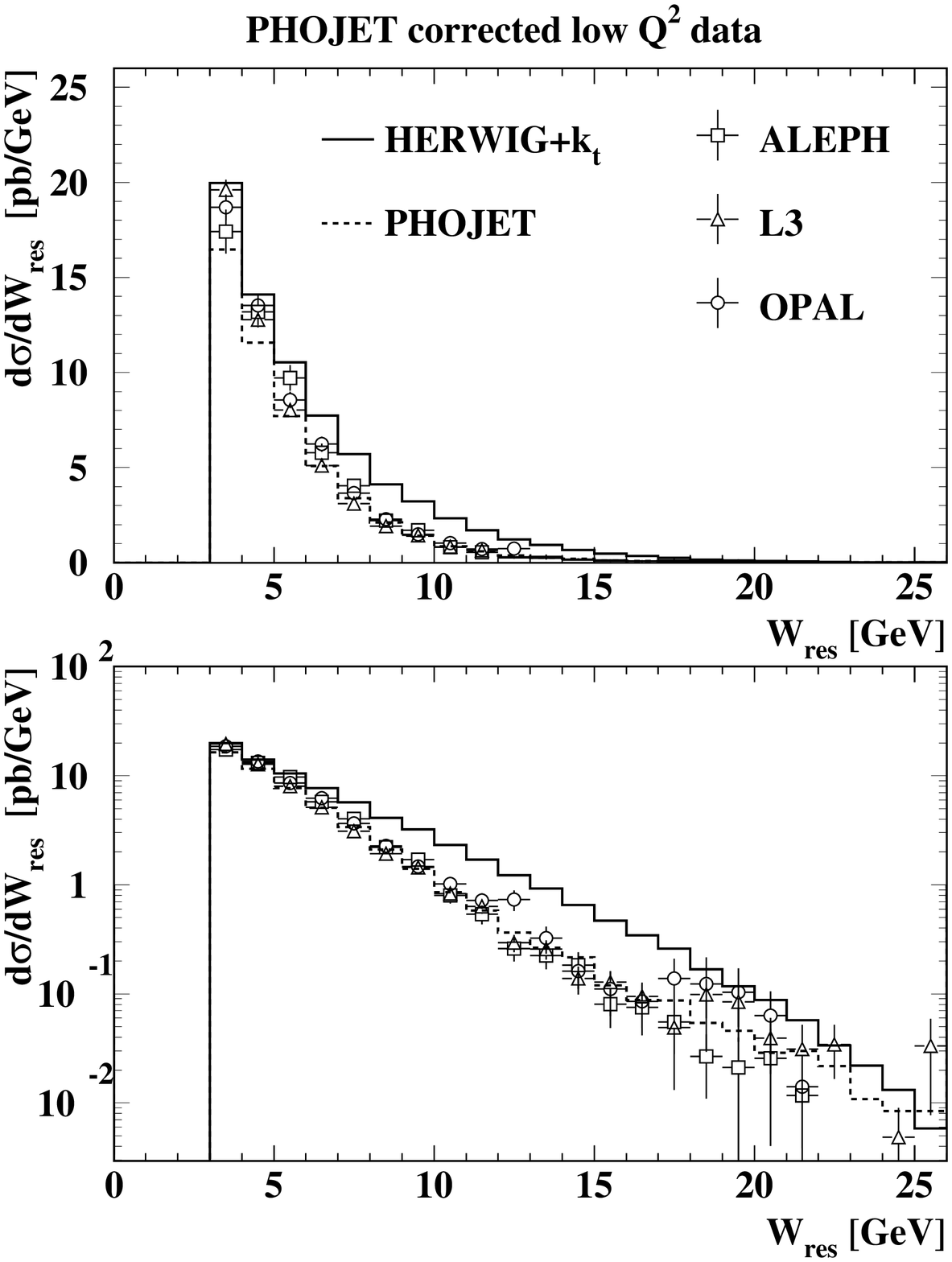,width=0.49\linewidth}
\epsfig{file=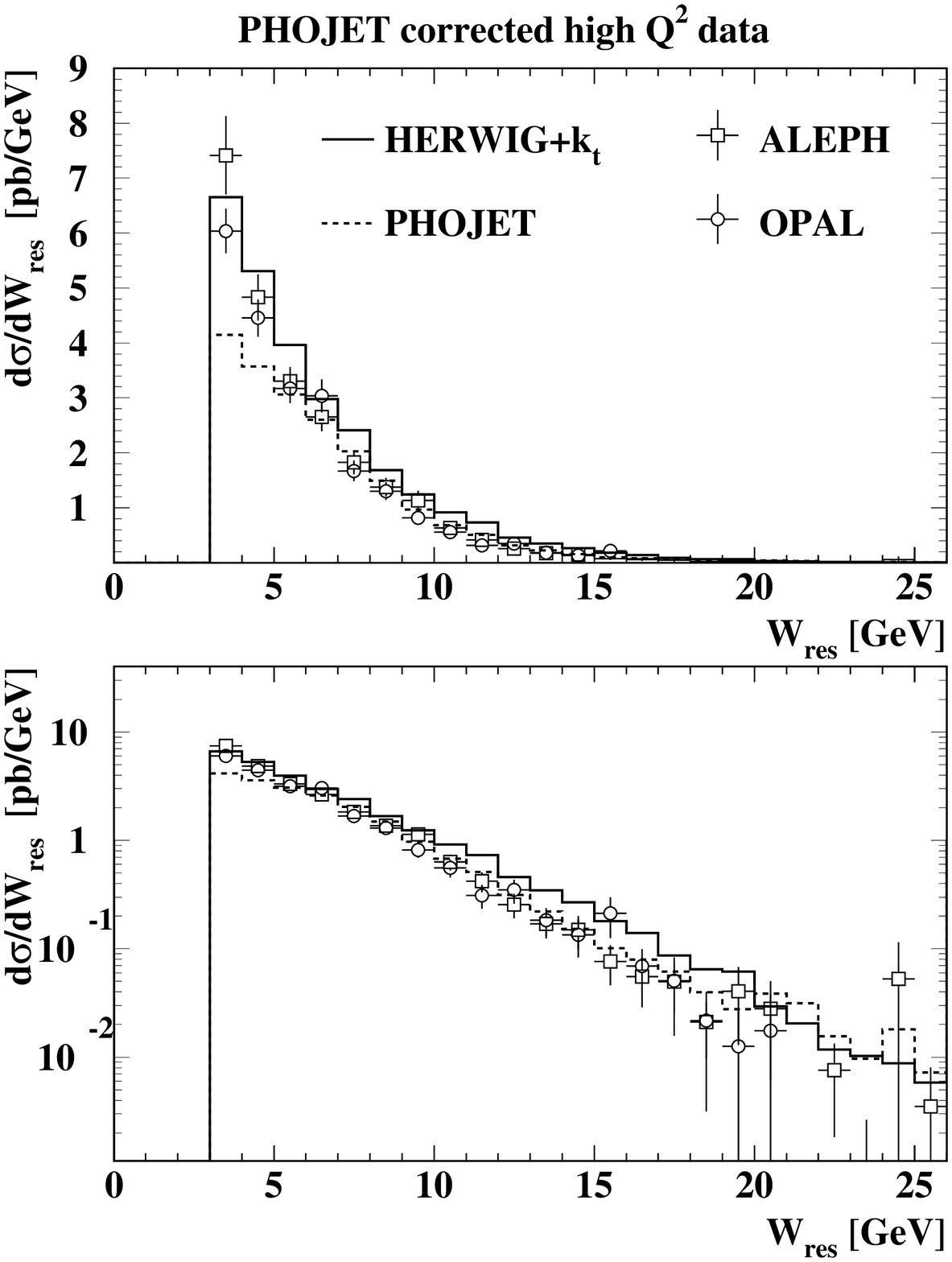,width=0.49\linewidth}
\caption{\label{fig:data2} 
         The \Wres distributions from ALEPH, L3 and OPAL
         for the low-\qsq (left) and high-\qsq region (right), corrected with 
         the PHOJET model on a linear scale (top) and on a log scale (bottom).
        }
\end{center}
\end{figure}

\begin{figure}
\begin{center}
\epsfig{file=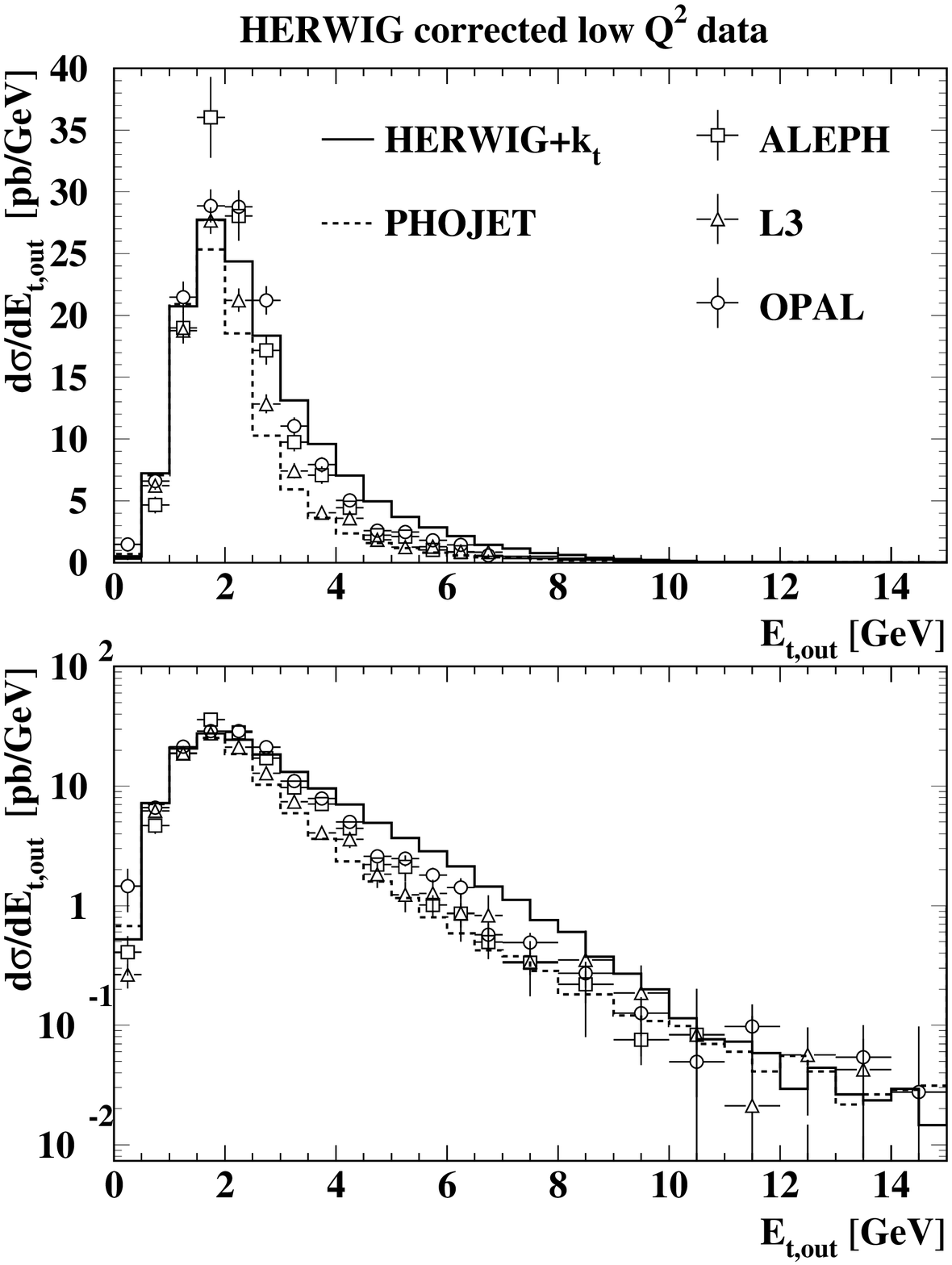,width=0.49\linewidth}
\epsfig{file=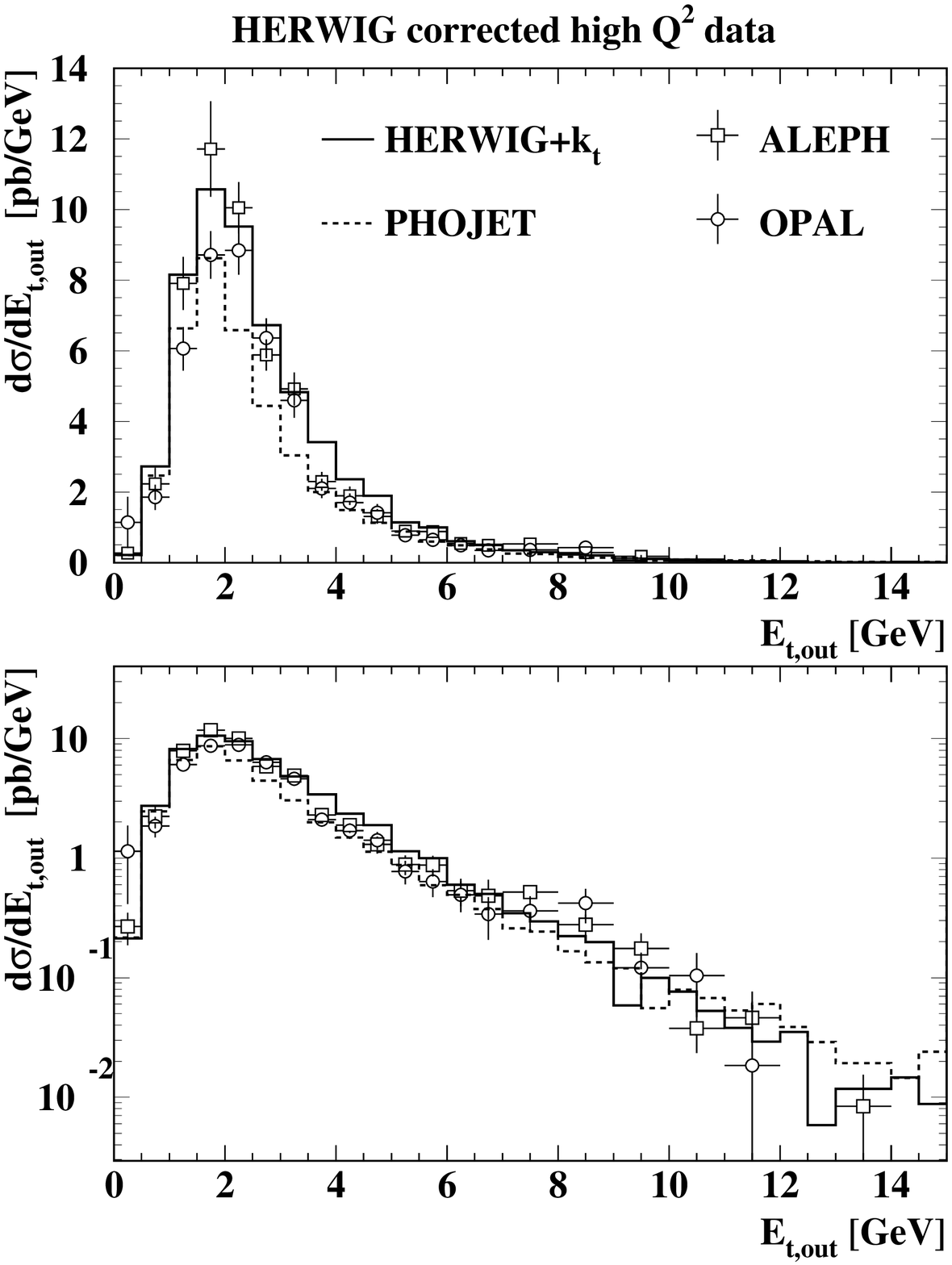,width=0.49\linewidth}
\caption{\label{fig:data5} 
         The \etout distributions from ALEPH, L3 and OPAL
         for the low-\qsq (left) and high-\qsq region (right), corrected with 
         the HERWIG+\kt model on a linear scale (top) and on a 
         log scale (bottom).
        }
\end{center}
\end{figure}

\begin{figure}
\begin{center}
\vspace{-0.5cm}
\epsfig{file=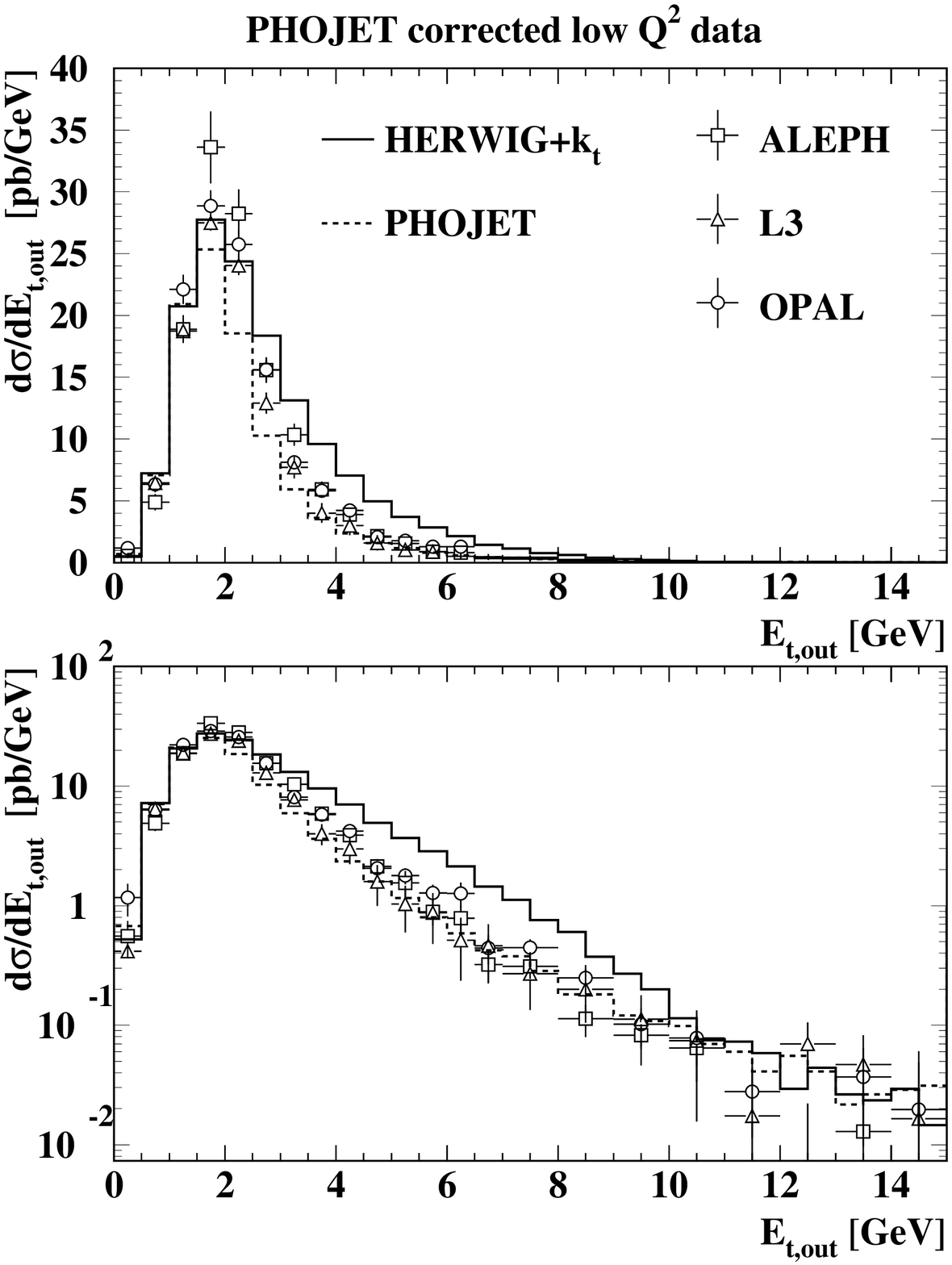,width=0.49\linewidth}
\epsfig{file=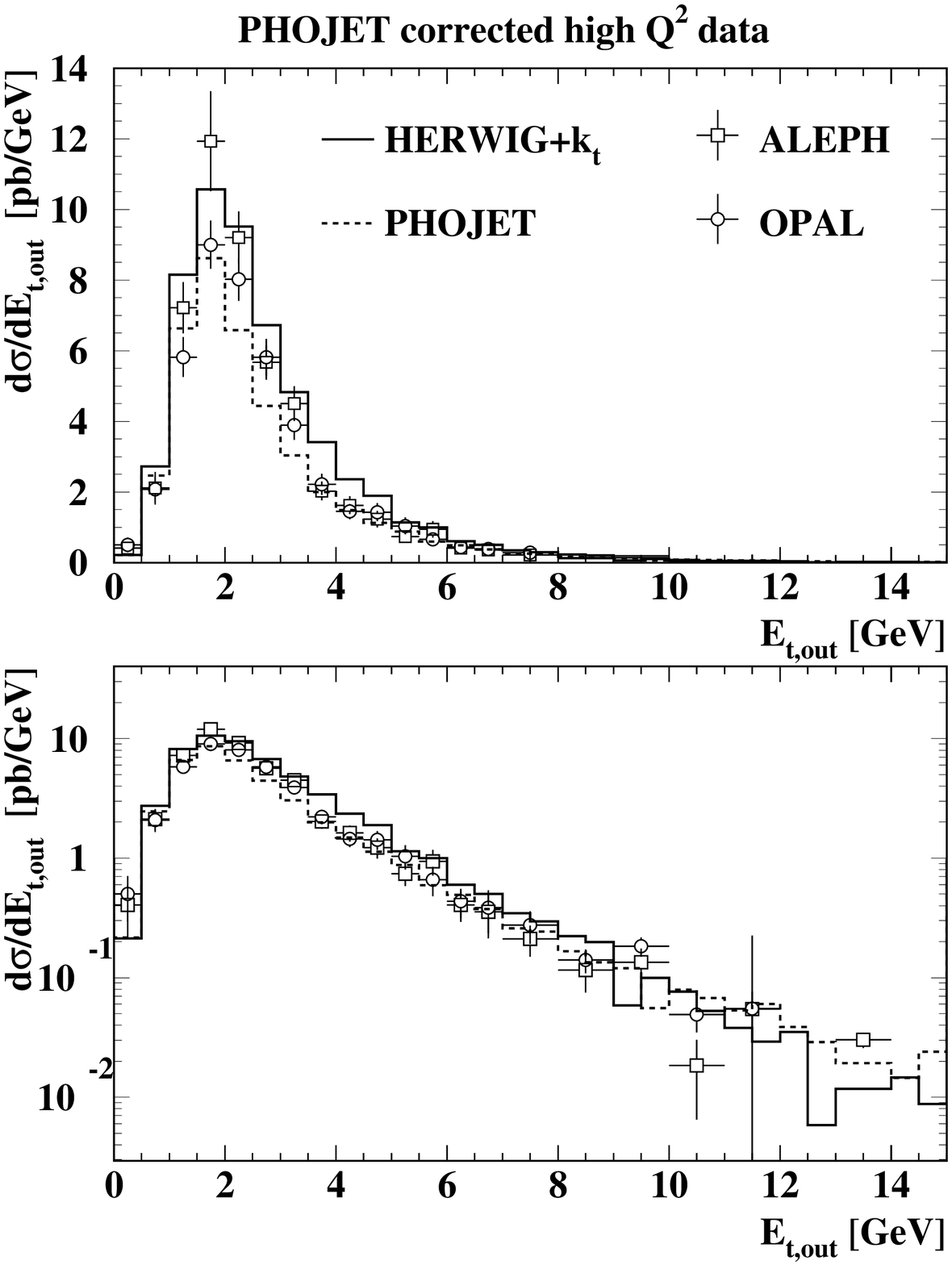,width=0.49\linewidth}
\caption{\label{fig:data6} 
         The \etout  distributions from ALEPH, L3 and OPAL
         for the low-\qsq (left) and high-\qsq region (right), corrected with 
         the PHOJET model on a linear scale (top) and on a log scale (bottom).
        }
\end{center}
\end{figure}

\begin{figure}
\begin{center}
\epsfig{file=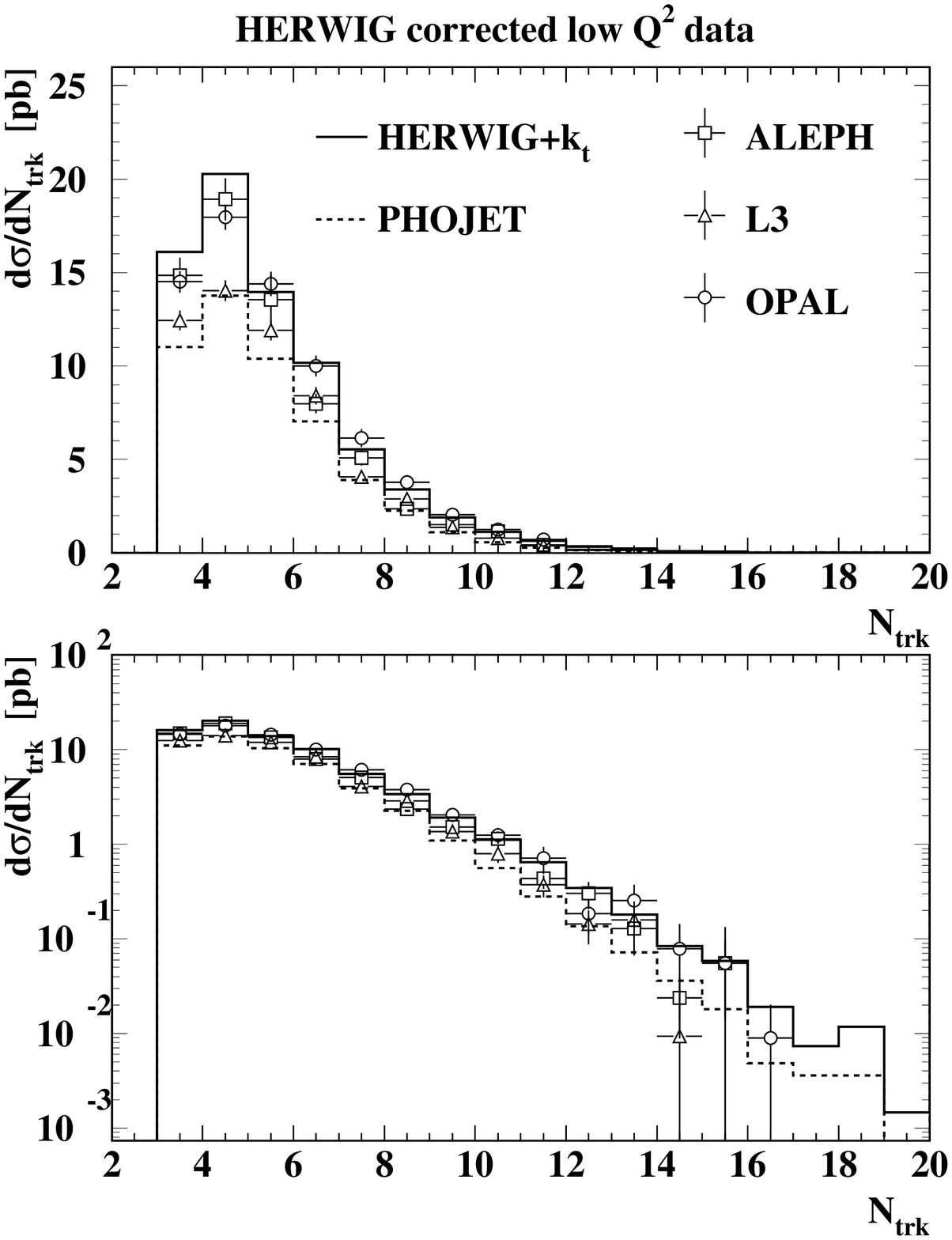,width=0.49\linewidth}
\epsfig{file=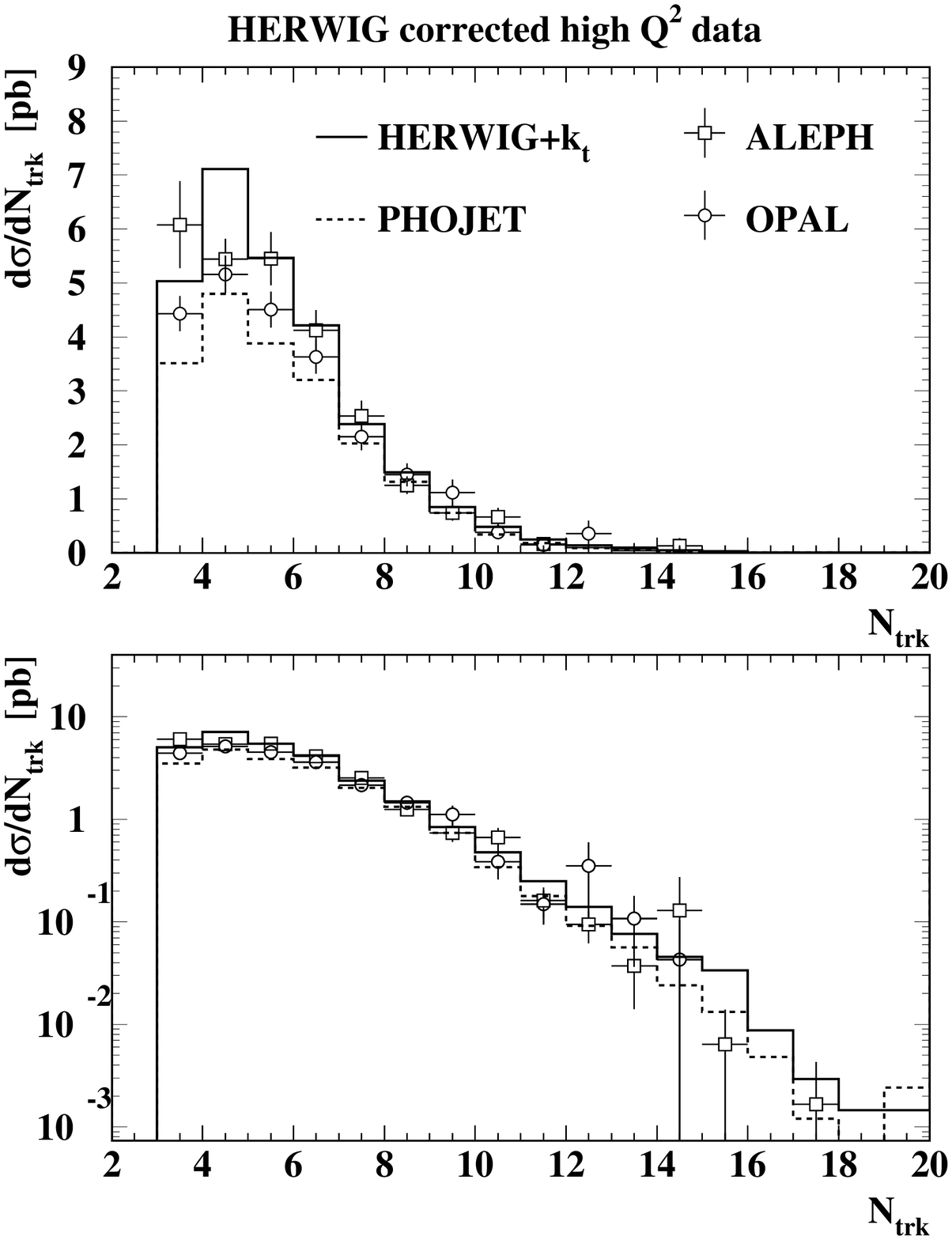,width=0.49\linewidth}
\caption{\label{fig:data7} 
         The \nch  distributions from ALEPH, L3 and OPAL
         for the low-\qsq (left) and high-\qsq region (right), corrected with 
         the HERWIG+\kt model on a linear scale (top) and on a 
         log scale (bottom).
        }
\end{center}
\end{figure}

\begin{figure}
\begin{center}
\vspace{-0.5cm}
\epsfig{file=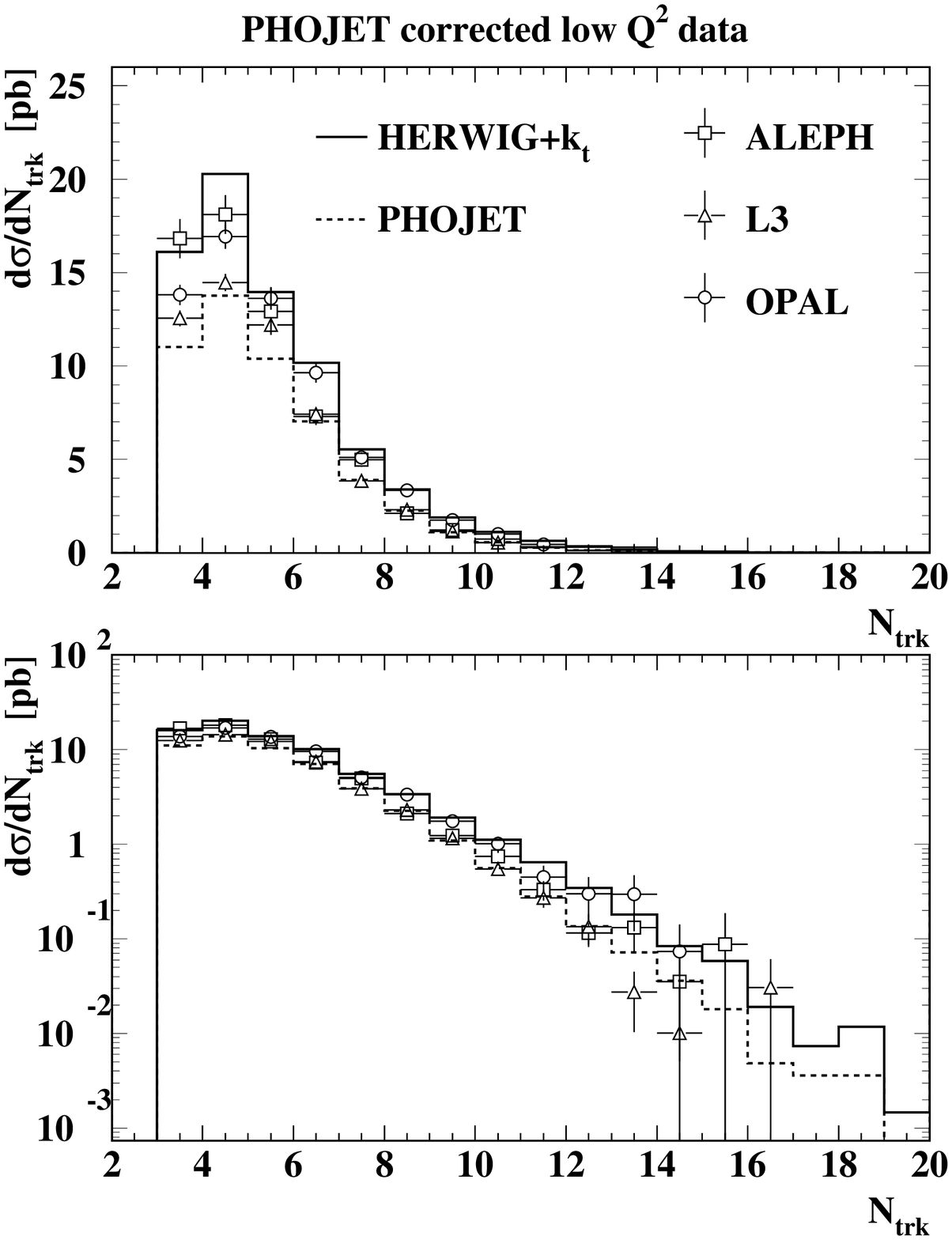,width=0.49\linewidth}
\epsfig{file=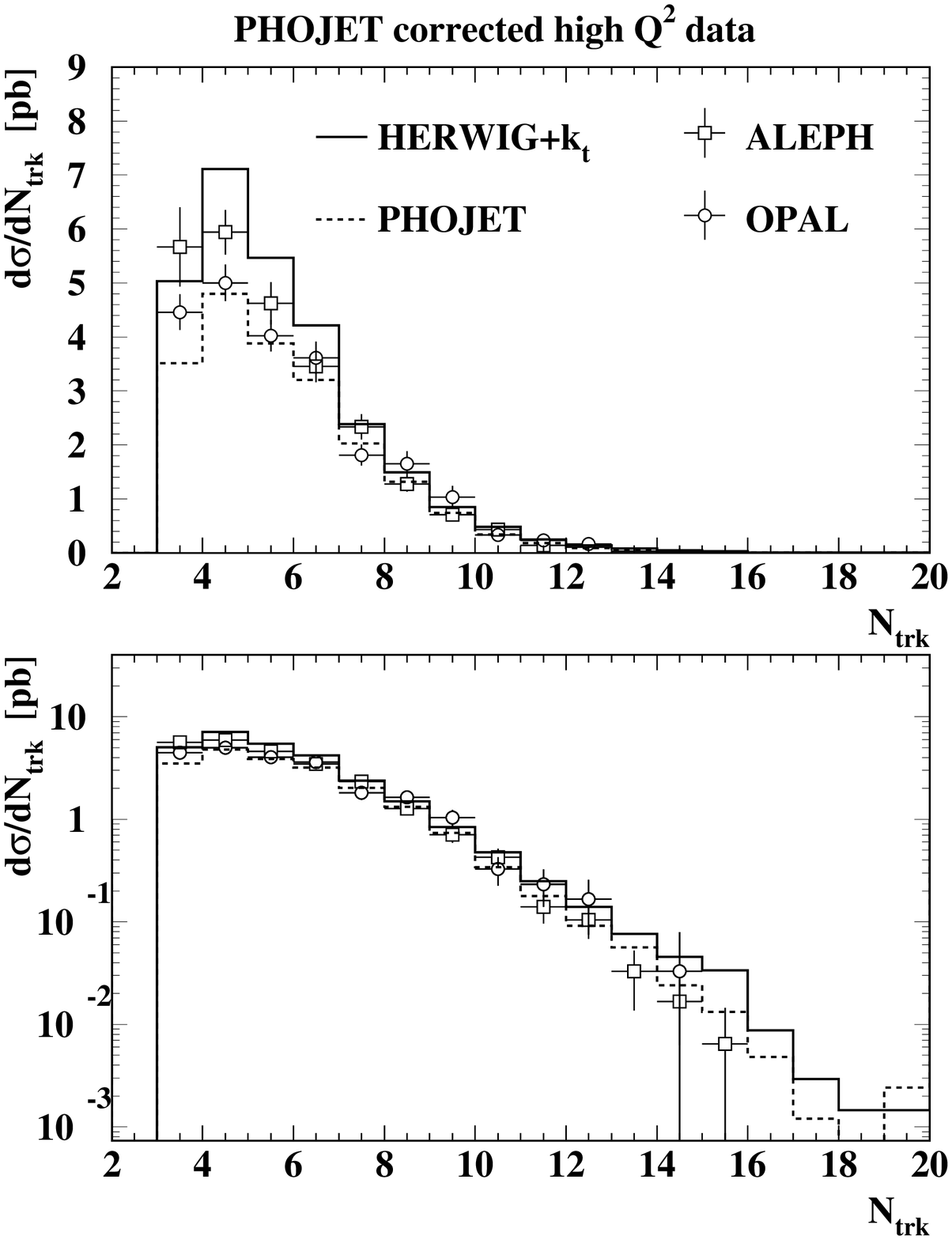,width=0.49\linewidth}
\caption{\label{fig:data8} 
         The \nch  distributions from ALEPH, L3 and OPAL
         for the low-\qsq (left) and high-\qsq region (right), corrected with 
         the PHOJET model on a linear scale (top) and on a log scale (bottom).
        }
\end{center}
\end{figure}

\begin{figure}
\begin{center}
\epsfig{file=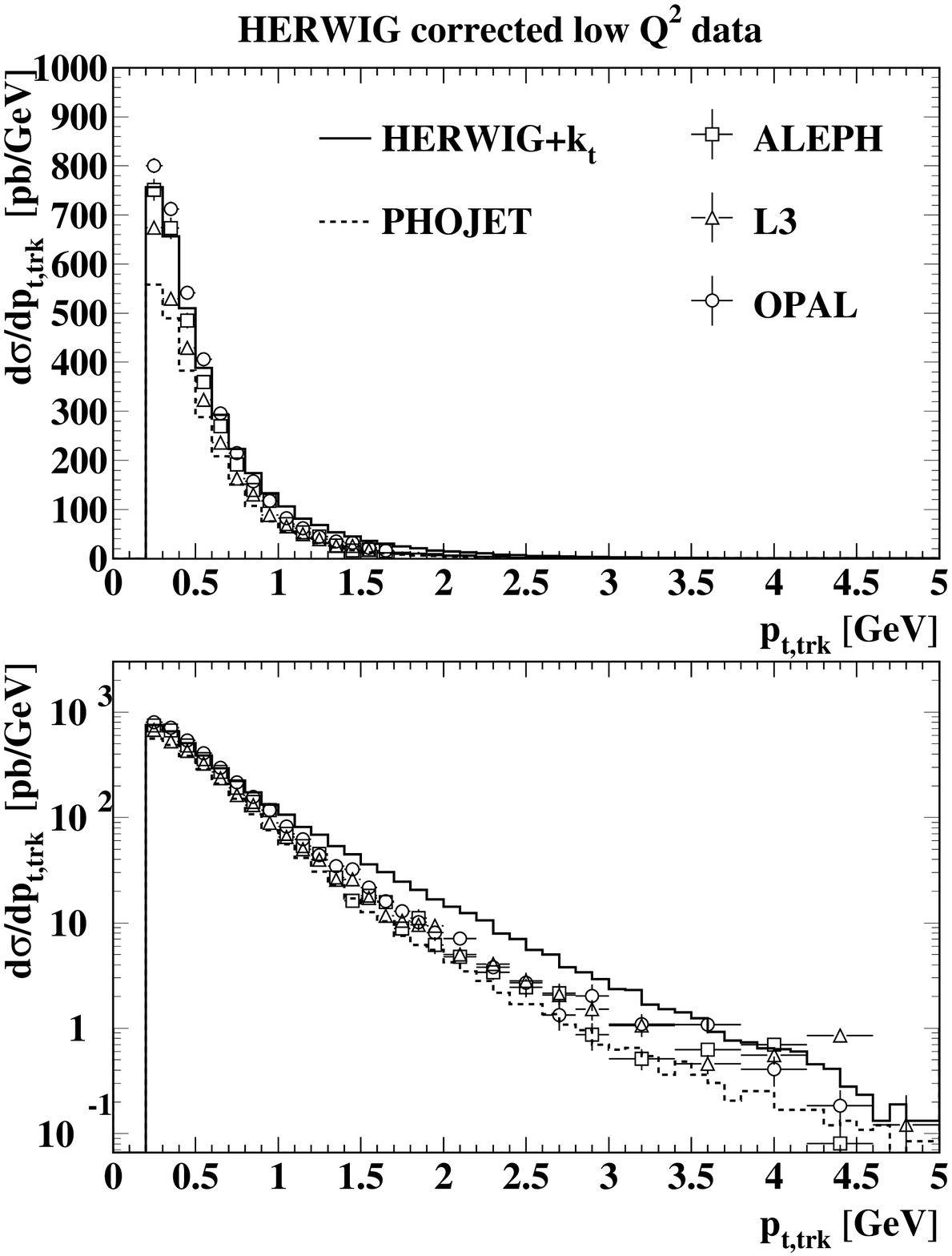,width=0.49\linewidth}
\epsfig{file=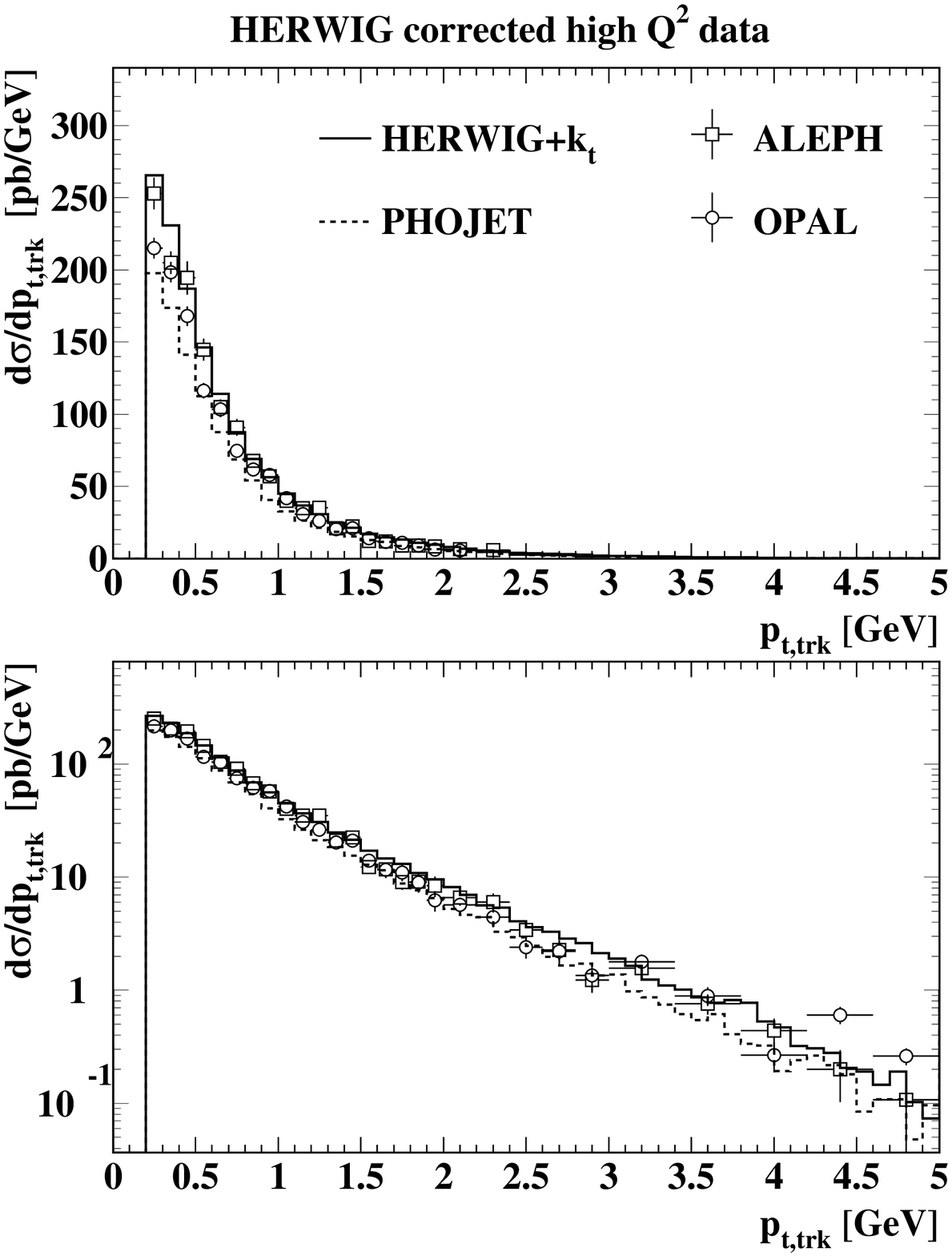,width=0.49\linewidth}
\caption{\label{fig:data9} 
         The \ptch  distributions from ALEPH, L3 and OPAL
         for the low-\qsq (left) and high-\qsq region (right), corrected with 
         the HERWIG+\kt model on a linear scale (top) and on a 
         log scale (bottom).
        }
\end{center}
\end{figure}

\begin{figure}
\begin{center}
\vspace{-0.5cm}
\epsfig{file=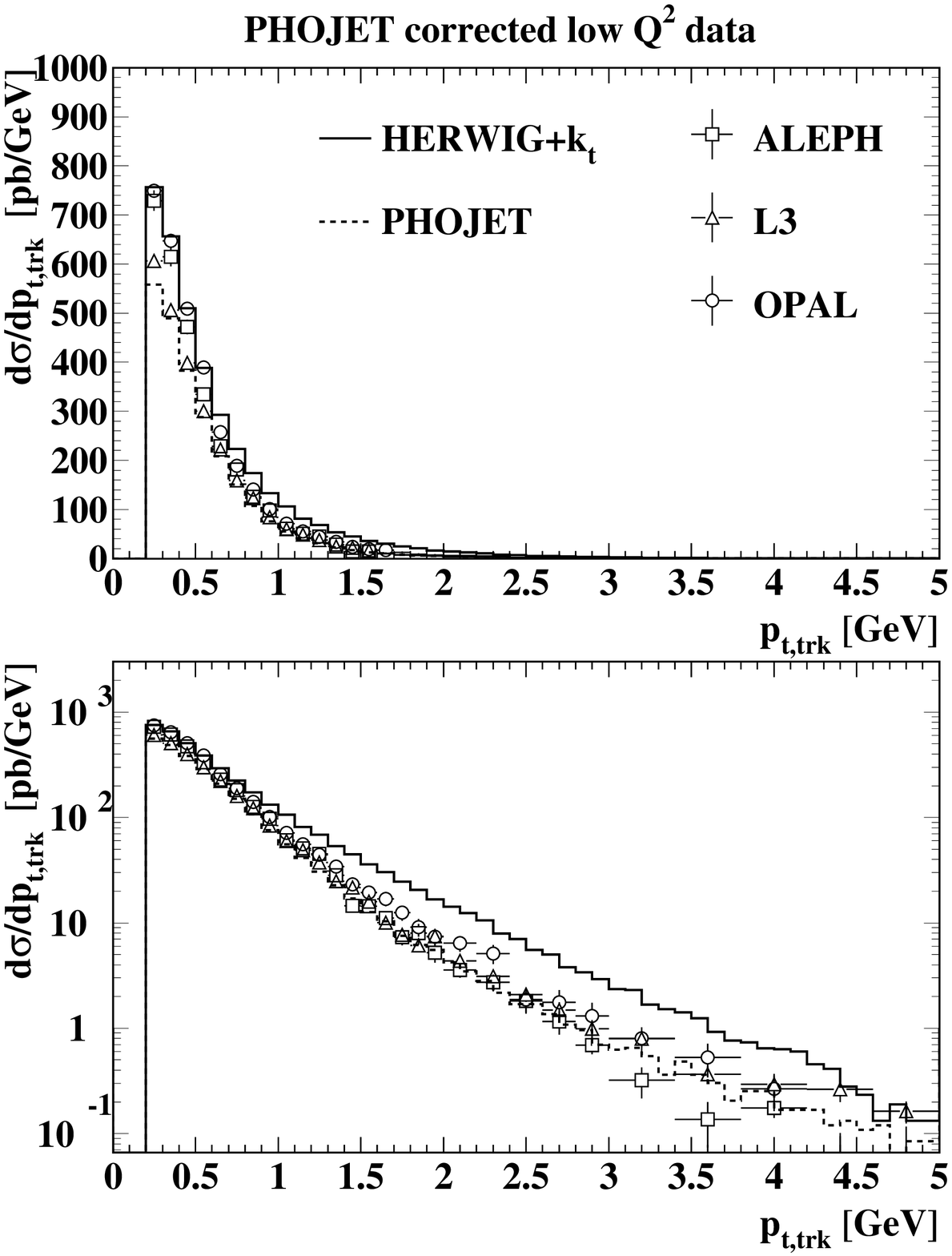,width=0.49\linewidth}
\epsfig{file=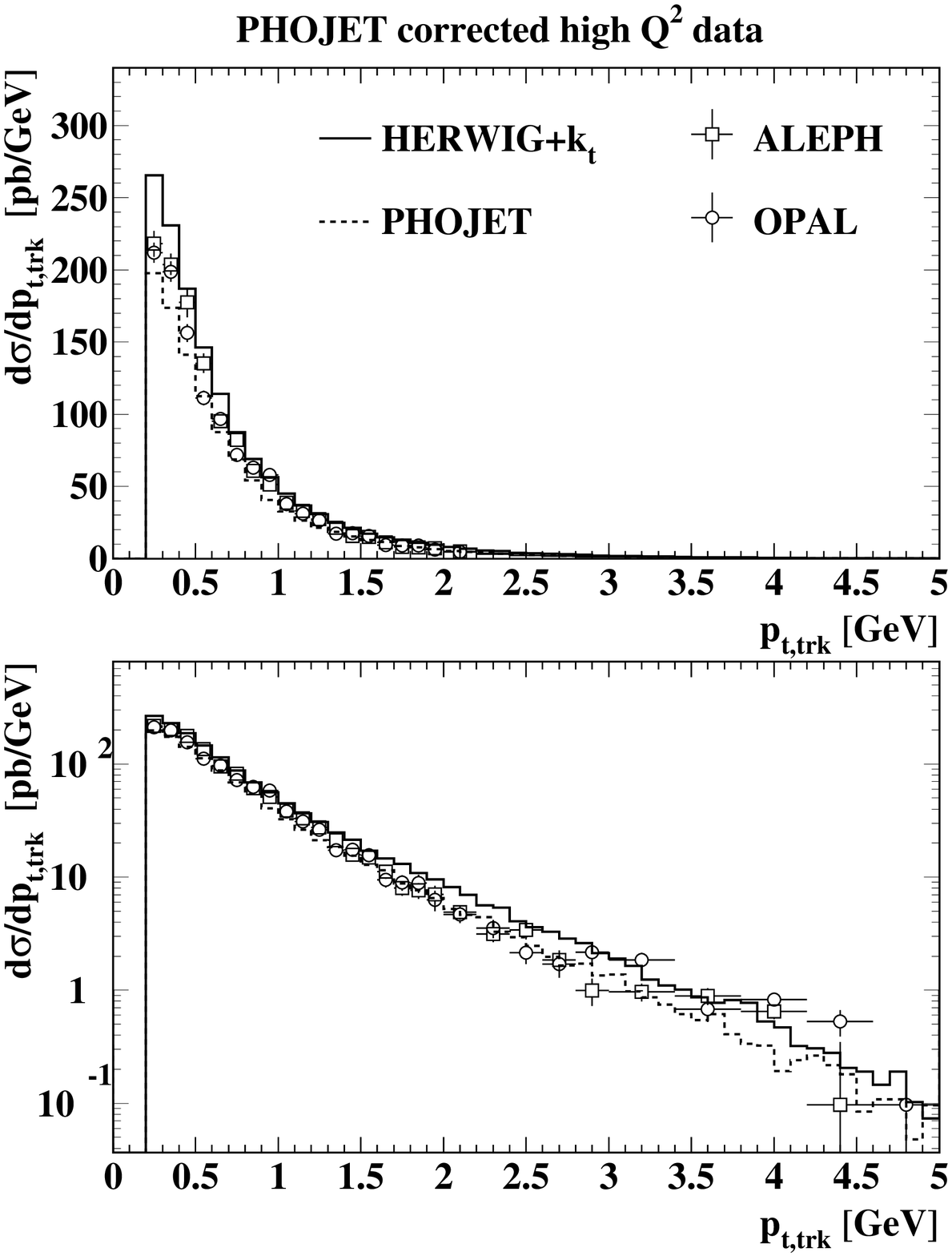,width=0.49\linewidth}
\caption{\label{fig:data10} 
         The \ptch  distributions from ALEPH, L3 and OPAL
         for the low-\qsq (left) and high-\qsq region (right), corrected with 
         the PHOJET model on a linear scale (top) and on a log scale (bottom).
        }
\end{center}
\end{figure}

 The Figures~\ref{fig:data1}--\ref{fig:data10} show the corrected 
 differential cross-sections, calculated in the kinematical range
 described in Section~\ref{sec:evsel},
 for the data compared with the HERWIG+\kt and PHOJET predictions.
 Figures~\ref{fig:data1},~\ref{fig:data5},~\ref{fig:data7} 
 and~\ref{fig:data9} show the data on a linear and logarithmic scale,
 corrected with the HERWIG+\kt model, while 
 Figures~\ref{fig:data2},~\ref{fig:data6},~\ref{fig:data8} 
 and~\ref{fig:data10} show the same data corrected with PHOJET.  
 For all distributions the errors shown are the 
 quadratic sum of the statistical errors of the measured quantity 
 and the statistical errors of the correction factors.
 As an example, Table~\ref{tab:Wall} shows the results for the \Wres 
 distribution for the three experiments, listing the statistical error on 
 the data and the statistical error on the correction factors, $f$, 
 separately. 
 \par
 The experimental results from ALEPH, L3 and OPAL agree with each 
 other within errors for large regions in most of the variables studied.
 However, there are also regions where they significantly differ from 
 each other, for example, in the region of \Wres$<10$~GeV, \etout $<5$~GeV, 
 for low charged multiplicities and low \ptch in the low-\qsq region.
 The level of agreement also depends on the Monte Carlo model used for
 correction. 
 The agreement between the experiments is slightly better for the data
 corrected with PHOJET, than for the data corrected with HERWIG+\kt.
 In the combination of the data, the differences between the 
 measured distributions of the different experiments will be used as an 
 estimate of the systematic error.
 \par
 There are significant differences between the Monte Carlo distributions
 and the data, particularly in the low-\qsq region. 
 The PHOJET distributions lie consistently below HERWIG+\kt, especially
 at the low end of the distributions, while the agreement of the tails 
 in the high-\qsq region is reasonable.
 In general the PHOJET predictions agree reasonably well with the data, 
 corrected with PHOJET, for large values of the variables.
 However, there are large differences in the low-\qsq region
 between the data, corrected with HERWIG+\kt, and the HERWIG+\kt predictions
 in all the distributions.  
 For the \Wres, \etout and \ptch distributions (except in the region of
 low values,
 where the data are inconsistent) the differences between the experiments
 are much less than the HERWIG+\kt$-$PHOJET differences.
 \par
\begin{figure}
\begin{center}
\epsfig{file=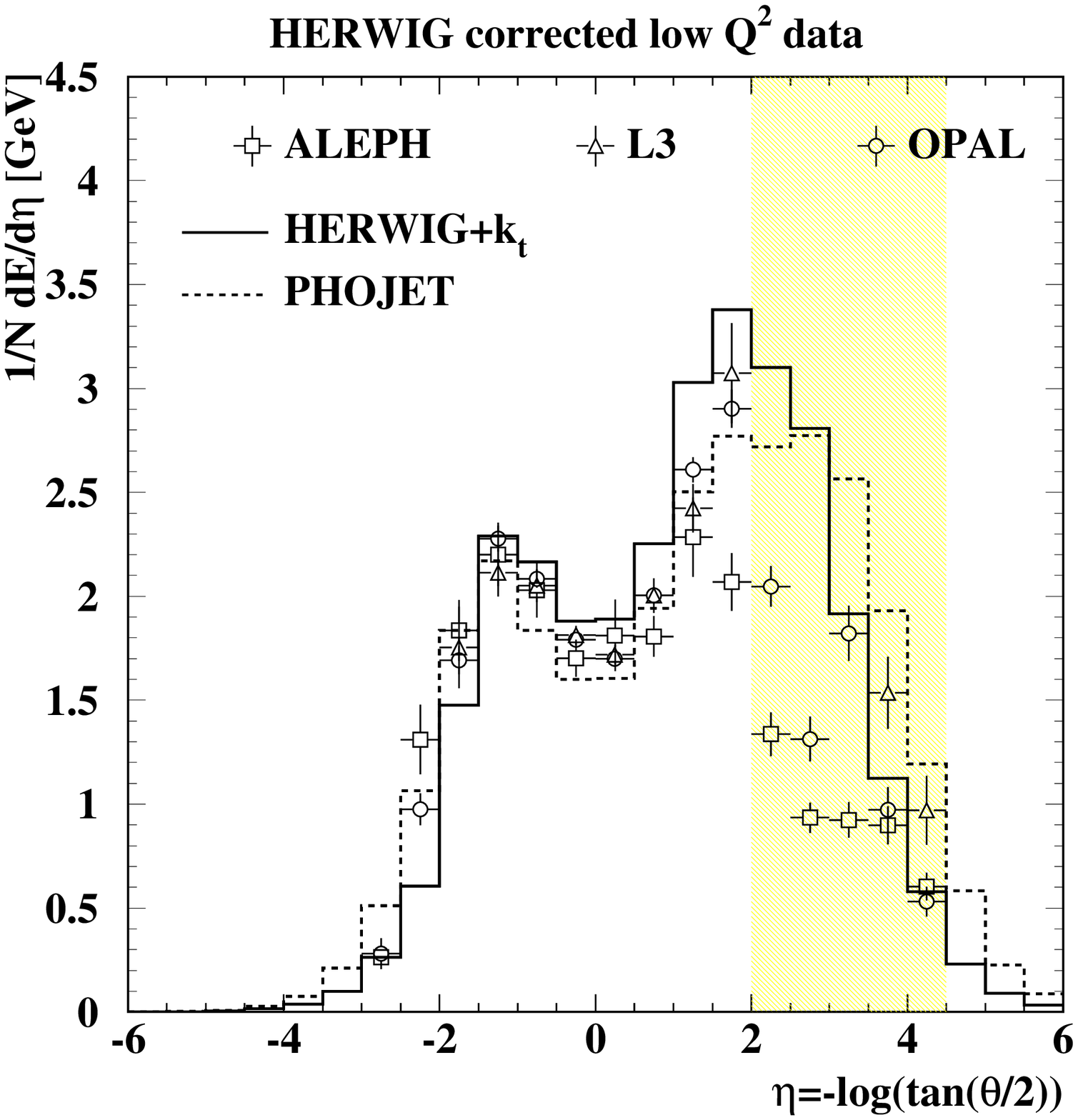,width=0.49\linewidth}
\epsfig{file=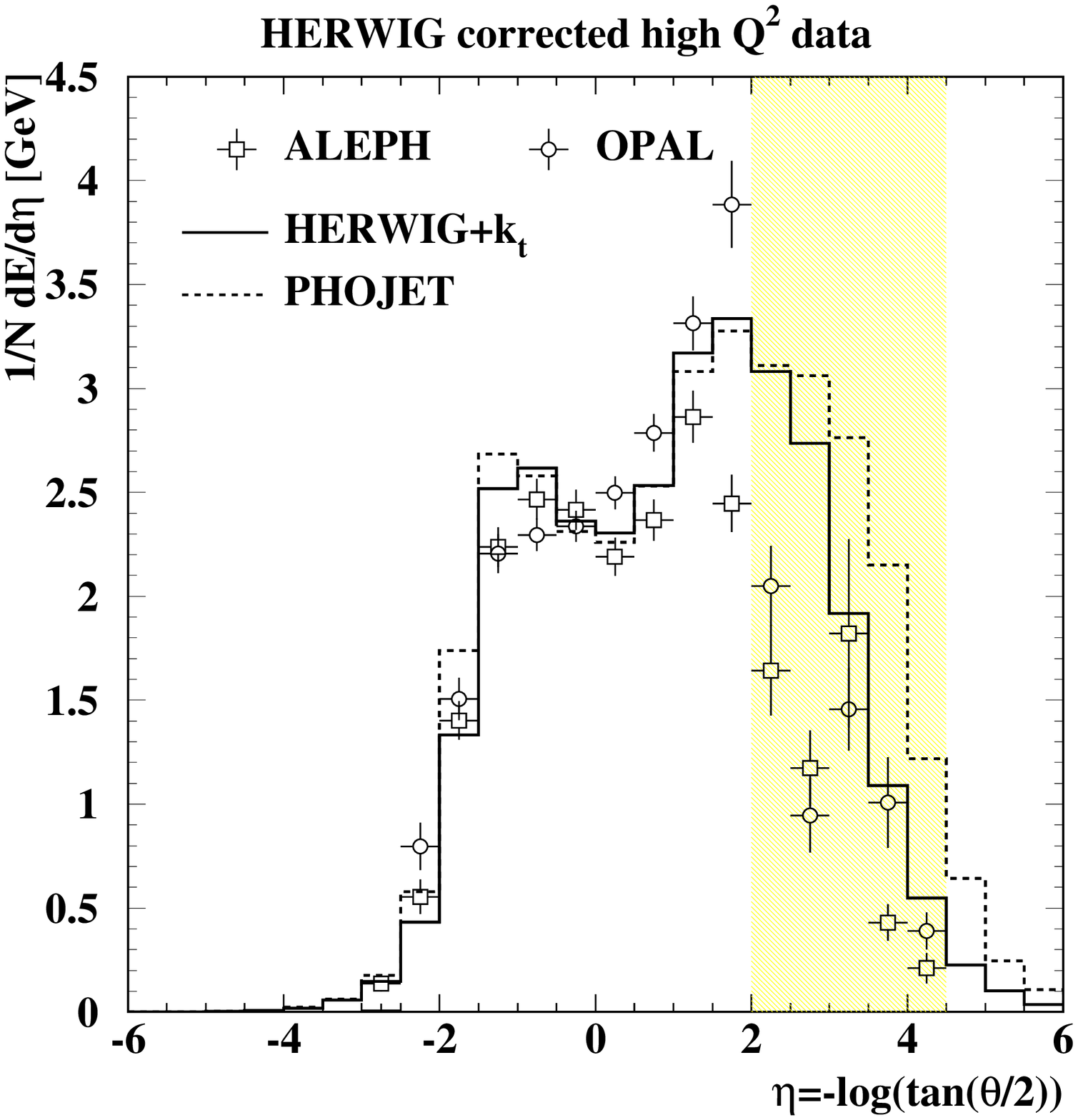,width=0.49\linewidth}
\caption{\label{fig:data11} 
         The hadronic energy flow from ALEPH, L3 and OPAL
         for the low-\qsq (left) and high-\qsq region (right), corrected with 
         the HERWIG+\kt model.
         The shaded band indicates the forward region of the experiments.
        }
\end{center}
\end{figure}
\begin{figure}
\begin{center}
\epsfig{file=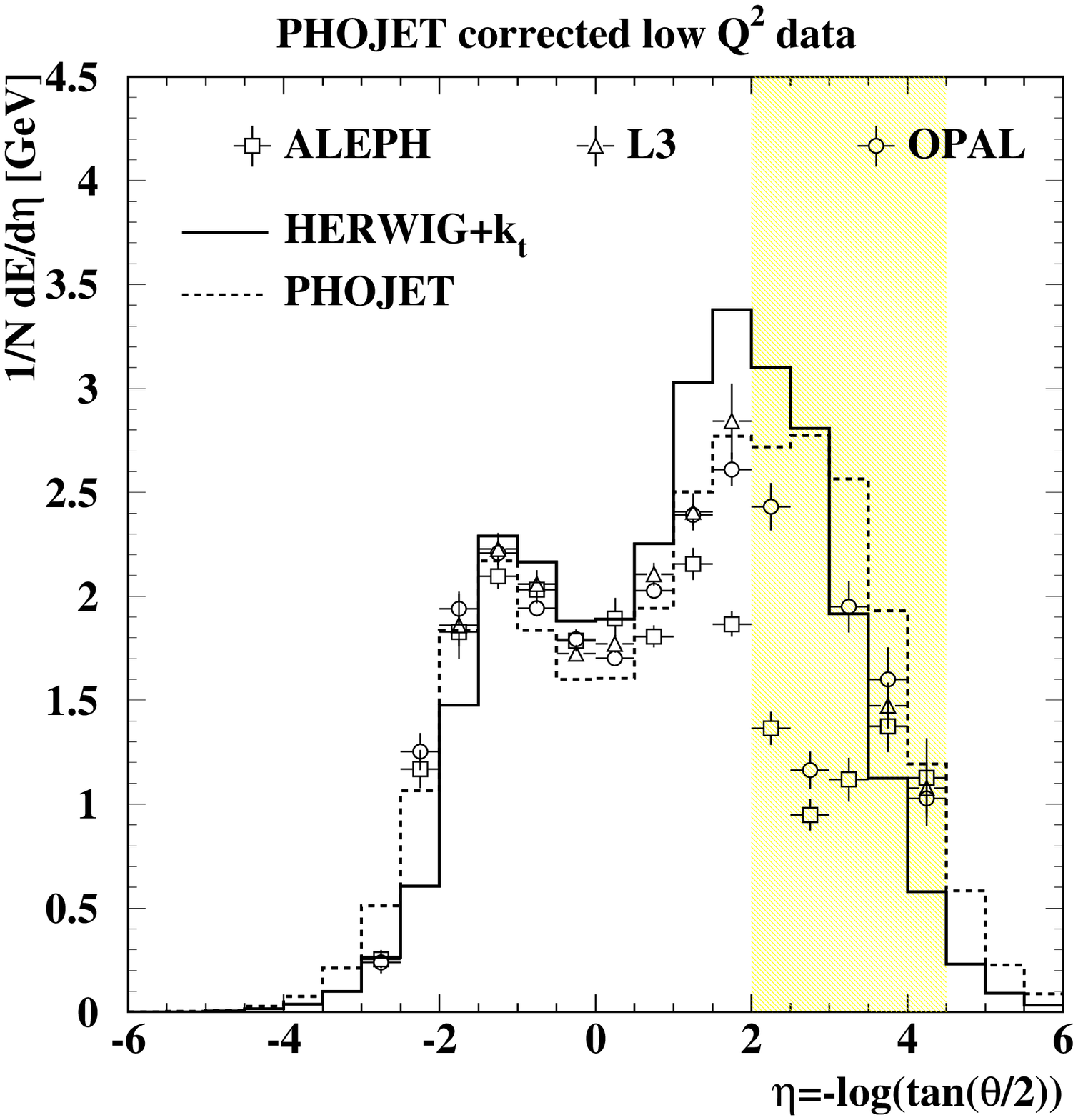,width=0.49\linewidth}
\epsfig{file=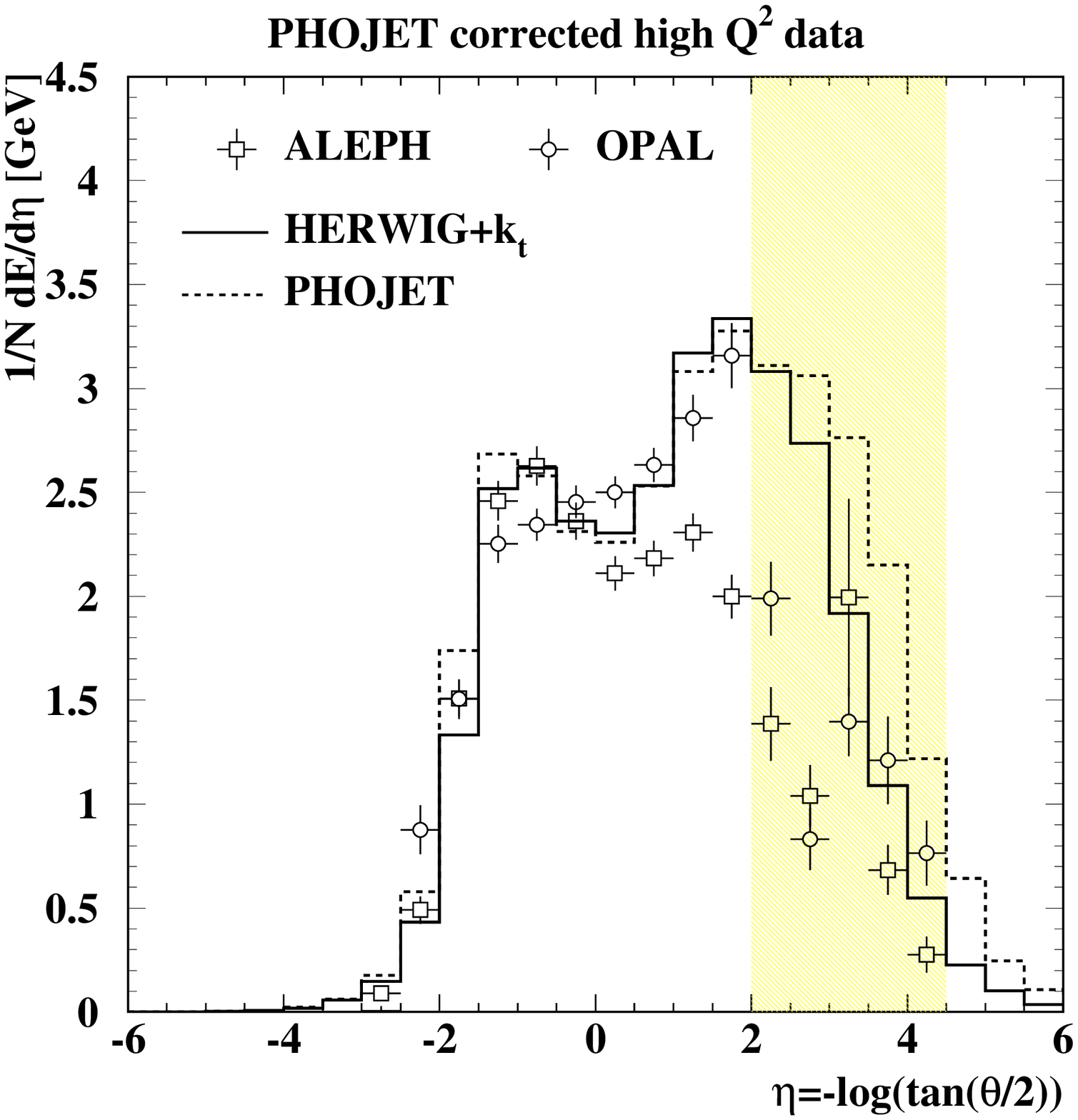,width=0.49\linewidth}
\caption{\label{fig:data12} 
         The hadronic energy flow from ALEPH, L3 and OPAL
         for the low-\qsq (left) and high-\qsq region (right), corrected with 
         the PHOJET model.
         The shaded band indicates the forward region of the experiments.
        }
\end{center}
\end{figure}

 Figures~\ref{fig:data11} and~\ref{fig:data12} show the corrected energy
 flow as a function of pseudorapidity in the low-\qsq and high-\qsq region 
 for the individual experiments compared to the HERWIG+\kt and PHOJET 
 predictions.
 Following the energy flow analyses at HERA, the statistical errors for 
 the energy flow are taken as $\sqrt{\sum E^2}/\Nevt$.
 In Figure~\ref{fig:data11} the data are corrected with the HERWIG+\kt
 model and in Figure~\ref{fig:data12} with the PHOJET model. 
 In the central region of the detector, between $-1.5<\eta<1.5$, the
 data lie between the HERWIG+\kt and the PHOJET predictions in the
 low-\qsq region, whereas in the high-\qsq region the two MC
 predictions lie closer to each other than the data of OPAL and ALEPH.
 These corrected energy flows in the central region are stable against
 changes in the requirements for the hadronic final state, e.g. 
 changes to the quality cuts of the accepted tracks and clusters. 
 \par
 For $\eta>1.5$ the data from the various experiments vary much more
 than the statistical errors. Furthermore, it has been found that
 the corrected energy flows are not very stable against variations of 
 the event selection cuts such as the anti-tag condition or a maximum 
 energy cut on the energy deposited in the forward detectors.
 The observed changes are much larger at the detector level than at
 the hadron level, leading to the conclusion that the modelling of
 the energy response to the hadronic energy in the forward region is
 poor. 
 This can be understood since the sub-detectors covering this region
 have poor energy resolution for hadronic energies 
 and no particle identification.
 Therefore the uncertainty of the hadronic energy flow in the forward region
 is larger than is indicated by the spread of the experiments and it is 
 difficult to draw firm conclusions on the description by the Monte Carlo
 models.  
 However, the data appear to lie consistently below the prediction of either
 generator.
%
%
 \subsection{Modification of the HERWIG model}
 As discussed above,
 the distribution of transverse momentum \kt of the photon remnant 
 with respect to the direction of the incoming photon has been altered,
 motivated by the findings in photoproduction at HERA.
 At LEP, the modification was initially studied as a possible improvement
 of the agreement between the HERWIG prediction and the high-\qsq data of 
 OPAL and ALEPH. 
 While HERWIG+\kt seems to reasonably describe the data in the high-\qsq 
 region for low and high values of \etout, it overestimates the distribution 
 for medium values of \etout, as shown in Figure~\ref{fig:data5}.
 Even though the description of the energy flow is improved 
 with the HERWIG+\kt generator, it fails to accurately describe the
 data over a wide range of the pseudorapidity $\eta$.
 \par
 While the fixed limit of $\ktsqm=25$~GeV$^2$ is adequate for the high-\qsq 
 region, in the low-\qsq region it introduces too much transverse momentum, 
 which is most clearly seen in the transverse momentum distribution 
 of the tracks in Figure~\ref{fig:data9}. 
 Therefore the dynamical adjustment, HERWIG+\kt(dyn), as discussed
 in Section~\ref{sec:MC}, has been introduced.
 As will be seen in the next section, when comparing with the
 combined LEP data, this change leads to an improved description of
 the data also for the low-\qsq region.
%
%
 \subsection{Combined data}
 In this section the results from the individual experiments discussed
 in Section~\ref{sec:indiv} are combined. 
 In large ranges of the phase space the individual results agree within
 the statistical precision quoted, however there are also significant
 differences as discussed above. 
 These differences are expected because the previous analyses of the
 individual LEP experiments~\cite{OPAL,DELPHI,L3C-9803L3C-9804,ALEPH}
 showed that the systematic errors, which are not included above,
 dominate.
 Because the Monte Carlo models do not sufficiently well resemble the 
 data, evaluating the experimental systematic errors of the measurements
 based on these models will not be very reliable. 
 On the other hand a combined result is desirable in order to serve as a 
 guidance for the model builders to improve on their Monte Carlo programs.
 Therefore, in a first attempt to make a combined measurement, the 
 experimental systematic error is taken from the spread of the measured
 results, wherever they are significantly different on the basis of the 
 statistical error alone. In this case the purely statistical error of the
 combined result is inflated as discussed below.
 \par
\begin{figure}
\begin{center}
\epsfig{file=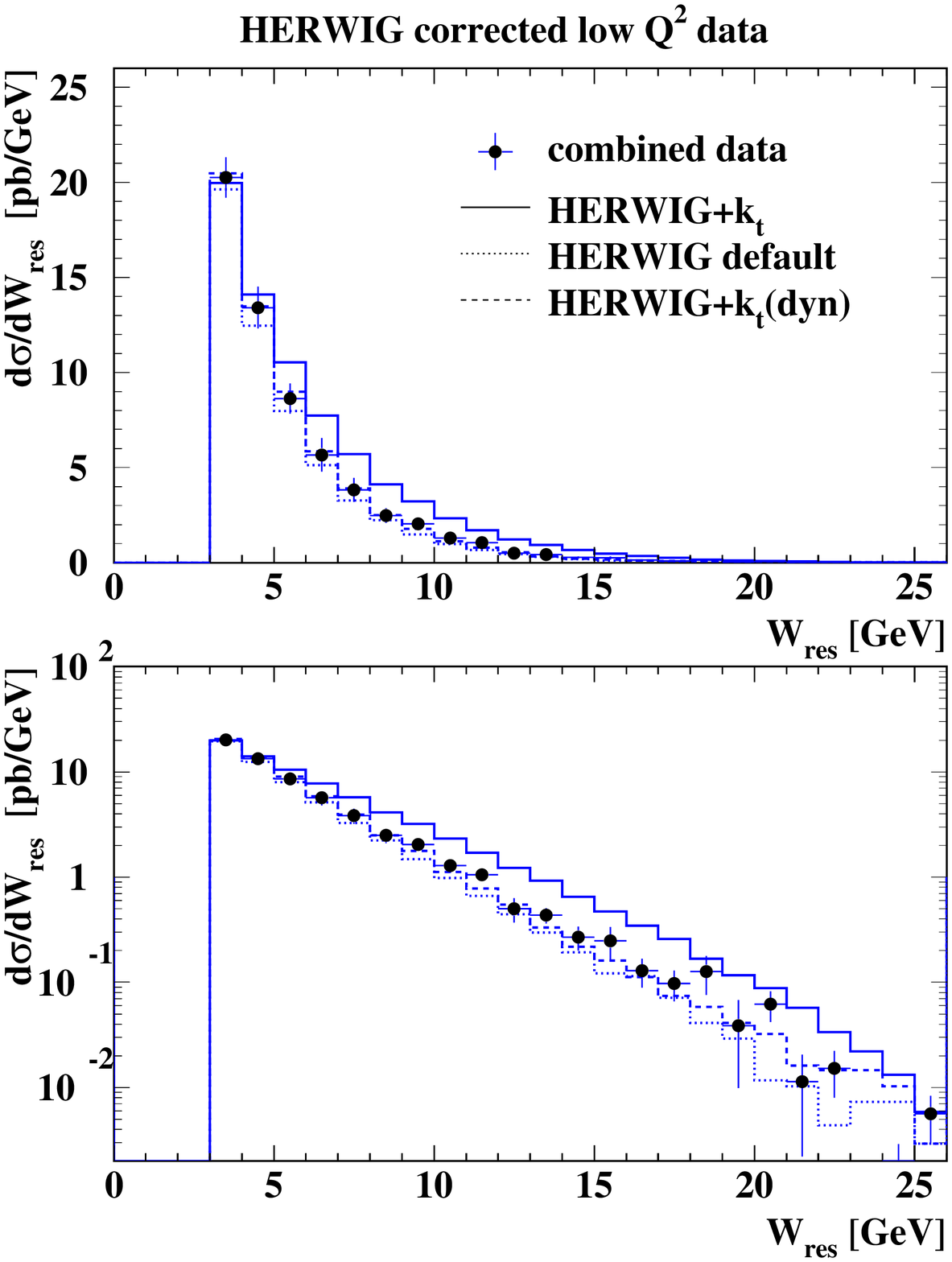,width=0.49\linewidth}
\epsfig{file=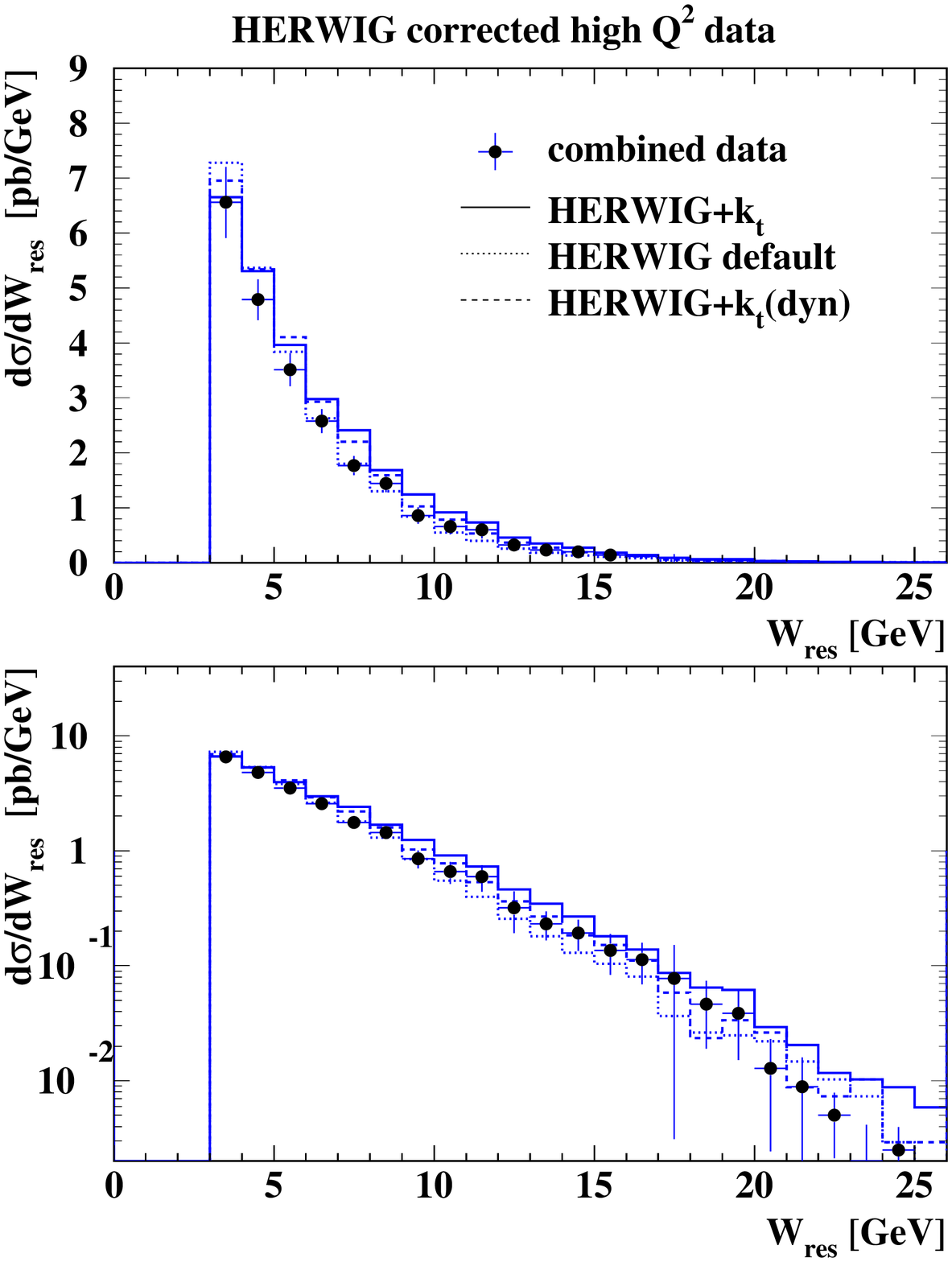,width=0.49\linewidth}
\caption{\label{fig:newkt1} 
         The combined \Wres distribution from ALEPH, L3 and OPAL
         for the low-\qsq (left) and high-\qsq region (right), corrected with 
         the HERWIG+\kt model on a linear scale (top) and on a 
         log scale (bottom).
         The data are compared to three different model assumptions 
         of the HERWIG+\kt model.
        }
\end{center}
\end{figure}
\begin{figure}
\begin{center}
\vspace{-1.cm}
\epsfig{file=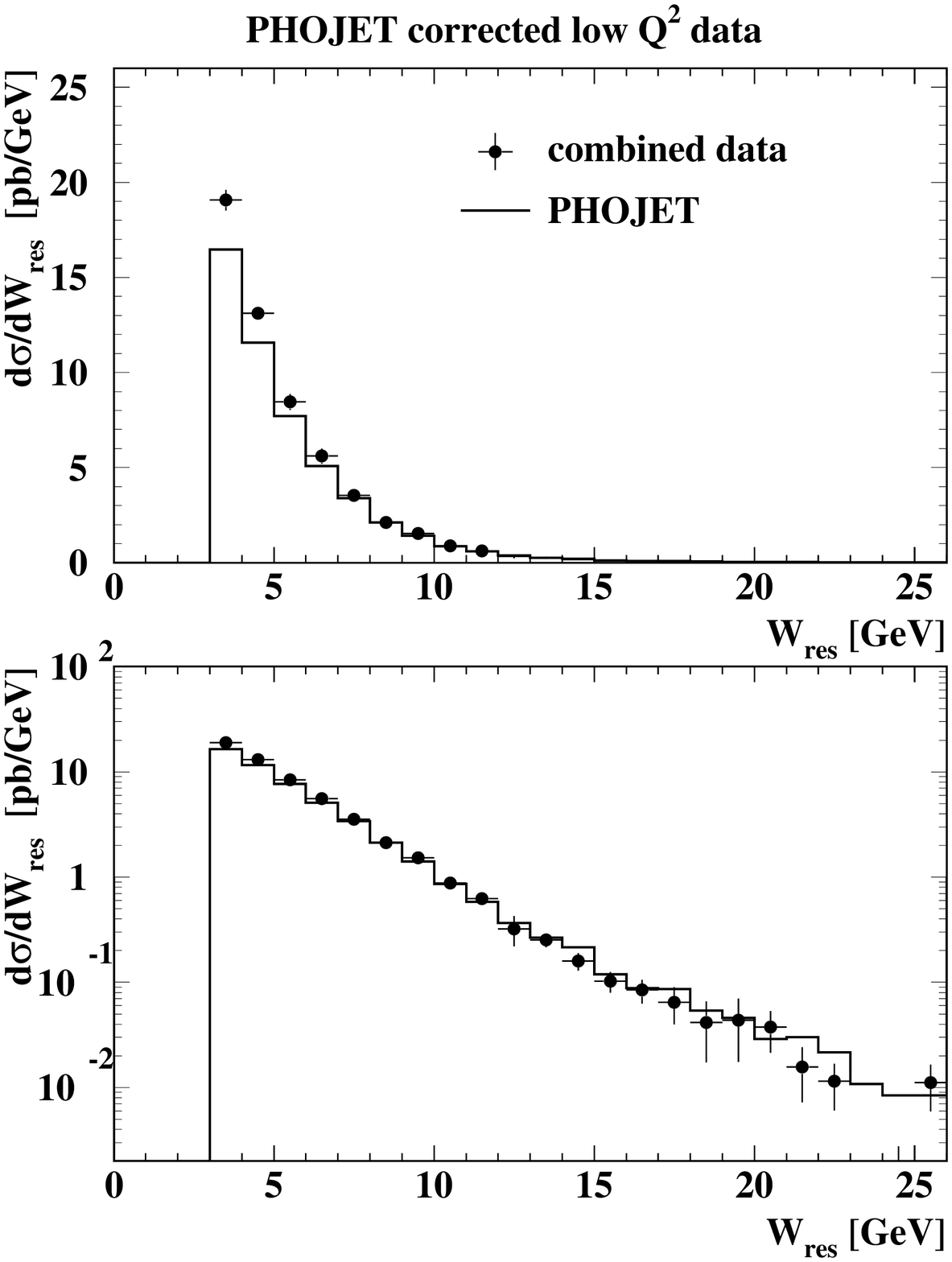,width=0.49\linewidth}
\epsfig{file=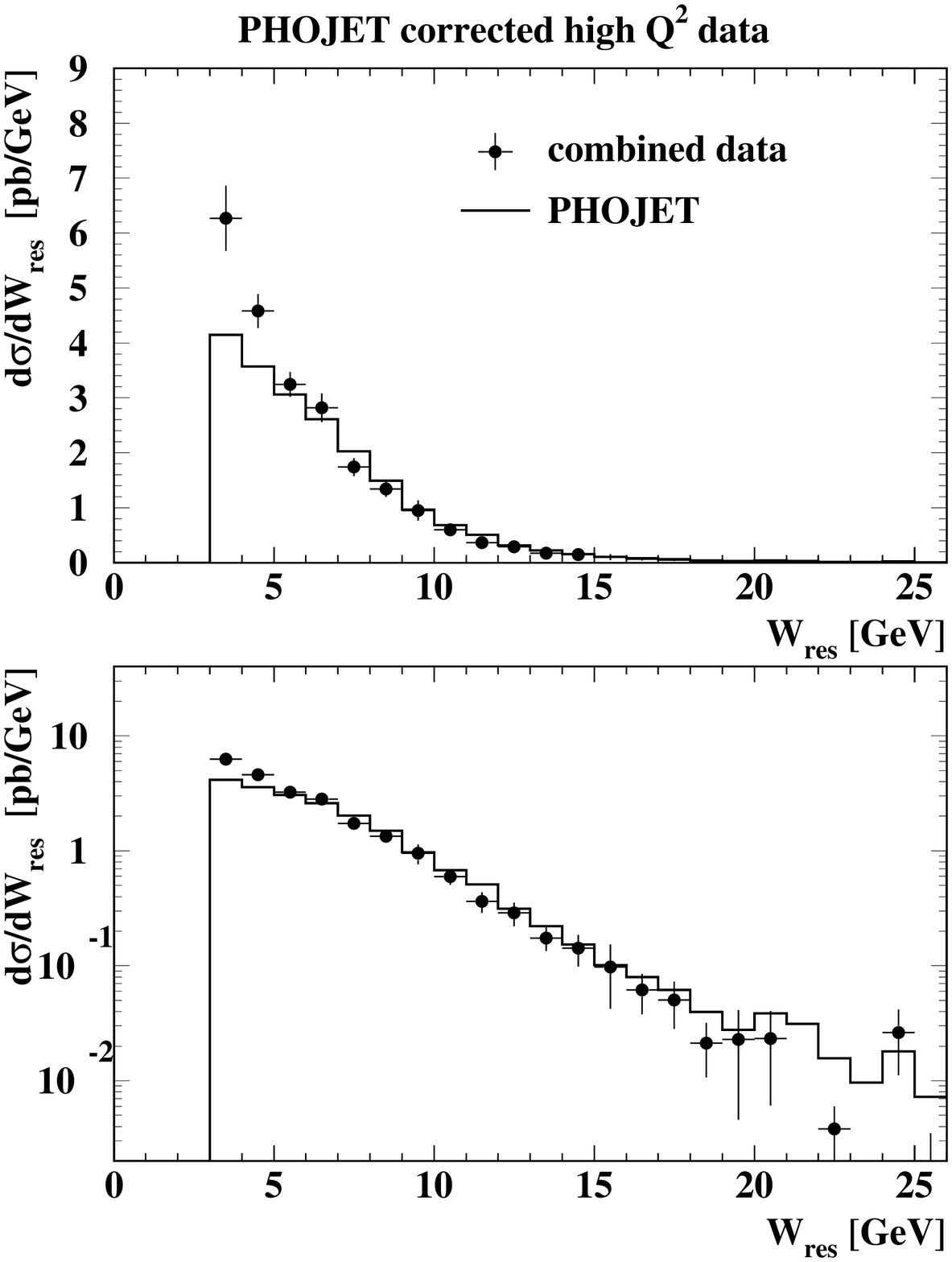,width=0.49\linewidth}
\caption{\label{fig:ph1} 
         The combined \Wres distribution from ALEPH, L3 and OPAL
         for the low-\qsq (left) and high-\qsq region (right), corrected with 
         and compared to the PHOJET model.
        }
\end{center}
\end{figure}
\begin{figure}
\begin{center}
\epsfig{file=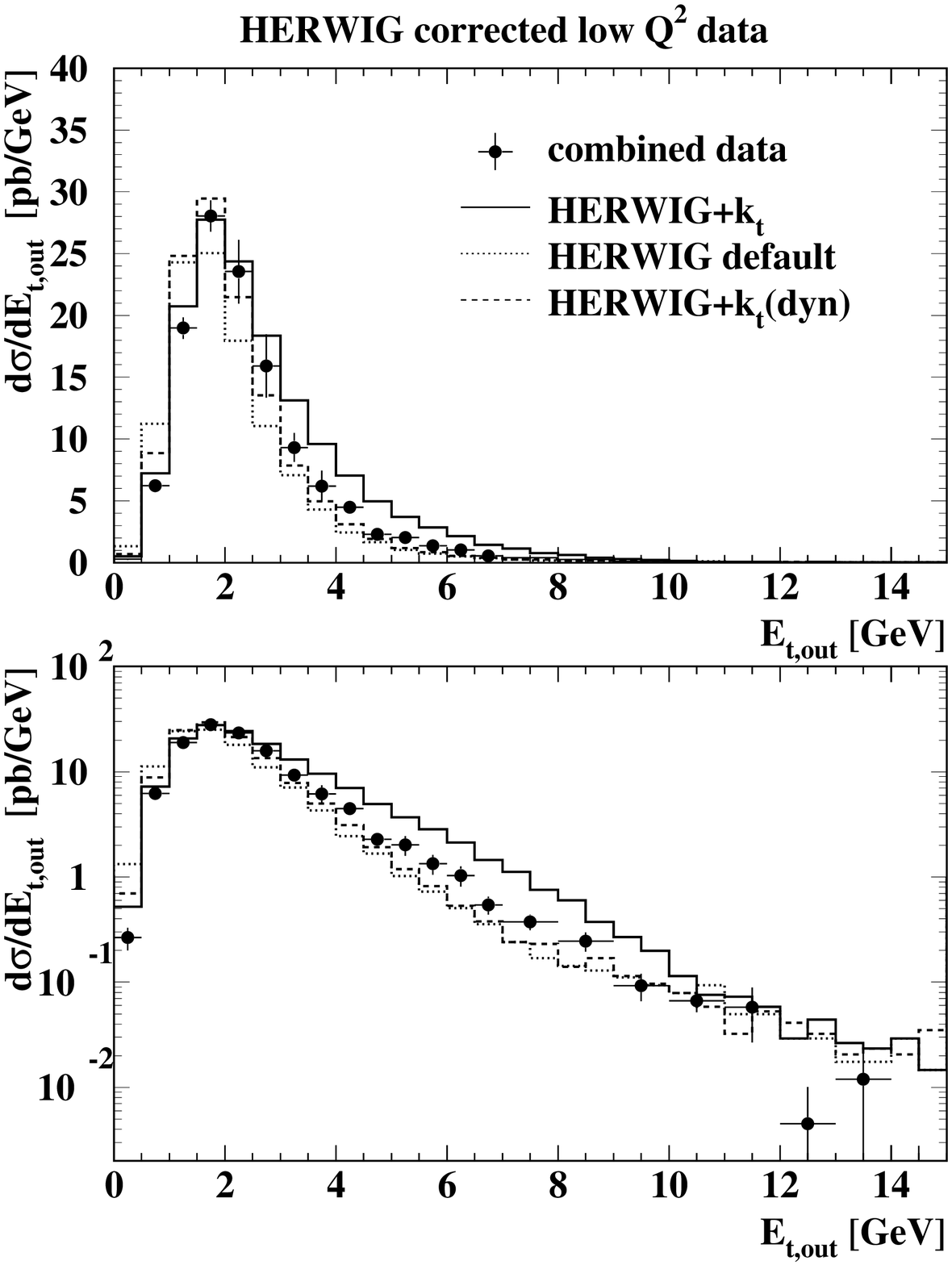,width=0.49\linewidth}
\epsfig{file=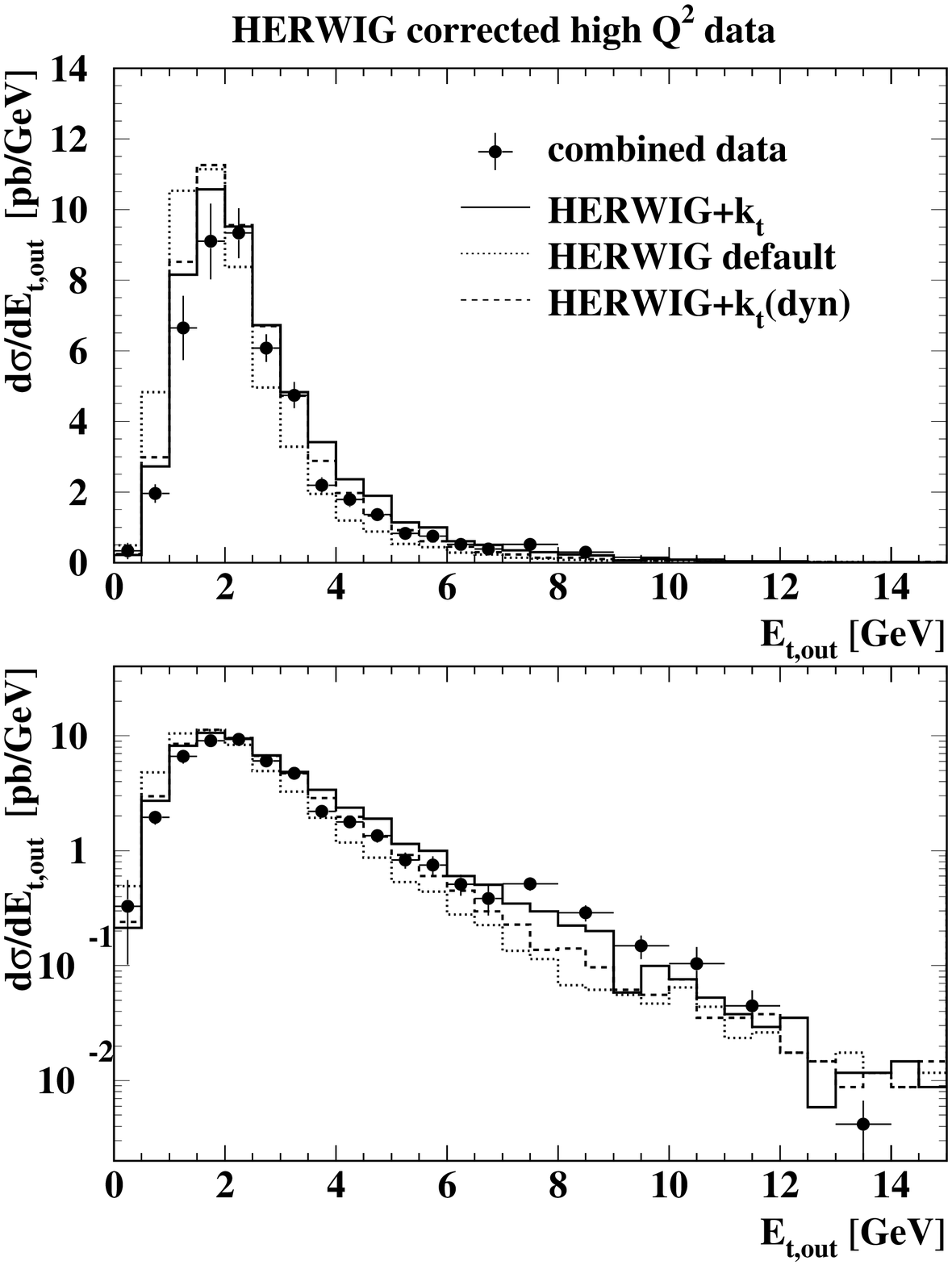,width=0.49\linewidth}
\caption{\label{fig:newkt2} 
         The combined \etout distribution from ALEPH, L3 and OPAL
         for the low-\qsq (left) and high-\qsq region(right), corrected with 
         the HERWIG+\kt model on a linear scale (top) and on a 
         log scale (bottom).
         The data are compared to three different model assumptions 
         of the HERWIG+\kt model.
        }
\end{center}
\end{figure}
\begin{figure}
\begin{center}
\vspace{-1.cm}
\epsfig{file=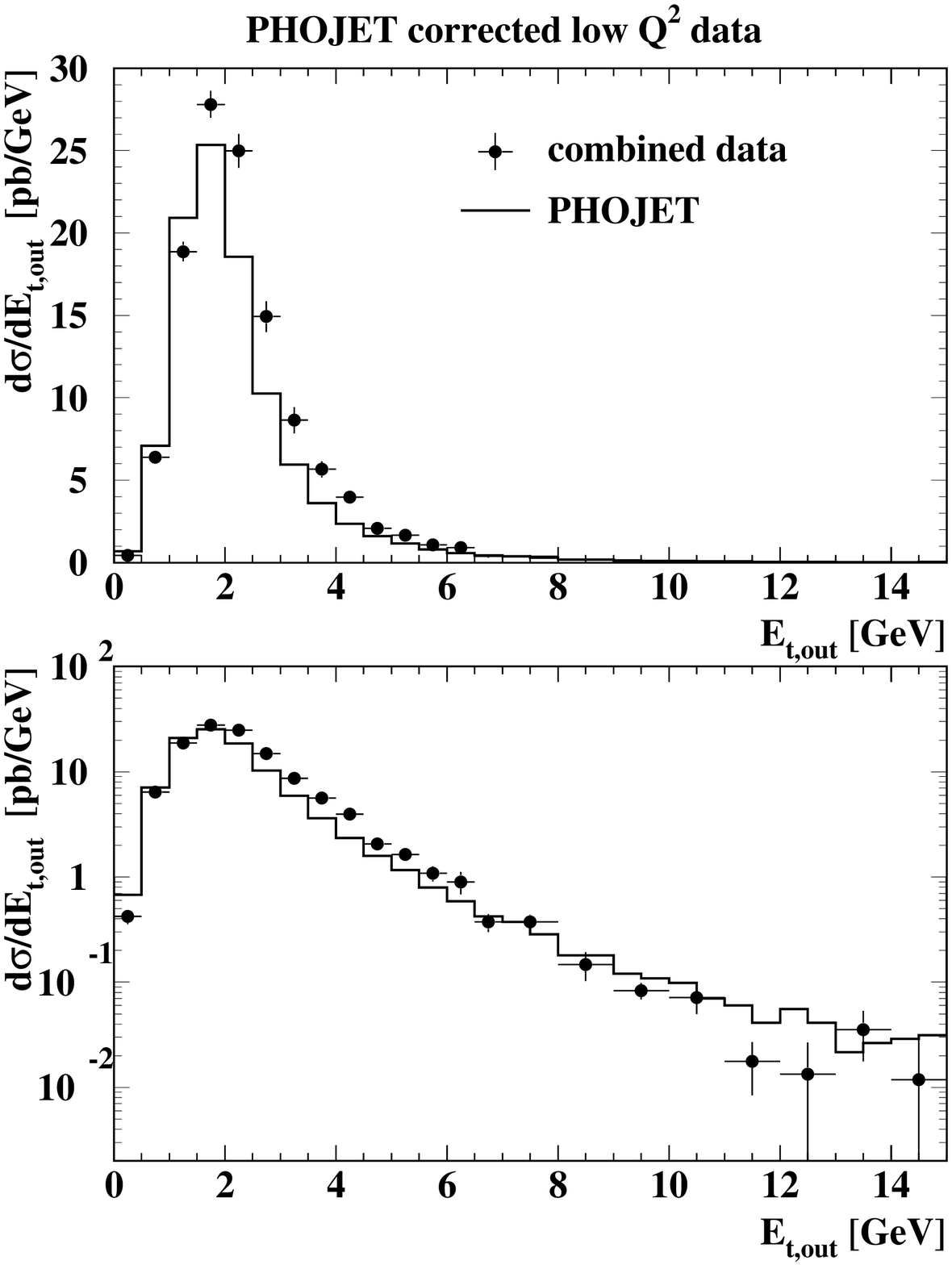,width=0.49\linewidth}
\epsfig{file=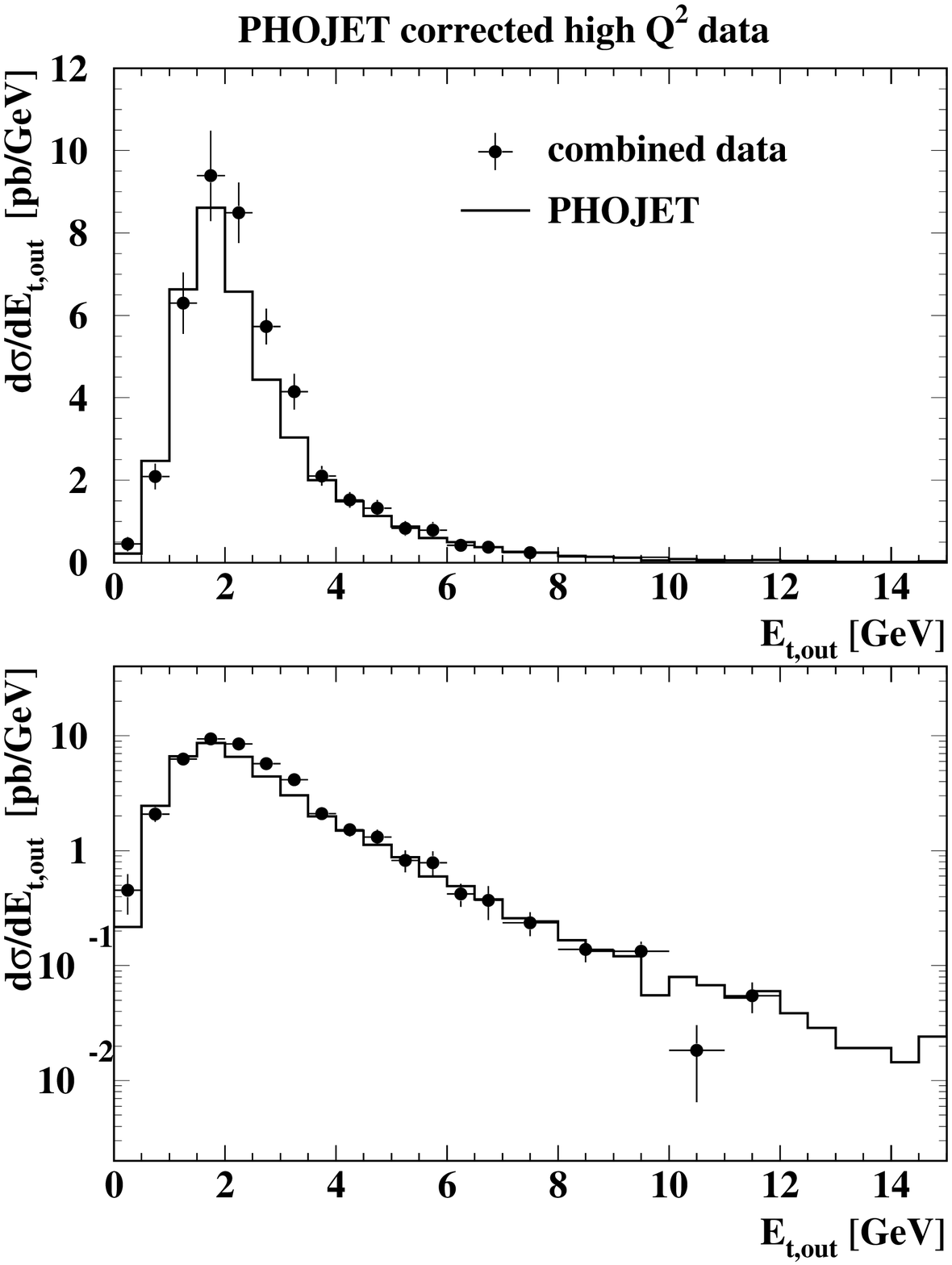,width=0.49\linewidth}
\caption{\label{fig:ph2} 
         The combined \etout distribution from ALEPH, L3 and OPAL
         for the low-\qsq (left) and high-\qsq region(right), corrected with 
         and compared to the PHOJET model.
        }
\end{center}
\end{figure}
\begin{figure}
\begin{center}
\epsfig{file=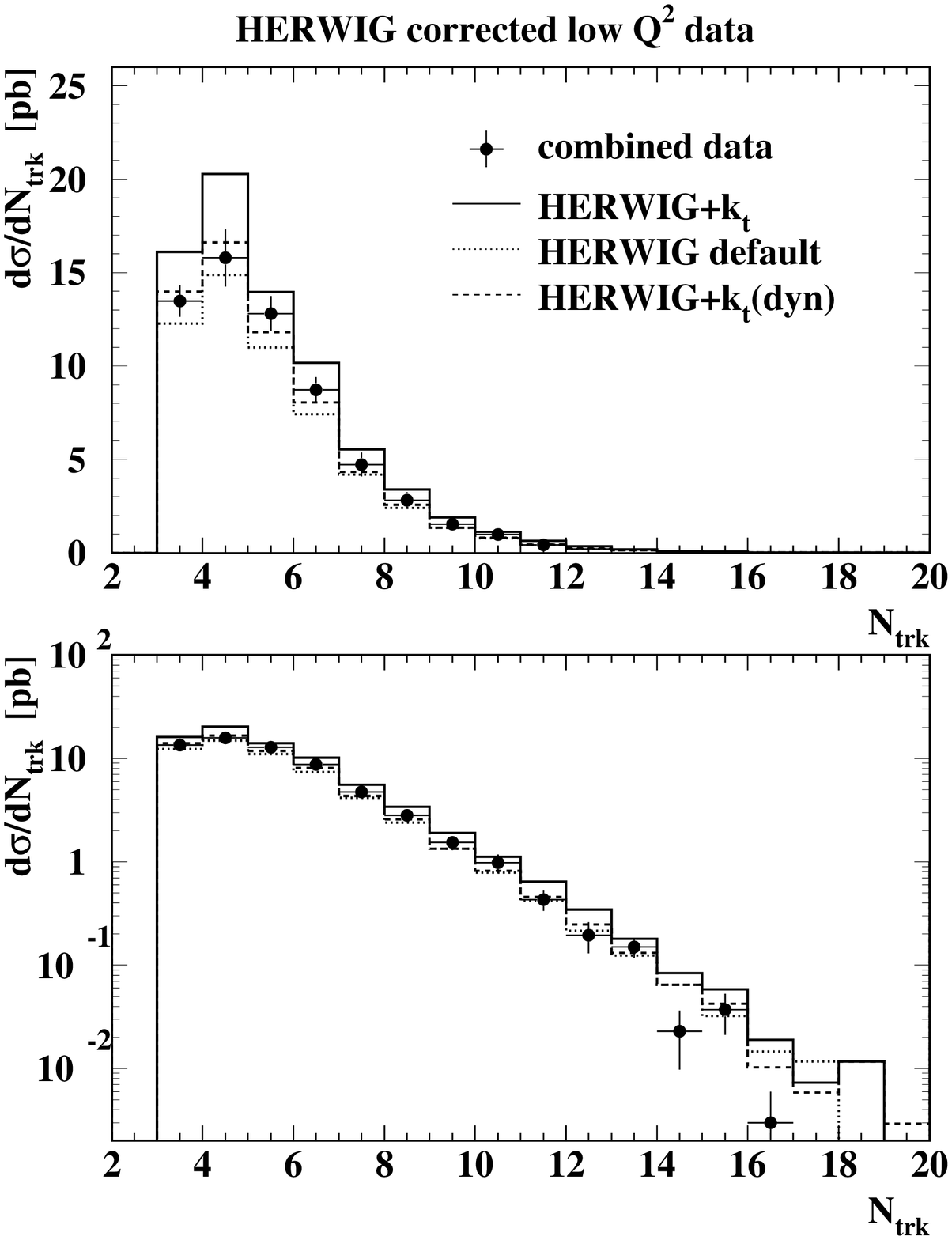,width=0.49\linewidth}
\epsfig{file=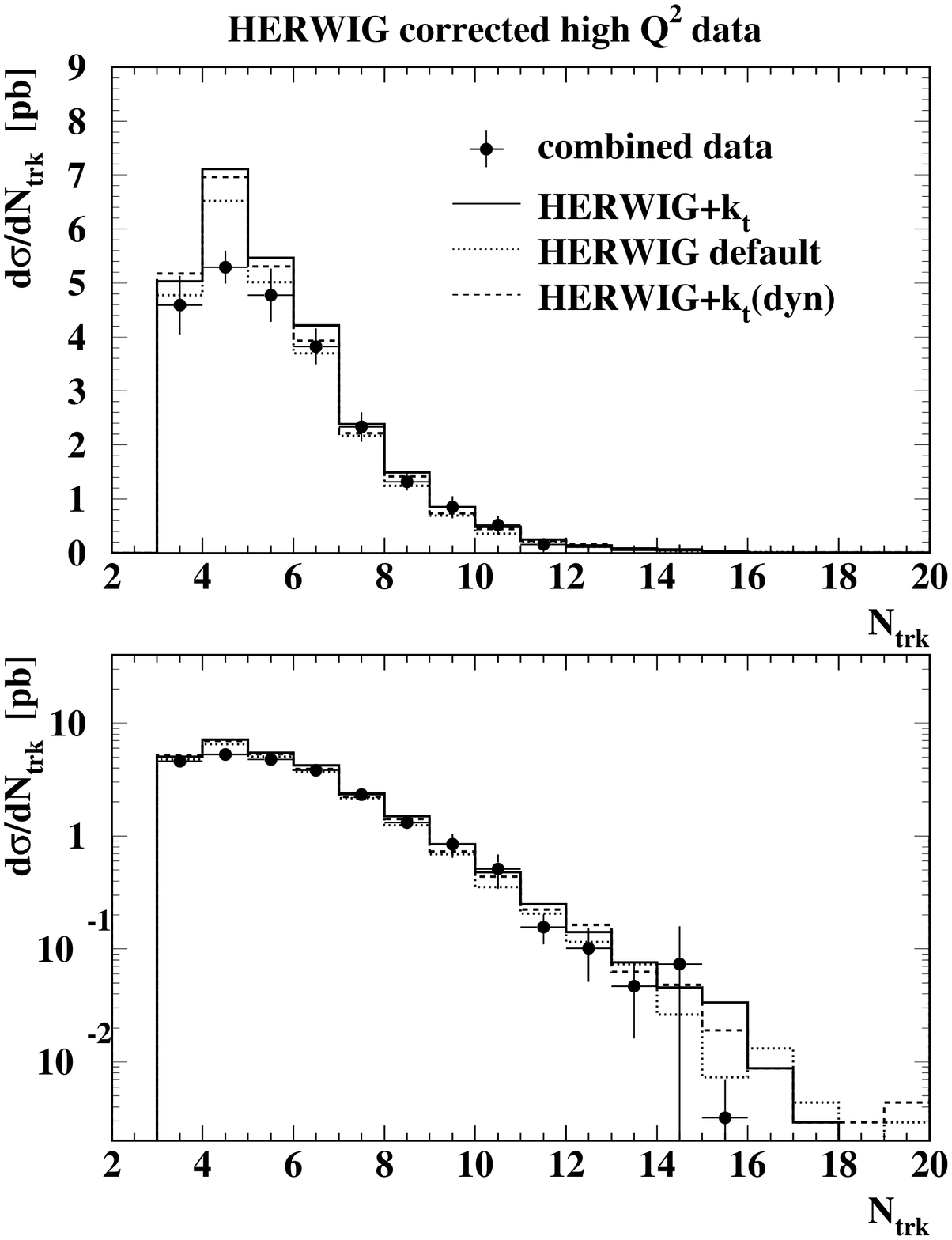,width=0.49\linewidth}
\caption{\label{fig:newkt3} 
         The combined \nch distribution from ALEPH, L3 and OPAL
         for the low-\qsq (left) and high-\qsq region (right), corrected with 
         the HERWIG+\kt model on a linear scale (top) and on a 
         log scale (bottom).
         The data are compared to three different model assumptions 
         of the HERWIG+\kt model.
        }
\end{center}
\end{figure}
\begin{figure}
\begin{center}
\vspace{-1.cm}
\epsfig{file=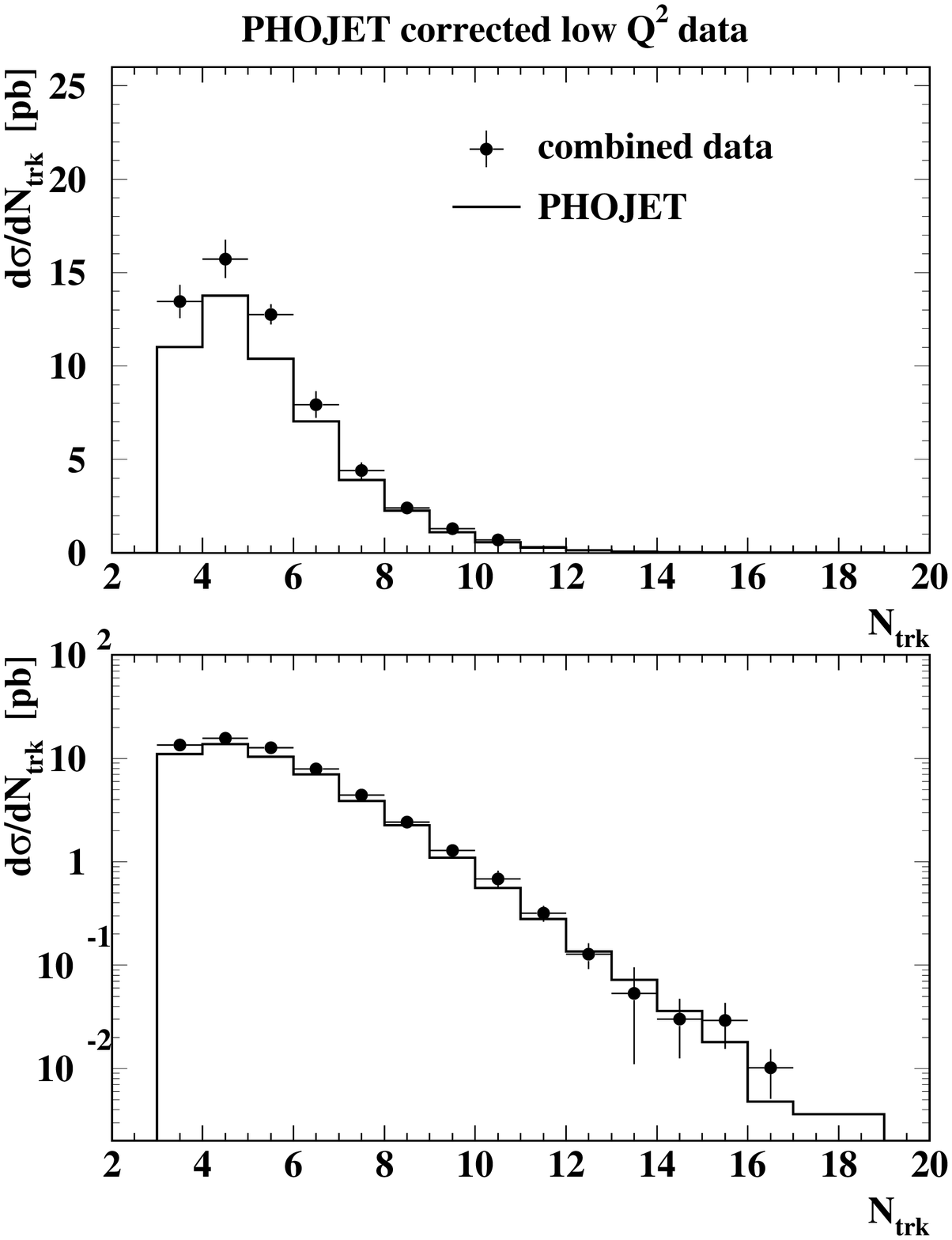,width=0.49\linewidth}
\epsfig{file=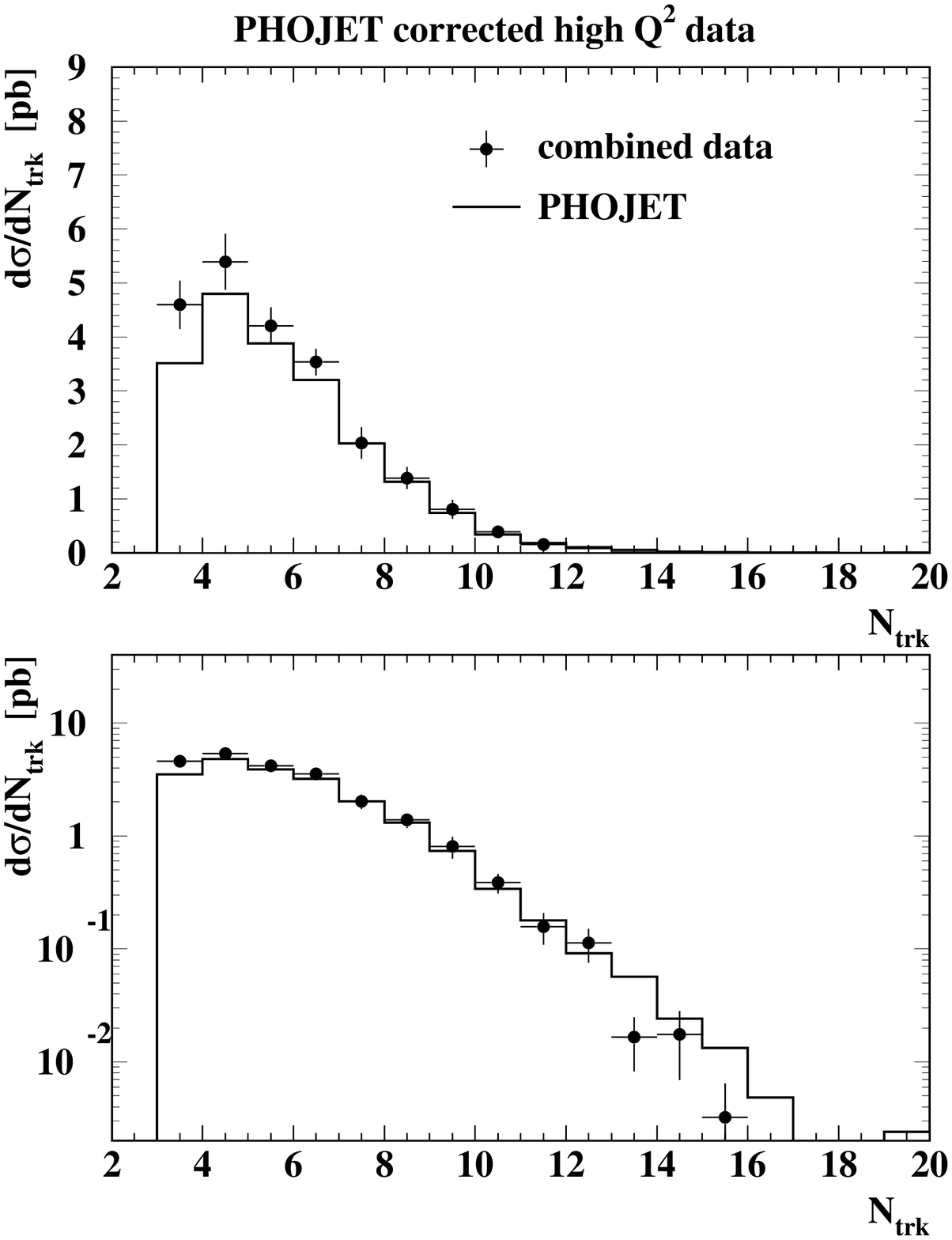,width=0.49\linewidth}
\caption{\label{fig:ph3} 
         The combined \nch distribution from ALEPH, L3 and OPAL
         for the low-\qsq (left) and high-\qsq region (right), corrected with 
         and compared to the PHOJET model.
        }
\end{center}
\end{figure}
\begin{figure}
\begin{center}
\epsfig{file=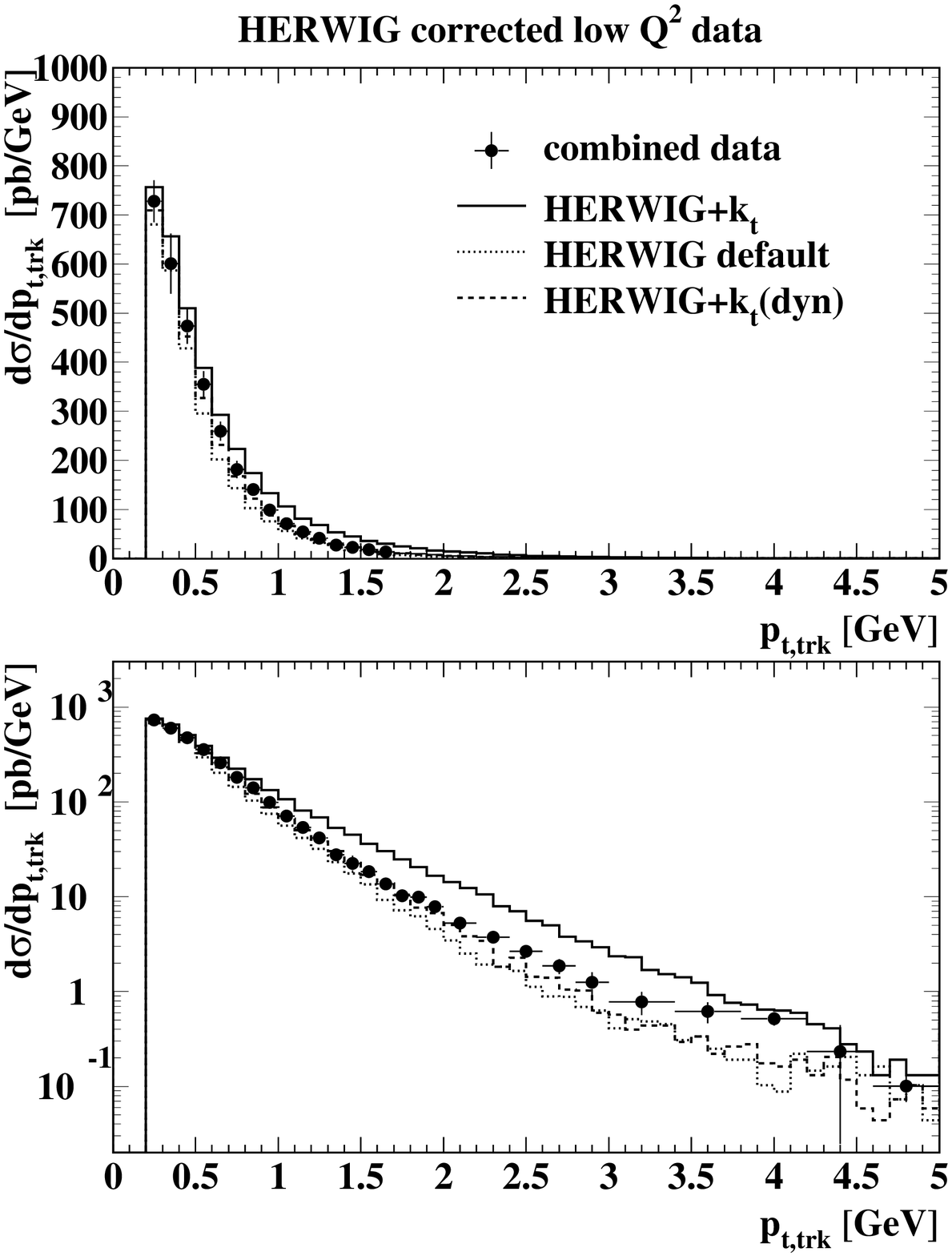,width=0.49\linewidth}
\epsfig{file=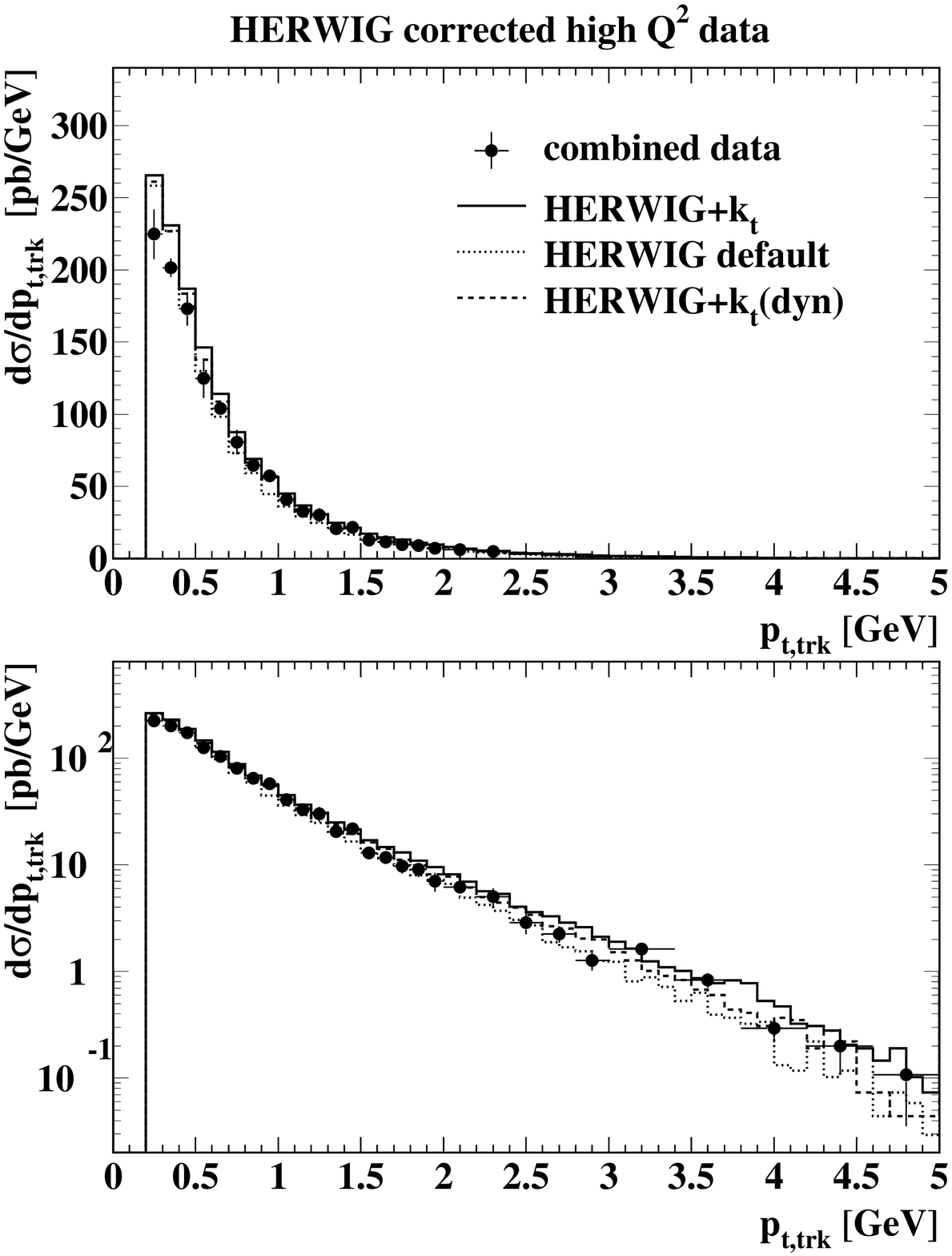,width=0.49\linewidth}
\caption{\label{fig:newkt4} 
         The combined \ptch distribution from ALEPH, L3 and OPAL
         for the low-\qsq (left) and high-\qsq region (right), corrected with 
         the HERWIG+\kt model on a linear scale (top) and on a 
         log scale (bottom).
         The data are compared to three different model assumptions 
         of the HERWIG+\kt model.
        }
\end{center}
\end{figure}
\begin{figure}
\begin{center}
\vspace{-1.cm}
\epsfig{file=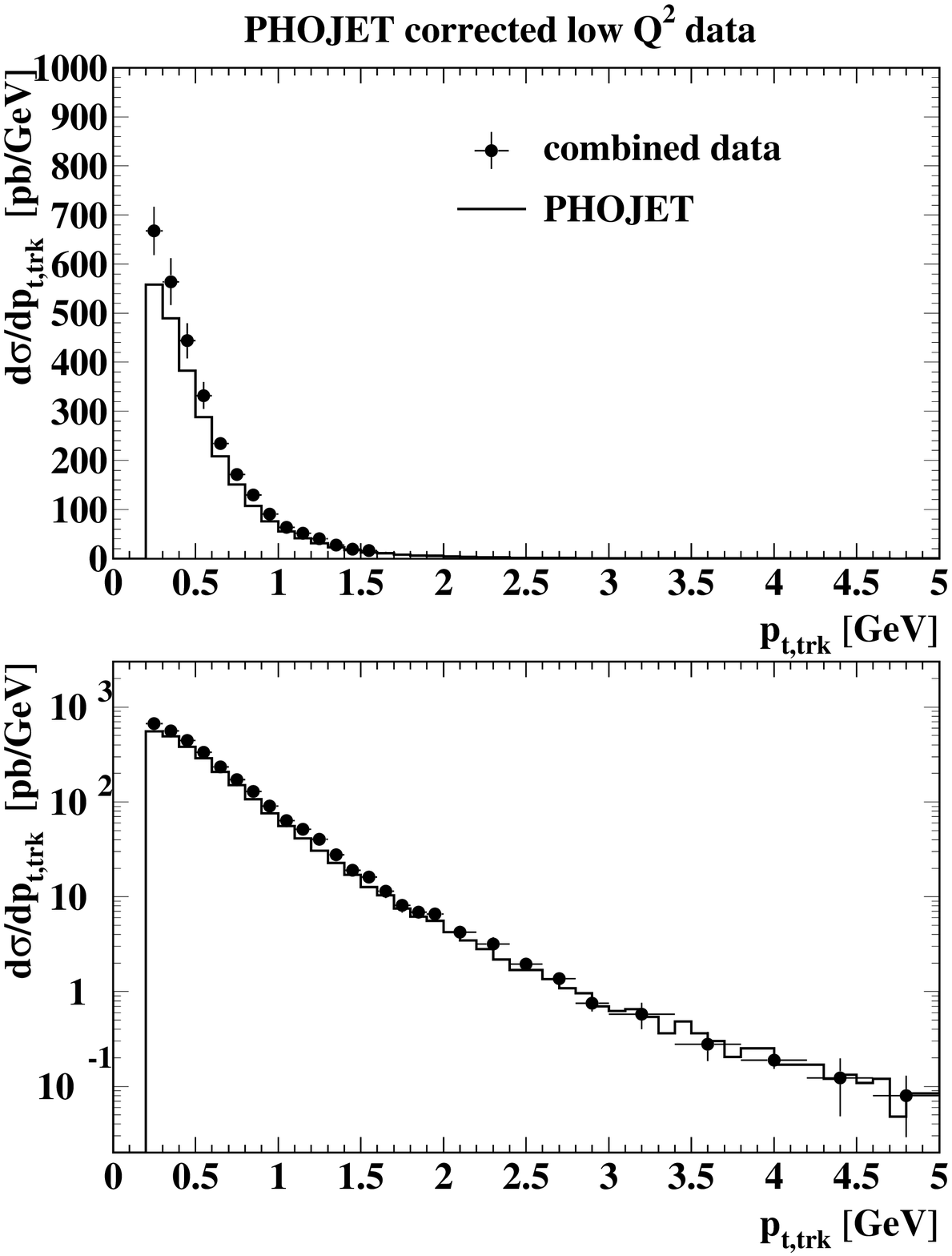,width=0.49\linewidth}
\epsfig{file=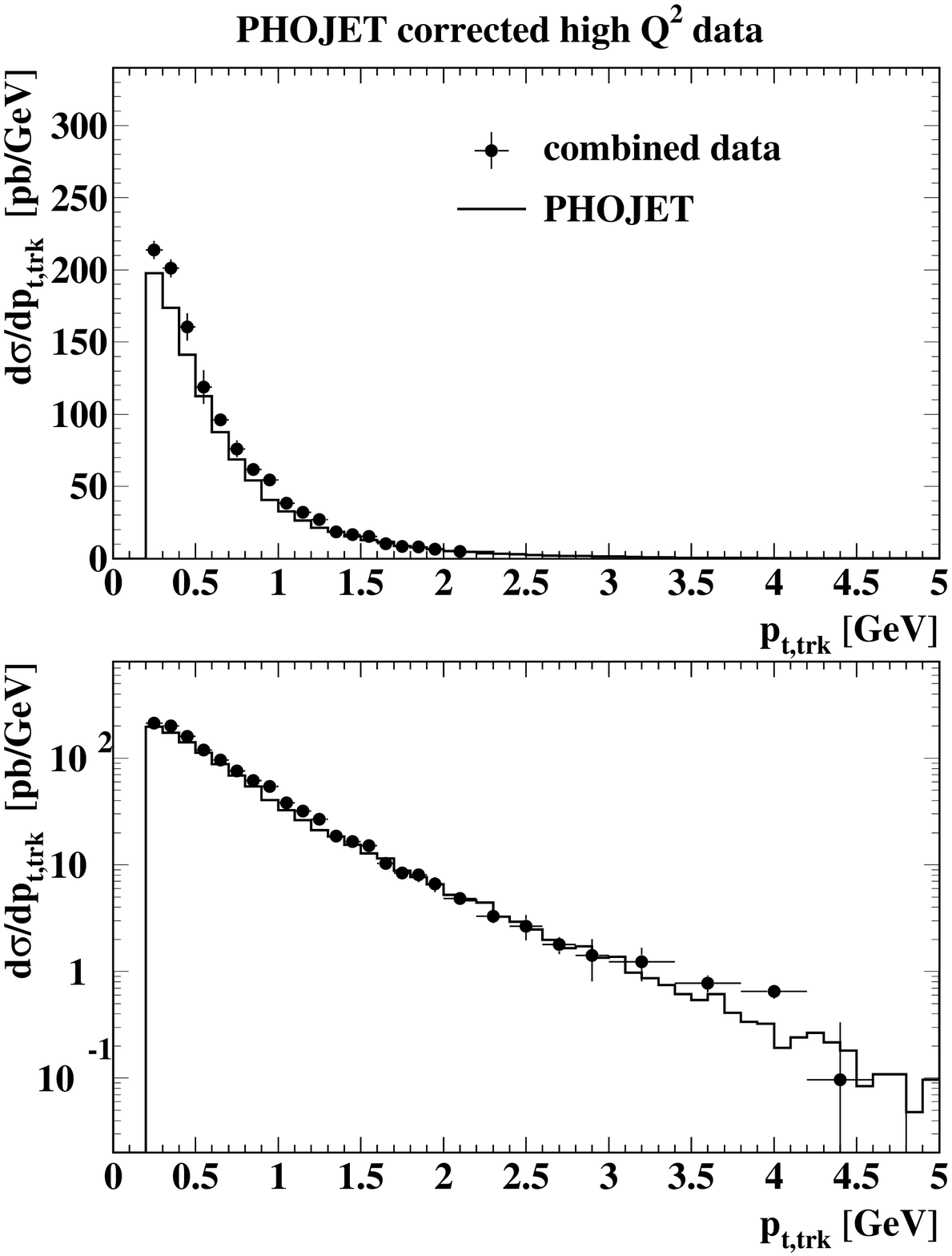,width=0.49\linewidth}
\caption{\label{fig:ph4} 
         The combined \ptch distribution from ALEPH, L3 and OPAL
         for the low-\qsq (left) and high-\qsq region (right), corrected with 
         and compared to the PHOJET model.
        }
\end{center}
\end{figure}
\begin{figure}
\begin{center}
\epsfig{file=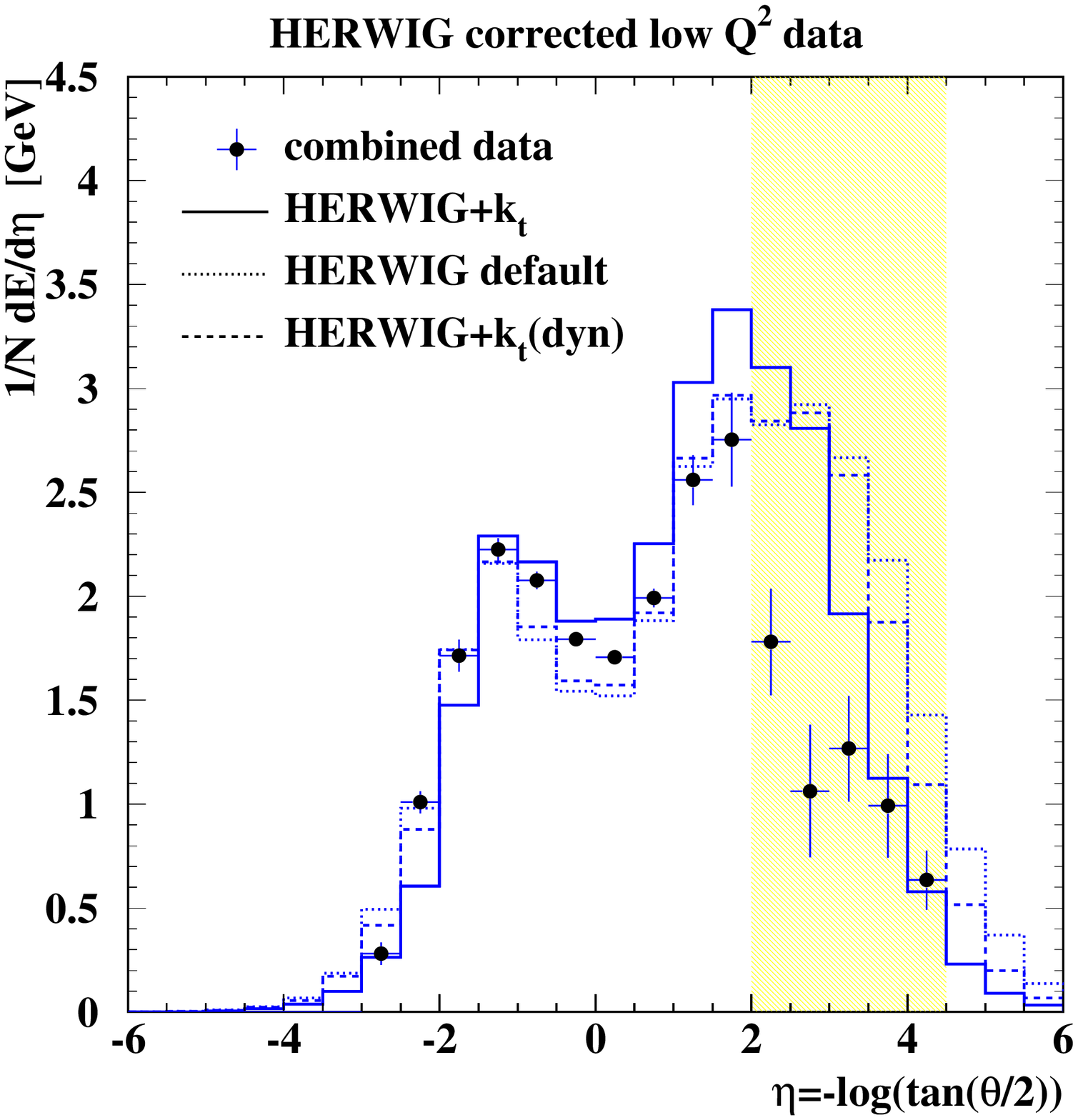,width=0.49\linewidth}
\epsfig{file=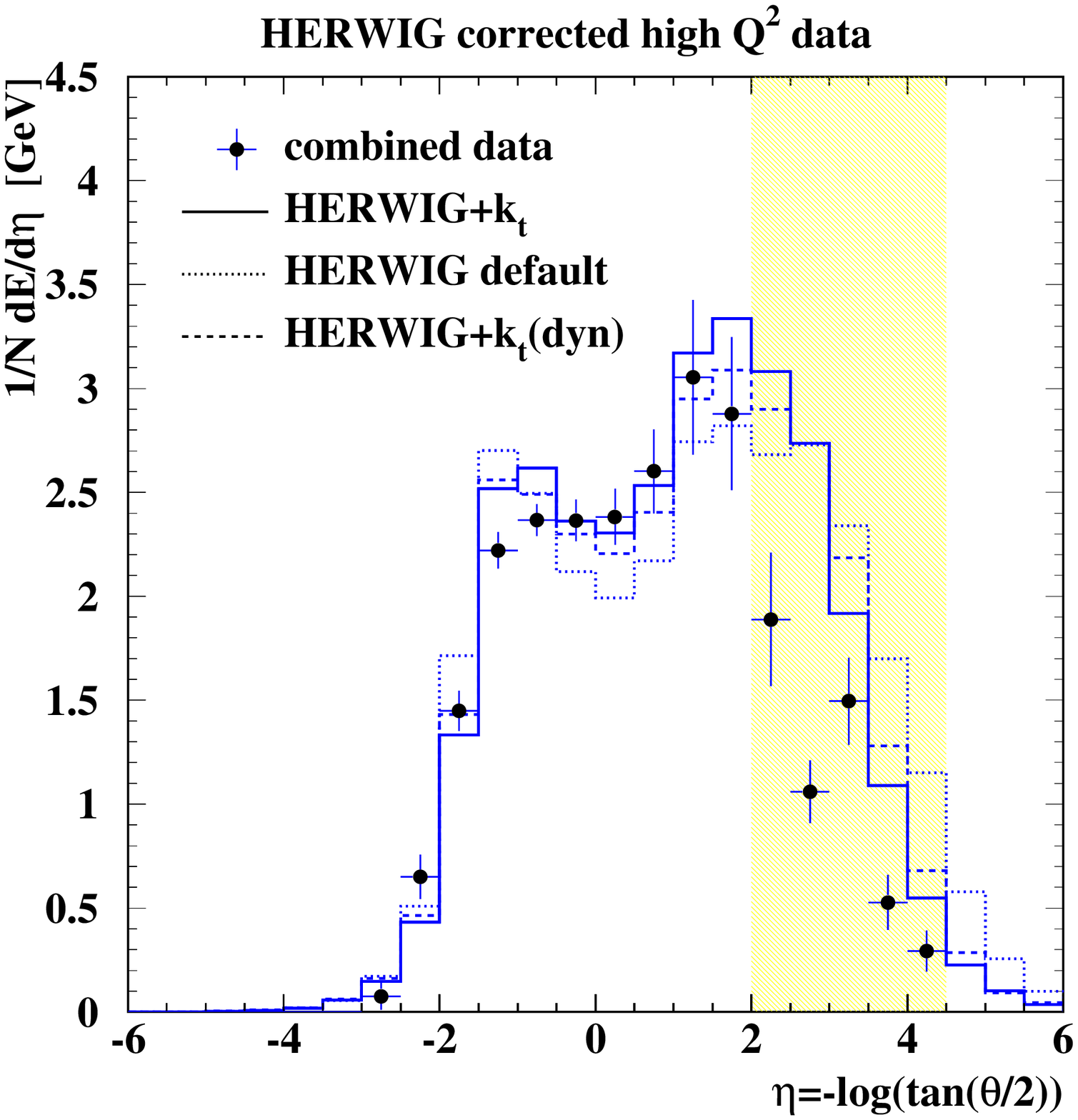,width=0.49\linewidth}
\caption{\label{fig:newkt5} 
         The combined hadronic energy flow from ALEPH, L3 and OPAL
         for the low-\qsq (left) and high-\qsq region (right), corrected with 
         the HERWIG+\kt model.
         The data are compared to three different model assumptions 
         of the HERWIG+\kt model. 
         The shaded band indicates the forward region of the experiments.
        }
\end{center}
\end{figure}
\begin{figure}
\begin{center}
\epsfig{file=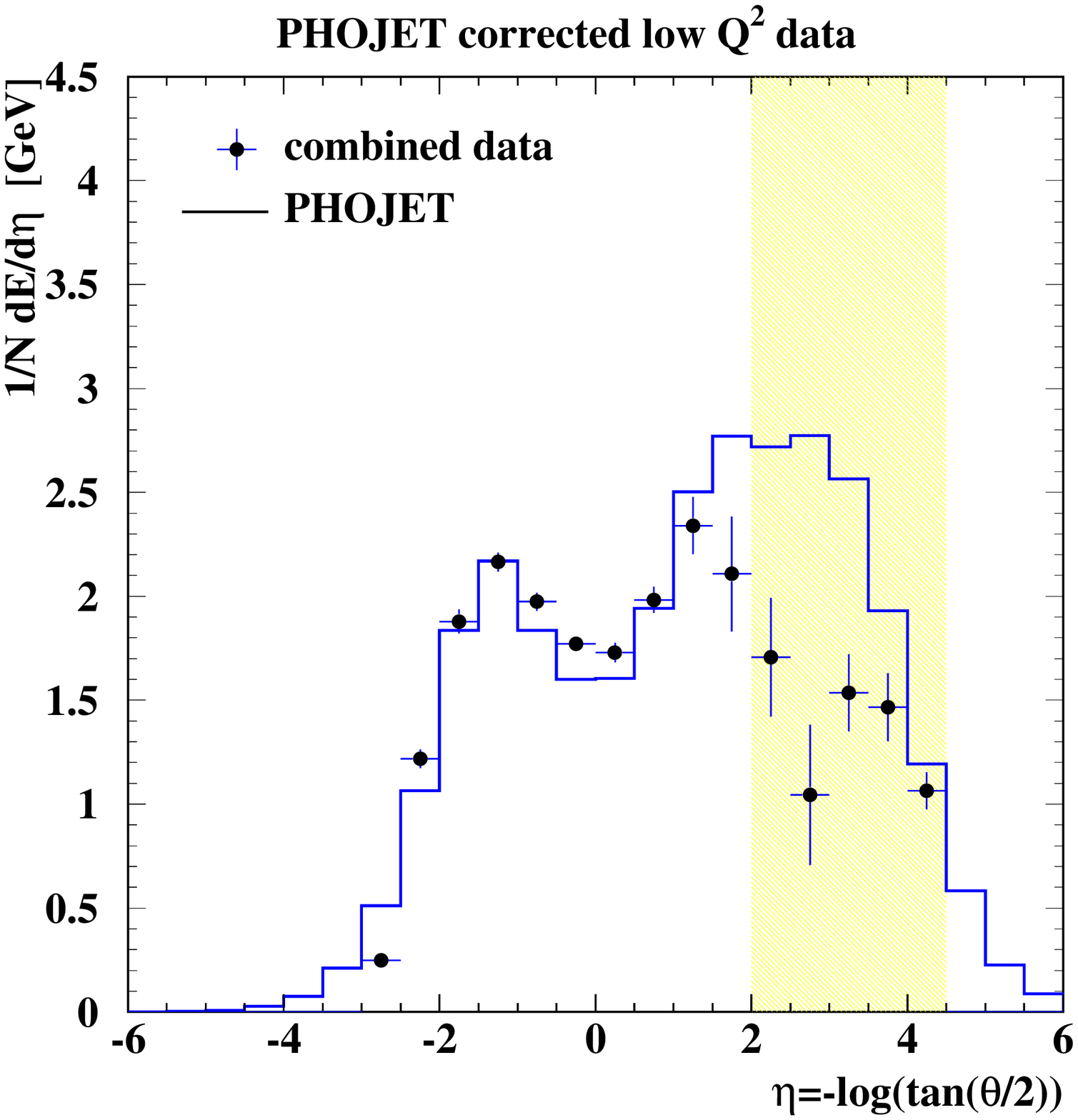,width=0.49\linewidth}
\epsfig{file=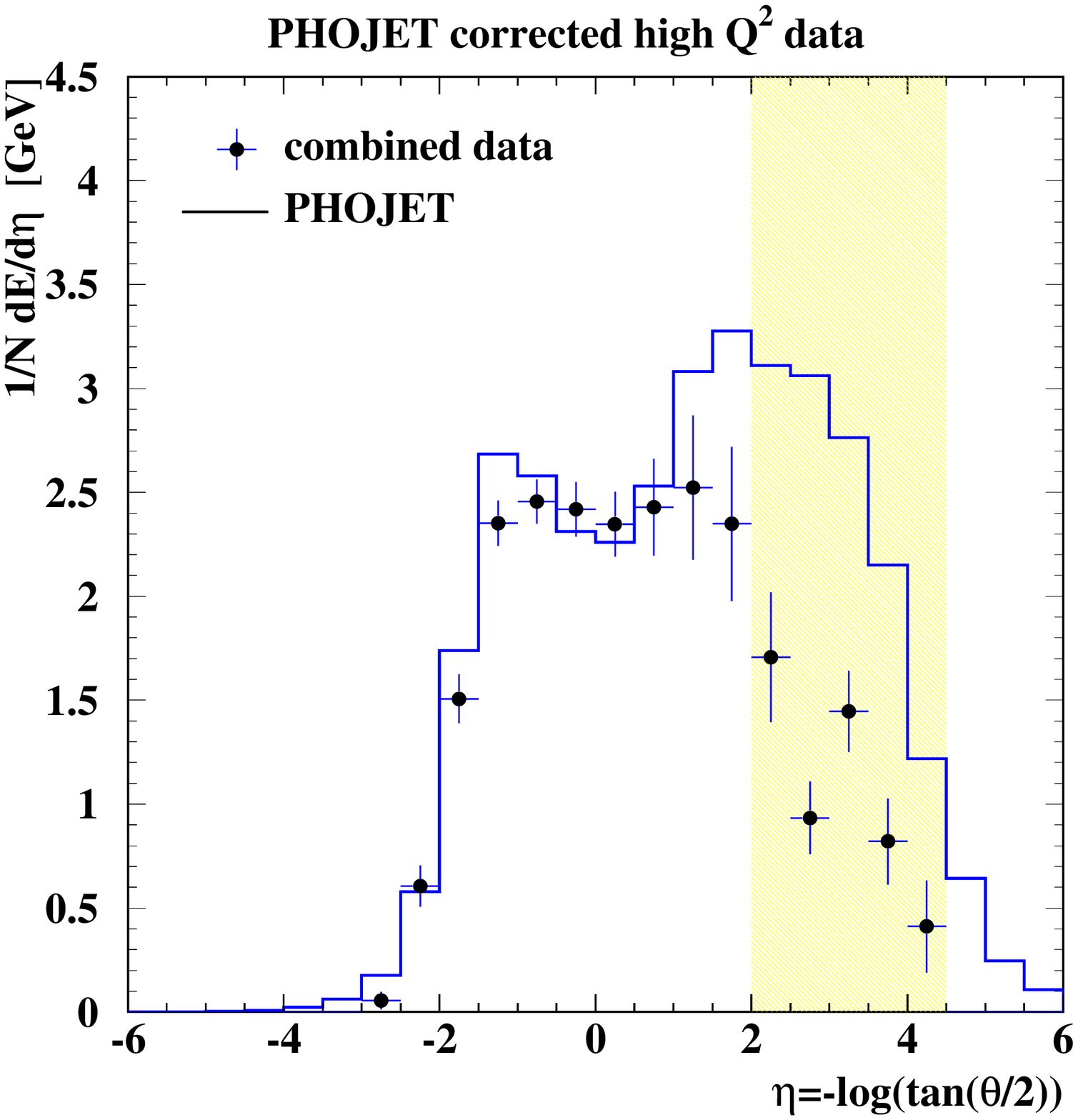,width=0.49\linewidth}
\caption{\label{fig:ph5} 
         The combined hadronic energy flow from ALEPH, L3 and OPAL
         for the low-\qsq (left) and high-\qsq region (right), corrected with 
         the PHOJET model.
         The data are compared to the PHOJET model.
         The shaded band indicates the forward region of the experiments.
        }
\end{center}
\end{figure}

 The combined distributions from ALEPH, L3 and OPAL are shown
 in comparison to PHOJET and various predictions from HERWIG
 in Figures~\ref{fig:newkt1}--\ref{fig:ph5}. 
 The measured values are listed in Tables~\ref{tab:Wave}--\ref{tab:pefave}.
 The combination of the data follows the procedure recommended by the 
 Particle Data Group in section 4.2.2 of~\cite{PDG-9801}.
 Since this is a crucial point of the analysis and because specific
 choices have to be made in the combination, the procedure is 
 briefly discussed below.
 \par
 The combined value for a given bin is calculated as the weighted
 average of the measurements of the individual experiments using
 the statistical errors to calculate both the weights and the error 
 of the combined value.
 To calculate the average bin content $\overline{x}$ and its error 
 $\sigma_{\overline{x}}$ from the individual contents $x_{\rm i}$ and
 their statistical errors $\sigma_{\rm i}$ the following procedure is applied.
 The average content is calculated from 
 $\overline{x}=\sum w_{\rm i}\cdot x_{\rm i}/\sum w_{\rm i}$, 
 using the weights $w_{\rm i}= 1/\sigma_{\rm i}^2$.
 The sum runs over
 all experiments $i=1,\ldots,\Nex$ with non-zero entries in that bin, and
 the error $\sigma_{\overline{x}}$ is taken to be $1/\sqrt{\sum w_{\rm i}}$.
 The $\chi^2$ of the average is calculated from 
 $\chi^2=\sum w_{\rm i}(\overline{x}-x_{\rm i})^2$.
 If in the tail of a distribution, for example, some experiments 
 measure zero in a particular bin, then $\overline{x}$
 and $\chi^2$ are scaled by the ratio of the number of experiments
 with nonzero entries to the number of experiments being averaged
 for that bin.
 If $\chi^2$ is larger than $\Nex-1$, the error is increased by a factor
 $S=\scale$, thereby taking the spread of the experiments as an estimate
 of the experimental systematic uncertainty.
 Finally, to obtain the errors quoted the uncertainties due to the correction 
 factors are added in quadrature.
 Since the same data sets are used by each experiment 
 to calculate the correction factors, the corresponding errors 
 are strongly correlated between the experiments.
 To take this into account, the error on the correction factors is included
 by taking the smallest quoted error of the individual experiments
 as an estimate of this systematic error, which is assumed to be 
 100$\%$ correlated amongst the experiments. 
 \par
 Also in the combination of the results of the individual experiments
 the \flow distribution is treated slightly differently from the other 
 distributions. 
 There is a large scatter in the measured values of the different 
 experiments, especially in the forward region, as can be seen from
 Figures~\ref{fig:data11} and~\ref{fig:data12}.
 As a consequence there is also a large scatter in the scale factors
 listed in Tables~\ref{tab:hefave} and~\ref{tab:pefave}.
 To avoid the combination procedure manufacturing artificially
 small errors for bins where the measurements happen to coincide, 
 the scale factor applied to obtain the combined measurement is taken
 as the average of the individual scale factors from
 three neighbouring bins centered around the bin under study.
 \par
 It is apparent from 
 Figures~\ref{fig:newkt1},~\ref{fig:newkt2},~\ref{fig:newkt3},~\ref{fig:newkt4}
 and~\ref{fig:newkt5}, that the new HERWIG+\kt(dyn) with the dynamical 
 cut-off lies much closer to the low-\qsq averaged data than the version
 of HERWIG+\kt using the fixed cut-off. 
 However, the difference between HERWIG+\kt(dyn) and HERWIG default is 
 rather small.
 \par
 In the case of the energy flow none of the various HERWIG models
 is able to accurately describe the data. 
 This suggests that even though the new HERWIG+\kt(dyn) better describes most
 of the data distributions, the energy flow is still not well understood.
 \par
 The PHOJET model, 
 Figures~\ref{fig:ph1},~\ref{fig:ph2},~\ref{fig:ph3},~\ref{fig:ph4} 
 and~\ref{fig:ph5}, describes the
 data reasonably well in both regions of \qsq, but underestimates
 the cross-section near the lower limit of the distributions.
 This is understood as a consequence of the high transverse momentum 
 cut-off of 2.5~GeV for the scattered partons in the hard scattering
 matrix element.
 Below this cut-off only soft events are generated. Due to the
 different \qsq dependences of the soft and hard components in PHOJET, this
 leads to a suppression of the low-$W$ events~\cite{RalphPrivatComm}.
 \par
 The energy flow corrected with HERWIG+\kt and with PHOJET
 Figures~\ref{fig:newkt5} and~\ref{fig:ph5}, mostly agree with each
 other within errors, except in a few bins in the forward region. 
 In this region the data corrected with PHOJET lie
 below the data corrected with HERWIG+\kt in the region of the peak at
 $\eta\simeq 2$ and above the flow corrected with HERWIG+\kt in the
 region of $\eta>2.5$.
%
%
\section{Conclusion}
 For the first time the results of deep inelastic electron-photon 
 scattering from three of the LEP experiments have been combined and 
 compared to predictions from the PHOJET and the HERWIG models.
 It is found that the data from the ALEPH, L3 and OPAL experiments 
 agree within statistical errors except near the edges of
 the distributions. 
 Where the spread is larger than expected
 from the statistical errors, as, for example, for low charged
 multiplicities, this difference is taken as an estimate of the
 detector dependent systematic uncertainty of the measurement.
 \par 
 In the comparison of the data with the HERWIG+\kt model the most striking
 discrepancy is seen in the distributions of the low-\qsq region, 
 where the HERWIG+\kt model systematically overestimates the data.
 This discrepancy is found to be mainly due to the fixed cut-off for the 
 intrinsic transverse momentum of the quarks in the photon
 in the HERWIG+\kt model.
 By dynamically adjusting the cut-off according to the kinematics of
 the individual event in the HERWIG+\kt(dyn) model the description
 of the data is significantly improved, particularly in the low-\qsq
 region.
 \par
 The PHOJET model describes the data reasonably well in both regions of
 \qsq, but underestimate the cross-section near the lower limit of the
 distributions, due to the high transverse momentum cut-off for the 
 scattered partons in the hard scattering matrix element.
 \par
 The energy flow of the data lies between the predictions of the HERWIG 
 and PHOJET Monte Carlo models in the central regions of the detectors. 
 In the forward region the Monte Carlo predictions lie systematically above 
 the data.
 It should be noted, however, that it is difficult to assess the systematic
 errors in this region because of the poor resolution of the hadronic
 energy measured in the electromagnetic luminosity monitors.
 \par
 The method of combining the data of several of the LEP experiments
 has proven useful to detect shortcomings of Monte Carlo models
 in the description of these data. 
 For the HERWIG Monte Carlo this investigation already demonstrated 
 that the changed distribution of transverse momentum \kt of the photon
 remnant with respect to the direction of the incoming photon,
 the HERWIG+\kt(dyn) model, gives a better description of the LEP data.
 As the data distributions are corrected to the hadron level,
 they can be directly compared to the predictions of the 
 Monte Carlo models, without the need of detector simulation,
 and thus can be used more easily by Monte Carlo model builders.
%
%
\section*{Acknowledgements}
 We wish to thank R.~Engel and M.H.~Seymour for valuable discussions
 and useful advice concerning the use of the Monte Carlo models.
 \clearpage
%
%

\clearpage
%
%
%
\begin{table}
\begin{center}
\begin{tabular}{|c|c|c|c|} \hline
             & ALEPH          & L3             & OPAL     \\ \hline
 \Wres [GeV] &\dsdw [pb/GeV] &\dsdw [pb/GeV] &\dsdw [pb/GeV]\\ 
\hline
 $ 3-  4$&22.47$\pm$1.378$\pm$0.713&19.21$\pm$0.405$\pm$0.541&21.72$\pm$0.535$\pm$0.595\\
 $ 4-  5$&13.54$\pm$0.560$\pm$0.510&12.23$\pm$0.332$\pm$0.420&15.54$\pm$0.457$\pm$0.514\\
 $ 5-  6$&9.807$\pm$0.553$\pm$0.438&7.690$\pm$0.270$\pm$0.311&9.725$\pm$0.352$\pm$0.359\\
 $ 6-  7$&6.455$\pm$0.441$\pm$0.340&4.752$\pm$0.210$\pm$0.222&7.448$\pm$0.325$\pm$0.339\\
 $ 7-  8$&4.039$\pm$0.239$\pm$0.247&3.083$\pm$0.181$\pm$0.177&5.184$\pm$0.266$\pm$0.270\\
 $ 8-  9$&2.579$\pm$0.181$\pm$0.192&2.035$\pm$0.146$\pm$0.136&3.317$\pm$0.213$\pm$0.202\\
 $ 9- 10$&2.240$\pm$0.177$\pm$0.200&1.898$\pm$0.155$\pm$0.158&2.059$\pm$0.167$\pm$0.141\\
 $10- 11$&1.051$\pm$0.118$\pm$0.106&1.391$\pm$0.143$\pm$0.144&1.590$\pm$0.155$\pm$0.136\\
 $11- 12$&0.962$\pm$0.126$\pm$0.124&1.021$\pm$0.117$\pm$0.122&1.203$\pm$0.135$\pm$0.119\\
 $12- 13$&0.424$\pm$0.073$\pm$0.059&0.429$\pm$0.067$\pm$0.054&0.873$\pm$0.113$\pm$0.100\\
 $13- 14$&0.431$\pm$0.073$\pm$0.068&0.525$\pm$0.087$\pm$0.088&0.374$\pm$0.071$\pm$0.048\\
 $14- 15$&0.390$\pm$0.083$\pm$0.087&0.205$\pm$0.053$\pm$0.037&0.298$\pm$0.072$\pm$0.050\\
 $15- 16$&0.230$\pm$0.066$\pm$0.063&0.467$\pm$0.101$\pm$0.128&0.181$\pm$0.061$\pm$0.036\\
 $16- 17$&0.099$\pm$0.030$\pm$0.022&0.156$\pm$0.042$\pm$0.038&0.205$\pm$0.064$\pm$0.050\\
 $17- 18$&0.118$\pm$0.042$\pm$0.038&0.057$\pm$0.040$\pm$0.016&0.116$\pm$0.037$\pm$0.027\\
 $18- 19$&0.115$\pm$0.054$\pm$0.053&0.195$\pm$0.056$\pm$0.082&0.094$\pm$0.044$\pm$0.031\\
 $19- 20$&0.019$\pm$0.014$\pm$0.008&0.107$\pm$0.032$\pm$0.042&0.097$\pm$0.037$\pm$0.036\\
 $20- 21$&0.000$\pm$0.000$\pm$0.000&0.077$\pm$0.031$\pm$0.039&0.145$\pm$0.055$\pm$0.075\\
 $21- 22$&0.007$\pm$0.007$\pm$0.003&0.024$\pm$0.013$\pm$0.011&0.054$\pm$0.031$\pm$0.032\\
\hline
\end{tabular}
\caption{\label{tab:Wall}
         The individual differential cross-section \dsdw in the 
         low-\qsq region for the ALEPH, L3 and OPAL data at $\sqrt{s}=91$ GeV,
         calculated in the kinematical range defined in the text.
         The data have been corrected with HERWIG+\kt. 
         The first error listed is
         the statistical error on the data only, the second one is the
         statistical error arising from the correction factors, $f$.
        }
\end{center}
\end{table}
%
%
\begin{table}
\begin{center}
\begin{tabular}{|c|c|c|c||c|c|c|}
\hline
 & \multicolumn{3}{c||}{low-\qsq} & \multicolumn{3}{c|}{high-\qsq}\\
\hline
 HERWIG+\kt & MC & Data &     & MC & Data & \\
\hline
\Wres [GeV] & \multicolumn{2}{c|}{$\langle\dsdw\rangle$ [pb/GeV]} &
              $S$ 
            & \multicolumn{2}{c|}{$\langle\dsdw\rangle$ [pb/GeV]} &
              $S$ \\ \hline
 $ 3-  4$ & 19.97 & 20.25 $\pm$ 1.064 & 2.887 & 6.651 & 6.557 $\pm$ 0.646 & 2.142\\
 $ 4-  5$ & 14.10 & 13.41 $\pm$ 1.097 & 4.146 & 5.304 & 4.787 $\pm$ 0.370 & 1.301\\
 $ 5-  6$ & 10.53 & 8.622 $\pm$ 0.792 & 3.628 & 3.961 & 3.510 $\pm$ 0.302 & 1.448\\
 $ 6-  7$ & 7.732 & 5.669 $\pm$ 0.875 & 5.115 & 2.979 & 2.575 $\pm$ 0.219 & 1.006\\
 $ 7-  8$ & 5.718 & 3.832 $\pm$ 0.626 & 4.680 & 2.411 & 1.764 $\pm$ 0.175 & 0.419\\
 $ 8-  9$ & 4.106 & 2.485 $\pm$ 0.385 & 3.534 & 1.682 & 1.444 $\pm$ 0.165 & 0.204\\
 $ 9- 10$ & 3.222 & 2.051 $\pm$ 0.172 & 1.028 & 1.243 & 0.854 $\pm$ 0.152 & 1.841\\
 $10- 11$ & 2.323 & 1.292 $\pm$ 0.195 & 2.040 & 0.915 & 0.658 $\pm$ 0.140 & 1.813\\
 $11- 12$ & 1.707 & 1.053 $\pm$ 0.127 & 0.957 & 0.733 & 0.595 $\pm$ 0.154 & 1.537\\
 $12- 13$ & 1.224 & 0.499 $\pm$ 0.129 & 2.570 & 0.459 & 0.318 $\pm$ 0.125 & 2.530\\
 $13- 14$ & 0.926 & 0.433 $\pm$ 0.071 & 0.953 & 0.345 & 0.231 $\pm$ 0.065 & 1.322\\
 $14- 15$ & 0.653 & 0.270 $\pm$ 0.069 & 1.366 & 0.267 & 0.193 $\pm$ 0.059 & 0.967\\
 $15- 16$ & 0.469 & 0.247 $\pm$ 0.087 & 1.726 & 0.180 & 0.136 $\pm$ 0.053 & 1.362\\
 $16- 17$ & 0.345 & 0.129 $\pm$ 0.040 & 1.202 & 0.139 & 0.114 $\pm$ 0.045 & 0.455\\
 $17- 18$ & 0.259 & 0.098 $\pm$ 0.032 & 0.874 & 0.086 & 0.078 $\pm$ 0.075 & 1.995\\
 $18- 19$ & 0.168 & 0.127 $\pm$ 0.051 & 1.023 & 0.064 & 0.047 $\pm$ 0.028 & 0.359\\
 $19- 20$ & 0.117 & 0.039 $\pm$ 0.029 & 2.134 & 0.061 & 0.039 $\pm$ 0.024 & 0.388\\
 $20- 21$ & 0.087 & 0.062 $\pm$ 0.020 & 1.117 & 0.029 & 0.013 $\pm$ 0.010 & 0.561\\
 $21- 22$ & 0.057 & 0.011 $\pm$ 0.009 & 1.287 & 0.020 & 0.009 $\pm$ 0.007 & 0.127\\
\hline
\end{tabular}
\caption{\label{tab:Wave}
         The combined differential cross-section $\langle\dsdw\rangle$
         calculated in the kinematical range defined in the text 
         for the low-\qsq and high-\qsq regions.
         The data are corrected with the HERWIG+\kt model.
         The HERWIG+\kt prediction (MC) and the scale factor $S$ are 
         shown in addition.
        }
\end{center}
\end{table}
\begin{table}
\begin{center}
\begin{tabular}{|c|c|c|c||c|c|c|}
\hline
 & \multicolumn{3}{c||}{low-\qsq} & \multicolumn{3}{c|}{high-\qsq}\\
\hline
 PHOJET & MC & Data && MC & Data & \\
\hline
\Wres [GeV] &\multicolumn{2}{c|}{$\langle\dsdw\rangle$ [pb/GeV]} & $S$ 
            &\multicolumn{2}{c|}{$\langle\dsdw\rangle$ [pb/GeV]} & $S$ \\ \hline
 $ 3 -   4$ & 16.47 & 19.06 $\pm$ 0.545 & 1.555 & 4.145 & 6.266 $\pm$ 0.594 & 2.025 \\
 $ 4 -   5$ & 11.55 & 13.12 $\pm$ 0.341 & 0.978 & 3.571 & 4.580 $\pm$ 0.306 & 0.897 \\
 $ 5 -   6$ & 7.704 & 8.447 $\pm$ 0.432 & 1.975 & 3.058 & 3.244 $\pm$ 0.225 & 0.471 \\
 $ 6 -   7$ & 5.078 & 5.601 $\pm$ 0.394 & 2.266 & 2.606 & 2.819 $\pm$ 0.261 & 1.322 \\
 $ 7 -   8$ & 3.399 & 3.532 $\pm$ 0.290 & 2.271 & 2.028 & 1.740 $\pm$ 0.163 & 0.691 \\
 $ 8 -   9$ & 2.115 & 2.124 $\pm$ 0.142 & 1.270 & 1.490 & 1.339 $\pm$ 0.139 & 0.391 \\
 $ 9 -  10$ & 1.405 & 1.531 $\pm$ 0.114 & 1.103 & 0.968 & 0.950 $\pm$ 0.182 & 2.052 \\
 $10 -  11$ & 0.858 & 0.875 $\pm$ 0.087 & 1.220 & 0.680 & 0.598 $\pm$ 0.090 & 0.629 \\
 $11 -  12$ & 0.581 & 0.623 $\pm$ 0.072 & 1.193 & 0.509 & 0.363 $\pm$ 0.073 & 1.164 \\
 $12 -  13$ & 0.366 & 0.323 $\pm$ 0.104 & 3.259 & 0.314 & 0.288 $\pm$ 0.066 & 1.183 \\
 $13 -  14$ & 0.266 & 0.254 $\pm$ 0.039 & 0.985 & 0.220 & 0.175 $\pm$ 0.041 & 0.267 \\
 $14 -  15$ & 0.215 & 0.160 $\pm$ 0.030 & 0.600 & 0.152 & 0.143 $\pm$ 0.044 & 0.270 \\
 $15 -  16$ & 0.119 & 0.102 $\pm$ 0.023 & 0.939 & 0.101 & 0.098 $\pm$ 0.055 & 2.301 \\
 $16 -  17$ & 0.087 & 0.084 $\pm$ 0.022 & 0.422 & 0.079 & 0.061 $\pm$ 0.023 & 0.452 \\
 $17 -  18$ & 0.086 & 0.065 $\pm$ 0.025 & 1.281 & 0.061 & 0.051 $\pm$ 0.022 & 0.036 \\
 $18 -  19$ & 0.054 & 0.042 $\pm$ 0.024 & 1.922 & 0.039 & 0.021 $\pm$ 0.011 & 0.035 \\
 $19 -  20$ & 0.045 & 0.044 $\pm$ 0.027 & 1.895 & 0.027 & 0.023 $\pm$ 0.018 & 1.339 \\
 $20 -  21$ & 0.028 & 0.037 $\pm$ 0.016 & 0.948 & 0.038 & 0.023 $\pm$ 0.017 & 0.443 \\
 $21 -  22$ & 0.030 & 0.016 $\pm$ 0.008 & 0.693 & 0.031 & 0.000 $\pm$ 0.000 & 0.000 \\
\hline
\end{tabular}
\caption{\label{tab:Wavep}
         The combined differential cross-section $\langle\dsdw\rangle$
         calculated in the kinematical range defined in the text 
         for the low-\qsq and high-\qsq regions.
         The data are corrected with the PHOJET model.
         The PHOJET prediction (MC) and the scale factor $S$ are 
         shown in addition.
        }
\end{center}
\end{table}
\begin{table}
\begin{center}
\begin{tabular}{|c|c|c|c||c|c|c|}
\hline
 & \multicolumn{3}{c||}{low-\qsq} & \multicolumn{3}{c|}{high-\qsq}\\
\hline
 HERWIG+\kt & MC & Data && MC & Data & \\
\hline
\etout [GeV] &\multicolumn{2}{c|}{$\langle\dsde\rangle$ [pb/GeV]}& $S$ 
              &\multicolumn{2}{c|}{$\langle\dsde\rangle$ [pb/GeV]}& $S$ \\ \hline
  $0.0 -   0.5$ & 0.521 & 0.266 $\pm$ 0.064 & 3.015 & 0.213 & 0.330 $\pm$ 0.227 & 3.043 \\
  $0.5 -   1.0$ & 7.221 & 6.201 $\pm$ 0.413 & 1.929 & 2.726 & 1.955 $\pm$ 0.262 & 0.733 \\
  $1.0 -   1.5$ & 20.73 & 18.97 $\pm$ 0.875 & 2.113 & 8.155 & 6.647 $\pm$ 0.907 & 2.282 \\
  $1.5 -   2.0$ & 27.75 & 28.04 $\pm$ 1.259 & 2.132 & 10.56 & 9.095 $\pm$ 1.070 & 2.214 \\
  $2.0 -   2.5$ & 24.38 & 23.54 $\pm$ 2.567 & 5.679 & 9.519 & 9.330 $\pm$ 0.712 & 1.585 \\
  $2.5 -   3.0$ & 18.34 & 15.91 $\pm$ 2.566 & 6.627 & 6.723 & 6.075 $\pm$ 0.393 & 0.928 \\
  $3.0 -   3.5$ & 13.12 & 9.316 $\pm$ 1.170 & 3.804 & 4.831 & 4.739 $\pm$ 0.369 & 0.630 \\
  $3.5 -   4.0$ & 9.595 & 6.182 $\pm$ 1.275 & 4.722 & 3.405 & 2.196 $\pm$ 0.213 & 0.653 \\
  $4.0 -   4.5$ & 7.048 & 4.481 $\pm$ 0.473 & 1.744 & 2.360 & 1.782 $\pm$ 0.202 & 0.665 \\
  $4.5 -   5.0$ & 4.939 & 2.286 $\pm$ 0.249 & 1.214 & 1.894 & 1.357 $\pm$ 0.174 & 0.409 \\
  $5.0 -   5.5$ & 3.680 & 2.020 $\pm$ 0.431 & 2.122 & 1.142 & 0.830 $\pm$ 0.132 & 0.572 \\
  $5.5 -   6.0$ & 2.855 & 1.341 $\pm$ 0.289 & 2.149 & 0.998 & 0.750 $\pm$ 0.148 & 1.238 \\
  $6.0 -   6.5$ & 2.131 & 1.035 $\pm$ 0.225 & 1.593 & 0.600 & 0.513 $\pm$ 0.108 & 0.258 \\
  $6.5 -   7.0$ & 1.446 & 0.545 $\pm$ 0.106 & 0.661 & 0.503 & 0.385 $\pm$ 0.112 & 0.769 \\
  $7.0 -   8.0$ & 0.937 & 0.376 $\pm$ 0.062 & 1.113 & 0.321 & 0.517 $\pm$ 0.068 & 1.922 \\
  $8.0 -   9.0$ & 0.488 & 0.246 $\pm$ 0.052 & 0.596 & 0.211 & 0.288 $\pm$ 0.045 & 1.051 \\
  $9.0 -  10.0$ & 0.234 & 0.093 $\pm$ 0.027 & 0.990 & 0.079 & 0.149 $\pm$ 0.034 & 1.055 \\
 $10.0 -  11.0$ & 0.095 & 0.067 $\pm$ 0.015 & 0.894 & 0.064 & 0.104 $\pm$ 0.042 & 0.000 \\
 $11.0 -  12.0$ & 0.066 & 0.058 $\pm$ 0.031 & 0.783 & 0.034 & 0.045 $\pm$ 0.016 & 0.629 \\
 $12.0 -  13.0$ & 0.037 & 0.005 $\pm$ 0.006 & 1.011 & 0.020 & 0.000 $\pm$ 0.000 & 0.000 \\
 $13.0 -  14.0$ & 0.025 & 0.012 $\pm$ 0.014 & 1.700 & 0.012 & 0.004 $\pm$ 0.003 & 0.835 \\
 $14.0 -  15.0$ & 0.022 & 0.001 $\pm$ 0.002 & 0.516 & 0.012 & 0.000 $\pm$ 0.000 & 0.000 \\
\hline
\end{tabular}
\caption{\label{tab:heave}
         The combined differential cross-section $\langle\dsde\rangle$
         calculated in the kinematical range defined in the text 
         for the low-\qsq and high-\qsq regions.
         The data are corrected with the HERWIG+\kt model.
         The HERWIG+\kt prediction (MC) and the scale factor $S$ are 
         shown in addition.
        }
\end{center}
\end{table}
%
\begin{table}
\begin{center}
\begin{tabular}{|c|c|c|c||c|c|c|}
\hline
 & \multicolumn{3}{c||}{low-\qsq} & \multicolumn{3}{c|}{high-\qsq}  \\
\hline
 PHOJET & MC & Data & & MC & Data & \\
\hline
\etout [GeV] &\multicolumn{2}{c|}{$\langle\dsde\rangle$ [pb/GeV]}& $S$ 
             &\multicolumn{2}{c|}{$\langle\dsde\rangle$ [pb/GeV]}& $S$ \\ \hline
  $0.0 -  0.5$ & 0.674 & 0.421 $\pm$ 0.064 & 2.348 & 0.216 & 0.451 $\pm$ 0.172 & 0.583 \\
  $0.5 -  1.0$ & 7.089 & 6.401 $\pm$ 0.251 & 1.772 & 2.465 & 2.090 $\pm$ 0.313 & 0.058 \\
  $1.0 -  1.5$ & 20.92 & 18.87 $\pm$ 0.602 & 2.629 & 6.629 & 6.298 $\pm$ 0.749 & 1.853 \\
  $1.5 -  2.0$ & 25.34 & 27.80 $\pm$ 0.822 & 1.827 & 8.616 & 9.388 $\pm$ 1.092 & 2.115 \\
  $2.0 -  2.5$ & 18.55 & 24.99 $\pm$ 1.028 & 1.885 & 6.576 & 8.496 $\pm$ 0.735 & 1.706 \\
  $2.5 -  3.0$ & 10.25 & 14.93 $\pm$ 0.930 & 2.176 & 4.438 & 5.730 $\pm$ 0.433 & 0.303 \\
  $3.0 -  3.5$ & 5.931 & 8.639 $\pm$ 0.791 & 2.505 & 3.031 & 4.147 $\pm$ 0.434 & 1.352 \\
  $3.5 -  4.0$ & 3.614 & 5.654 $\pm$ 0.482 & 1.607 & 1.996 & 2.105 $\pm$ 0.238 & 0.665 \\
  $4.0 -  4.5$ & 2.343 & 3.968 $\pm$ 0.331 & 1.129 & 1.490 & 1.522 $\pm$ 0.192 & 0.724 \\
  $4.5 -  5.0$ & 1.589 & 2.063 $\pm$ 0.205 & 0.611 & 1.127 & 1.316 $\pm$ 0.205 & 0.808 \\
  $5.0 -  5.5$ & 1.160 & 1.644 $\pm$ 0.222 & 1.156 & 0.874 & 0.825 $\pm$ 0.179 & 1.404 \\
  $5.5 -  6.0$ & 0.797 & 1.081 $\pm$ 0.175 & 1.413 & 0.594 & 0.789 $\pm$ 0.200 & 1.408 \\
  $6.0 -  6.5$ & 0.585 & 0.900 $\pm$ 0.216 & 1.826 & 0.493 & 0.419 $\pm$ 0.095 & 0.220 \\
  $6.5 -  7.0$ & 0.423 & 0.373 $\pm$ 0.073 & 0.803 & 0.375 & 0.371 $\pm$ 0.122 & 0.178 \\
  $7.0 -  8.0$ & 0.330 & 0.376 $\pm$ 0.063 & 1.866 & 0.250 & 0.235 $\pm$ 0.055 & 0.758 \\
  $8.0 -  9.0$ & 0.181 & 0.147 $\pm$ 0.045 & 1.854 & 0.151 & 0.138 $\pm$ 0.031 & 0.710 \\
  $9.0 - 10.0$ & 0.114 & 0.083 $\pm$ 0.015 & 0.596 & 0.088 & 0.134 $\pm$ 0.028 & 0.000 \\
 $10.0 - 11.0$ & 0.084 & 0.072 $\pm$ 0.022 & 0.280 & 0.073 & 0.018 $\pm$ 0.012 & 0.000 \\
 $11.0 - 12.0$ & 0.051 & 0.018 $\pm$ 0.009 & 0.515 & 0.057 & 0.055 $\pm$ 0.016 & 0.000 \\
 $12.0 - 13.0$ & 0.048 & 0.013 $\pm$ 0.013 & 1.410 & 0.034 & 0.000 $\pm$ 0.000 & 0.000 \\
 $13.0 - 14.0$ & 0.024 & 0.036 $\pm$ 0.018 & 0.509 & 0.019 & 0.000 $\pm$ 0.000 & 0.000 \\
 $14.0 - 15.0$ & 0.030 & 0.012 $\pm$ 0.017 & 0.174 & 0.019 & 0.000 $\pm$ 0.000 & 0.000 \\
\hline
\end{tabular}
\caption{\label{tab:peave}
         The combined differential cross-section $\langle\dsde\rangle$
         calculated in the kinematical range defined in the text 
         for the low-\qsq and high-\qsq regions.
         The data are corrected with the PHOJET model.
         The PHOJET prediction (MC) and the scale factor $S$ are 
         shown in addition.
        }
\end{center}
\end{table}
%

\begin{table}
\begin{center}
\begin{tabular}{|c|c|c|c||c|c|c|}
\hline
 & \multicolumn{3}{c||}{low-\qsq} & \multicolumn{3}{c|}{high-\qsq} \\
\hline
 HERWIG+\kt & MC & Data & & MC & Data & \\
\hline
 \nch     & \multicolumn{2}{c|}{$\langle\dsdn\rangle$ [pb]}& $S$ 
          & \multicolumn{2}{c|}{$\langle\dsdn\rangle$ [pb]}& $S$ \\\hline
  3 & 16.10 & 13.48 $\pm$ 0.854 & 3.094 & 5.035 & 4.592 $\pm$ 0.542 & 2.196 \\
  4 & 20.26 & 15.78 $\pm$ 1.522 & 5.235 & 7.111 & 5.292 $\pm$ 0.300 & 0.727 \\
  5 & 13.96 & 12.80 $\pm$ 0.934 & 3.286 & 5.462 & 4.773 $\pm$ 0.492 & 2.090 \\
  6 & 10.16 & 8.732 $\pm$ 0.664 & 2.829 & 4.212 & 3.822 $\pm$ 0.329 & 1.395 \\
  7 & 5.546 & 4.729 $\pm$ 0.638 & 4.081 & 2.380 & 2.333 $\pm$ 0.271 & 1.464 \\
  8 & 3.396 & 2.821 $\pm$ 0.414 & 3.199 & 1.494 & 1.318 $\pm$ 0.160 & 1.082 \\
  9 & 1.904 & 1.540 $\pm$ 0.237 & 2.074 & 0.844 & 0.845 $\pm$ 0.204 & 2.072 \\
 10 & 1.121 & 0.979 $\pm$ 0.199 & 1.839 & 0.477 & 0.512 $\pm$ 0.171 & 2.069 \\
 11 & 0.642 & 0.431 $\pm$ 0.097 & 1.424 & 0.248 & 0.156 $\pm$ 0.046 & 0.229 \\
 12 & 0.346 & 0.195 $\pm$ 0.065 & 1.545 & 0.140 & 0.101 $\pm$ 0.050 & 2.046 \\
 13 & 0.180 & 0.150 $\pm$ 0.032 & 1.110 & 0.076 & 0.047 $\pm$ 0.031 & 1.490 \\
 14 & 0.083 & 0.023 $\pm$ 0.013 & 0.986 & 0.045 & 0.073 $\pm$ 0.085 & 1.203 \\
 15 & 0.058 & 0.037 $\pm$ 0.016 & 0.554 & 0.033 & 0.003 $\pm$ 0.004 & 0.500 \\
\hline
\end{tabular}
\caption{\label{tab:hnave}
         The combined differential cross-section $\langle\dsdn\rangle$
         calculated in the kinematical range defined in the text 
         for the low-\qsq and high-\qsq regions.
         The data are corrected with the HERWIG+\kt model.
         The HERWIG+\kt prediction (MC) and the scale factor $S$ are 
         shown in addition.
        }
\end{center}
\end{table}
\begin{table}
\begin{center}
\begin{tabular}{|c|c|c|c||c|c|c|}
\hline
 & \multicolumn{3}{c||}{low-\qsq} & \multicolumn{3}{c|}{high-\qsq} \\
\hline
 PHOJET & MC & Data & & MC & Data & \\
\hline
 \nch     & \multicolumn{2}{c|}{$\langle\dsdn\rangle$ [pb]}& $S$ 
          & \multicolumn{2}{c|}{$\langle\dsdn\rangle$ [pb]}& $S$ \\ \hline
  3 & 11.02 & 13.44 $\pm$ 0.893 & 3.457 & 3.514 & 4.594 $\pm$ 0.447 & 1.710 \\
  4 & 13.76 & 15.73 $\pm$ 1.022 & 3.556 & 4.800 & 5.392 $\pm$ 0.519 & 2.366 \\
  5 & 10.38 & 12.76 $\pm$ 0.536 & 1.890 & 3.878 & 4.207 $\pm$ 0.343 & 1.536 \\
  6 & 7.034 & 7.931 $\pm$ 0.713 & 3.801 & 3.202 & 3.537 $\pm$ 0.246 & 0.486 \\
  7 & 3.891 & 4.405 $\pm$ 0.443 & 3.180 & 2.025 & 2.031 $\pm$ 0.292 & 2.263 \\
  8 & 2.249 & 2.412 $\pm$ 0.321 & 3.147 & 1.319 & 1.385 $\pm$ 0.203 & 1.881 \\
  9 & 1.099 & 1.286 $\pm$ 0.172 & 2.151 & 0.735 & 0.805 $\pm$ 0.173 & 1.917 \\
 10 & 0.558 & 0.684 $\pm$ 0.136 & 2.243 & 0.340 & 0.387 $\pm$ 0.076 & 0.999 \\
 11 & 0.279 & 0.318 $\pm$ 0.054 & 1.171 & 0.179 & 0.158 $\pm$ 0.049 & 1.263 \\
 12 & 0.136 & 0.127 $\pm$ 0.035 & 1.373 & 0.091 & 0.113 $\pm$ 0.037 & 0.976 \\
 13 & 0.072 & 0.053 $\pm$ 0.042 & 2.829 & 0.056 & 0.017 $\pm$ 0.008 & 1.060 \\
 14 & 0.036 & 0.030 $\pm$ 0.017 & 0.990 & 0.024 & 0.018 $\pm$ 0.011 & 0.479 \\
 15 & 0.018 & 0.029 $\pm$ 0.014 & 1.054 & 0.013 & 0.003 $\pm$ 0.003 & 0.500 \\
 \hline
\end{tabular}
\caption{\label{tab:pnave}
         The combined differential cross-section $\langle\dsdn\rangle$
         calculated in the kinematical range defined in the text 
         for the low-\qsq and high-\qsq regions.
         The data are corrected with the PHOJET model.
         The PHOJET prediction (MC) and the scale factor $S$ are 
         shown in addition.
        }
\end{center}
\end{table}
\begin{table}
\begin{center}
\begin{tabular}{|c|c|c|c||c|c|c|}
\hline
 & \multicolumn{3}{c||}{low-\qsq} &\multicolumn{3}{c|}{high-\qsq} \\
\hline
 HERWIG+\kt & MC & Data & & MC & Data & \\
\hline
 \ptch     & \multicolumn{2}{c|}{$\langle\dsdp\rangle$ [pb/GeV]}& $S$ 
           & \multicolumn{2}{c|}{$\langle\dsdp\rangle$ [pb/GeV]}& $S$ \\ \hline
 $0.2 - 0.3$ & 757.2 & 728.1 $\pm$ 42.83 & 6.778 & 265.5 & 224.7 $\pm$ 17.29 & 3.641 \\
 $0.3 - 0.4$ & 656.1 & 600.9 $\pm$ 61.32 & 10.96 & 230.7 & 201.3 $\pm$ 6.290 & 0.850 \\
 $0.4 - 0.5$ & 509.7 & 473.4 $\pm$ 36.16 & 7.364 & 186.8 & 172.9 $\pm$ 11.53 & 2.412 \\
 $0.5 - 0.6$ & 388.7 & 355.3 $\pm$ 26.63 & 6.301 & 146.1 & 125.0 $\pm$ 13.71 & 3.996 \\
 $0.6 - 0.7$ & 292.4 & 259.5 $\pm$ 19.57 & 5.247 & 114.2 & 104.1 $\pm$ 4.512 & 0.312 \\
 $0.7 - 0.8$ & 223.5 & 181.6 $\pm$ 17.54 & 5.416 & 87.71 & 80.57 $\pm$ 8.489 & 3.043 \\
 $0.8 - 0.9$ & 173.9 & 140.3 $\pm$ 9.730 & 3.151 & 69.09 & 64.64 $\pm$ 4.254 & 1.411 \\
 $0.9 - 1.0$ & 132.8 & 98.77 $\pm$ 10.44 & 4.173 & 56.38 & 57.42 $\pm$ 3.638 & 0.200 \\
 $1.0 - 1.1$ & 106.3 & 71.11 $\pm$ 5.713 & 2.696 & 44.87 & 40.83 $\pm$ 2.890 & 0.575 \\
 $1.1 - 1.2$ & 80.99 & 53.99 $\pm$ 4.323 & 2.319 & 36.74 & 32.82 $\pm$ 2.939 & 1.399 \\
 $1.2 - 1.3$ & 68.34 & 41.70 $\pm$ 2.616 & 1.151 & 30.71 & 30.03 $\pm$ 4.872 & 3.061 \\
 $1.3 - 1.4$ & 53.41 & 27.85 $\pm$ 2.930 & 2.345 & 24.91 & 20.61 $\pm$ 2.116 & 0.283 \\
 $1.4 - 1.5$ & 44.89 & 22.39 $\pm$ 4.662 & 4.713 & 21.39 & 21.69 $\pm$ 2.352 & 0.503 \\
 $1.5 - 1.6$ & 36.04 & 18.35 $\pm$ 1.716 & 1.466 & 17.05 & 12.94 $\pm$ 1.441 & 1.035 \\
 $1.6 - 1.7$ & 30.36 & 13.68 $\pm$ 1.789 & 1.987 & 14.64 & 11.68 $\pm$ 1.488 & 0.070 \\
 $1.7 - 1.8$ & 24.71 & 10.28 $\pm$ 1.309 & 1.595 & 13.07 & 9.734 $\pm$ 1.403 & 1.220 \\
 $1.8 - 1.9$ & 20.61 & 9.879 $\pm$ 1.162 & 0.480 & 10.92 & 9.056 $\pm$ 1.368 & 0.123 \\
 $1.9 - 2.0$ & 16.67 & 7.824 $\pm$ 1.246 & 1.704 & 9.516 & 7.013 $\pm$ 1.385 & 1.310 \\
 $2.0 - 2.2$ & 13.29 & 5.296 $\pm$ 0.712 & 1.770 & 7.548 & 6.173 $\pm$ 0.787 & 0.916 \\
 $2.2 - 2.4$ & 9.261 & 3.743 $\pm$ 0.395 & 0.826 & 5.498 & 5.039 $\pm$ 0.973 & 1.574 \\
 $2.4 - 2.6$ & 6.303 & 2.671 $\pm$ 0.351 & 0.513 & 3.843 & 2.861 $\pm$ 0.624 & 1.740 \\
 $2.6 - 2.8$ & 4.399 & 1.865 $\pm$ 0.342 & 1.310 & 3.082 & 2.250 $\pm$ 0.396 & 0.096 \\
 $2.8 - 3.0$ & 3.155 & 1.248 $\pm$ 0.348 & 1.947 & 2.365 & 1.273 $\pm$ 0.253 & 0.306 \\
 $3.0 - 3.4$ & 1.962 & 0.778 $\pm$ 0.212 & 2.669 & 1.471 & 1.620 $\pm$ 0.163 & 0.982 \\
 $3.4 - 3.8$ & 1.087 & 0.617 $\pm$ 0.152 & 2.685 & 0.868 & 0.832 $\pm$ 0.129 & 0.785 \\
 $3.8 - 4.2$ & 0.651 & 0.516 $\pm$ 0.077 & 0.718 & 0.524 & 0.295 $\pm$ 0.078 & 1.555 \\
 $4.2 - 4.6$ & 0.344 & 0.234 $\pm$ 0.210 & 9.881 & 0.245 & 0.199 $\pm$ 0.090 & 0.000 \\
 $4.6 - 5.0$ & 0.146 & 0.101 $\pm$ 0.032 & 1.584 & 0.128 & 0.107 $\pm$ 0.072 & 0.000 \\
 \hline 
\end{tabular}
\caption{\label{tab:ppave}
         The combined differential cross-section $\langle\dsdp\rangle$
         calculated in the kinematical range defined in the text 
         for the low-\qsq and high-\qsq regions.
         The data are corrected with the HERWIG+\kt model.
         The HERWIG+\kt prediction (MC) and the scale factor $S$ are 
         shown in addition.
        }
\end{center}
\end{table}
\begin{table}
\begin{center}
\begin{tabular}{|c|c|c|c||c|c|c|}
\hline
 & \multicolumn{3}{c||}{low-\qsq} & \multicolumn{3}{c|}{high-\qsq} \\
\hline
 PHOJET & MC & Data & & MC & Data & \\
\hline
 \ptch     & \multicolumn{2}{c|}{$\langle\dsdp\rangle$ [pb/GeV]}& $S$ 
           & \multicolumn{2}{c|}{$\langle\dsdp\rangle$ [pb/GeV]}& $S$ \\\hline
 $0.2 - 0.3$ & 557.9 & 667.9 $\pm$ 49.08 & 8.647 & 197.6 & 213.8 $\pm$ 6.298 & 0.668 \\
 $0.3 - 0.4$ & 489.3 & 564.2 $\pm$ 47.42 & 9.111 & 173.6 & 201.1 $\pm$ 6.132 & 0.664 \\
 $0.4 - 0.5$ & 382.5 & 443.9 $\pm$ 35.90 & 7.937 & 141.2 & 160.3 $\pm$ 9.387 & 2.093 \\
 $0.5 - 0.6$ & 287.8 & 332.1 $\pm$ 27.61 & 7.165 & 112.3 & 118.8 $\pm$ 11.76 & 3.603 \\
 $0.6 - 0.7$ & 208.2 & 234.7 $\pm$ 11.73 & 3.522 & 87.50 & 95.93 $\pm$ 3.904 & 0.326 \\
 $0.7 - 0.8$ & 150.8 & 171.6 $\pm$ 10.26 & 3.371 & 68.66 & 76.02 $\pm$ 5.770 & 2.088 \\
 $0.8 - 0.9$ & 106.8 & 129.5 $\pm$ 6.614 & 2.390 & 54.16 & 61.77 $\pm$ 3.295 & 0.532 \\
 $0.9 - 1.0$ & 75.99 & 90.52 $\pm$ 6.095 & 2.669 & 40.57 & 54.56 $\pm$ 4.233 & 1.688 \\
 $1.0 - 1.1$ & 55.68 & 63.55 $\pm$ 3.961 & 2.163 & 32.59 & 38.27 $\pm$ 2.528 & 0.229 \\
 $1.1 - 1.2$ & 41.27 & 51.43 $\pm$ 2.551 & 1.361 & 26.29 & 32.04 $\pm$ 2.386 & 0.662 \\
 $1.2 - 1.3$ & 30.74 & 40.34 $\pm$ 2.822 & 1.720 & 21.22 & 26.86 $\pm$ 2.176 & 0.460 \\
 $1.3 - 1.4$ & 22.78 & 27.83 $\pm$ 2.839 & 2.432 & 18.44 & 18.60 $\pm$ 2.266 & 1.494 \\
 $1.4 - 1.5$ & 17.09 & 19.12 $\pm$ 2.860 & 3.431 & 15.41 & 16.47 $\pm$ 1.612 & 0.839 \\
 $1.5 - 1.6$ & 12.58 & 16.07 $\pm$ 1.555 & 1.762 & 12.82 & 15.19 $\pm$ 1.650 & 0.368 \\
 $1.6 - 1.7$ & 10.34 & 11.44 $\pm$ 1.734 & 2.523 & 11.45 & 10.29 $\pm$ 1.270 & 1.136 \\
 $1.7 - 1.8$ & 7.549 & 8.146 $\pm$ 1.307 & 2.333 & 8.789 & 8.366 $\pm$ 1.044 & 0.731 \\
 $1.8 - 1.9$ & 6.152 & 6.864 $\pm$ 0.971 & 1.607 & 7.670 & 8.055 $\pm$ 1.137 & 0.833 \\
 $1.9 - 2.0$ & 5.550 & 6.588 $\pm$ 0.919 & 1.562 & 6.538 & 6.632 $\pm$ 1.074 & 0.490 \\
 $2.0 - 2.2$ & 3.853 & 4.250 $\pm$ 0.693 & 2.354 & 4.949 & 4.814 $\pm$ 0.574 & 0.251 \\
 $2.2 - 2.4$ & 2.499 & 3.178 $\pm$ 0.569 & 2.400 & 3.859 & 3.298 $\pm$ 0.455 & 0.594 \\
 $2.4 - 2.6$ & 1.698 & 1.950 $\pm$ 0.246 & 0.540 & 2.703 & 2.664 $\pm$ 0.715 & 2.316 \\
 $2.6 - 2.8$ & 1.221 & 1.371 $\pm$ 0.211 & 1.124 & 1.818 & 1.784 $\pm$ 0.321 & 0.359 \\
 $2.8 - 3.0$ & 0.831 & 0.752 $\pm$ 0.137 & 1.628 & 1.536 & 1.410 $\pm$ 0.602 & 3.171 \\
 $3.0 - 3.4$ & 0.545 & 0.580 $\pm$ 0.179 & 2.801 & 0.990 & 1.239 $\pm$ 0.428 & 3.554 \\
 $3.4 - 3.8$ & 0.337 & 0.280 $\pm$ 0.095 & 2.388 & 0.545 & 0.779 $\pm$ 0.141 & 1.407 \\
 $3.8 - 4.2$ & 0.211 & 0.189 $\pm$ 0.036 & 2.066 & 0.274 & 0.649 $\pm$ 0.093 & 0.000 \\
 $4.2 - 4.6$ & 0.132 & 0.123 $\pm$ 0.075 & 2.943 & 0.187 & 0.097 $\pm$ 0.235 & 0.000 \\
 $4.6 - 5.0$ & 0.084 & 0.079 $\pm$ 0.050 & 3.143 & 0.090 & 0.009 $\pm$ 0.065 & 0.000 \\
\hline
\end{tabular}
\caption{\label{tab:hpave}
         The combined differential cross-section $\langle\dsdp\rangle$
         calculated in the kinematical range defined in the text 
         for the low-\qsq and high-\qsq regions.
         The data are corrected with the PHOJET model.
         The PHOJET prediction (MC) and the scale factor $S$ are 
         shown in addition.
        }
\end{center}
\end{table}
\begin{table}
\begin{center}
\begin{tabular}{|c|c|c|c||c|c|c|}
\hline
 & \multicolumn{3}{c||}{low-\qsq} & \multicolumn{3}{c|}{high-\qsq}  \\
\hline
 HERWIG+\kt & MC & Data & & MC & Data &  \\
\hline
 $\eta$   & \multicolumn{2}{c|}{$\langle\flow\rangle$[GeV]}& $S$ 
          & \multicolumn{2}{c|}{$\langle\flow\rangle$[GeV]}& $S$ \\\hline
 $-3.0 - -2.5$ & 0.264 & 0.281 $\pm$ 0.054 & 0.000 & 0.147 & 0.076 $\pm$ 0.066 & 2.947 \\
 $-2.5 - -2.0$ & 0.605 & 1.010 $\pm$ 0.054 & 2.364 & 0.432 & 0.651 $\pm$ 0.108 & 2.759 \\
 $-2.0 - -1.5$ & 1.477 & 1.715 $\pm$ 0.078 & 0.743 & 1.332 & 1.449 $\pm$ 0.096 & 1.119 \\
 $-1.5 - -1.0$ & 2.290 & 2.225 $\pm$ 0.056 & 1.451 & 2.516 & 2.221 $\pm$ 0.088 & 0.352 \\
 $-1.0 - -0.5$ & 2.164 & 2.077 $\pm$ 0.043 & 0.446 & 2.616 & 2.366 $\pm$ 0.076 & 2.040 \\
 $-0.5 - -0.0$ & 1.879 & 1.794 $\pm$ 0.033 & 0.885 & 2.362 & 2.365 $\pm$ 0.100 & 0.915 \\
 $\pz 0.0 - \pz 0.5$ & 1.891 & 1.707 $\pm$ 0.032 & 0.554 & 2.304 & 2.383 $\pm$ 0.135 & 3.555 \\
 $\pz 0.5 - \pz 1.0$ & 2.252 & 1.993 $\pm$ 0.045 & 1.492 & 2.533 & 2.602 $\pm$ 0.202 & 4.606 \\
 $\pz 1.0 - \pz 1.5$ & 3.028 & 2.559 $\pm$ 0.120 & 1.770 & 3.169 & 3.054 $\pm$ 0.372 & 3.627 \\
 $\pz 1.5 - \pz 2.0$ & 3.379 & 2.754 $\pm$ 0.226 & 4.235 & 3.337 & 2.879 $\pm$ 0.369 & 8.051 \\
 $\pz 2.0 - \pz 2.5$ & 3.100 & 1.780 $\pm$ 0.256 & 6.335 & 3.080 & 1.888 $\pm$ 0.321 & 1.702 \\
 $\pz 2.5 - \pz 3.0$ & 2.808 & 1.063 $\pm$ 0.319 & 3.643 & 2.735 & 1.059 $\pm$ 0.150 & 1.018 \\
 $\pz 3.0 - \pz 3.5$ & 1.915 & 1.267 $\pm$ 0.254 & 8.168 & 1.918 & 1.496 $\pm$ 0.209 & 0.825 \\
 $\pz 3.5 - \pz 4.0$ & 1.122 & 0.992 $\pm$ 0.249 & 4.124 & 1.089 & 0.528 $\pm$ 0.133 & 3.557 \\
 $\pz 4.0 - \pz 4.5$ & 0.579 & 0.634 $\pm$ 0.144 & 2.831 & 0.548 & 0.292 $\pm$ 0.100 & 1.992 \\
 \hline
\end{tabular}
\caption{\label{tab:hefave}
         The combined energy flow $\langle\flow\rangle$ 
         calculated in the kinematical range defined in the text 
         for the low-\qsq and high-\qsq regions.
         The data are corrected with the HERWIG+\kt model.
         The HERWIG+\kt prediction (MC) and the scale factor $S$ are 
         shown in addition.
        }
\end{center}
\end{table}
\begin{table}
\begin{center}
\begin{tabular}{|c|c|c|c||c|c|c|}
\hline
 & \multicolumn{3}{c||}{low-\qsq}  & \multicolumn{3}{c|}{high-\qsq}  \\
\hline
 PHOJET & MC & Data & & MC & Data & \\
\hline
 $\eta$   & \multicolumn{2}{c|}{$\langle\flow\rangle$[GeV]}& $S$ 
          & \multicolumn{2}{c|}{$\langle\flow\rangle$[GeV]}& $S$ \\ \hline
 $-3.0 -  -2.5$ & 0.511 & 0.248 $\pm$ 0.030 & 0.245 & 0.176 & 0.055 $\pm$ 0.043 & 2.748 \\
 $-2.5 -  -2.0$ & 1.064 & 1.218 $\pm$ 0.046 & 0.903 & 0.578 & 0.606 $\pm$ 0.098 & 4.332 \\
 $-2.0 -  -1.5$ & 1.836 & 1.879 $\pm$ 0.058 & 1.028 & 1.739 & 1.507 $\pm$ 0.117 & 0.041 \\
 $-1.5 -  -1.0$ & 2.170 & 2.165 $\pm$ 0.046 & 1.457 & 2.683 & 2.353 $\pm$ 0.109 & 2.151 \\
 $-1.0 -  -0.5$ & 1.835 & 1.974 $\pm$ 0.043 & 1.576 & 2.580 & 2.456 $\pm$ 0.107 & 3.213 \\
 $-0.5 -  -0.0$ & 1.601 & 1.771 $\pm$ 0.033 & 1.268 & 2.312 & 2.419 $\pm$ 0.131 & 1.085 \\
 $\pz 0.0 -  \pz 0.5$ & 1.604 & 1.730 $\pm$ 0.047 & 1.715 & 2.260 & 2.348 $\pm$ 0.156 & 4.633 \\
 $\pz 0.5 -  \pz 1.0$ & 1.941 & 1.983 $\pm$ 0.063 & 3.785 & 2.529 & 2.429 $\pm$ 0.234 & 5.302 \\
 $\pz 1.0 -  \pz 1.5$ & 2.503 & 2.340 $\pm$ 0.139 & 2.296 & 3.080 & 2.523 $\pm$ 0.348 & 5.272 \\
 $\pz 1.5 -  \pz 2.0$ & 2.770 & 2.108 $\pm$ 0.275 & 7.846 & 3.277 & 2.349 $\pm$ 0.371 & 7.974 \\
 $\pz 2.0 -  \pz 2.5$ & 2.718 & 1.707 $\pm$ 0.284 & 10.74 & 3.109 & 1.707 $\pm$ 0.312 & 2.817 \\
 $\pz 2.5 -  \pz 3.0$ & 2.772 & 1.046 $\pm$ 0.337 & 2.135 & 3.061 & 0.934 $\pm$ 0.175 & 1.052 \\
 $\pz 3.0 -  \pz 3.5$ & 2.565 & 1.536 $\pm$ 0.187 & 6.334 & 2.762 & 1.446 $\pm$ 0.196 & 1.249 \\
 $\pz 3.5 -  \pz 4.0$ & 1.929 & 1.467 $\pm$ 0.164 & 1.052 & 2.150 & 0.821 $\pm$ 0.207 & 2.550 \\
 $\pz 4.0 -  \pz 4.5$ & 1.193 & 1.065 $\pm$ 0.090 & 0.393 & 1.217 & 0.411 $\pm$ 0.223 & 3.190 \\
\hline
\end{tabular}
\caption{\label{tab:pefave}
         The combined energy flow $\langle\flow\rangle$
         calculated in the kinematical range defined in the text 
         for the low-\qsq and high-\qsq regions.
         The data are corrected with the PHOJET model.
         The PHOJET prediction (MC) and the scale factor $S$ are 
         shown in addition.
        }
\end{center}
\end{table}
%
%
%
%
%
%
%
%
%
%
%
%
\clearpage
%
%
%
%
\clearpage
%
%
%
%
\clearpage
%
%
%
%
\clearpage
%
%
%
%
%
\clearpage
%
%
\clearpage
%
%
\clearpage
%
%
\clearpage
%
%
%
%
%
%
\end{document}